\documentclass[useAMS,usenatbib]{mn2e}
\usepackage{graphicx}

\def\la{\raise.5ex\hbox{$<$}\kern-.8em\lower 1mm\hbox{$\sim$}}
\def\ma{\raise.5ex\hbox{$>$}\kern-.8em\lower 1mm\hbox{$\sim$}}

\def\kms{$\rm km\, s^{-1}$}
\def\cm3{$\rm cm^{-3}$}
\def\Ts{$\rm T_{*}$~}
\def\Vs{$\rm V_{s}$~}
\def\n0{$\rm n_{0}$}
\def\B0{$\rm B_{0}$}

\def\Ne{$\rm N_{e}$}

\def\erg{$\rm erg\, cm^{-2}\, s^{-1}$}
\def\mum{$\mu$m~}

\def\L12{L$_{12\mu m}$~}
\def\F12{F$_{12\mu m}$~}
\def\agr{a$_{gr}$}
\def\Hb{H${\beta}$~}
\def\Ha{H${\alpha}$~}
\def\Hg{H${\gamma}$~}

\def\ff{{\it ff}}

\def\RO3{R$_{[OIII]}$}

\title[SED of GRB, SLSN, SB and AGN hosts]{Continuum spectral energy distribution of  GRB, SLSN, SB and AGN host galaxies at intermediate redshifts 
}
\author[M. Contini]{ M. Contini
\\
School of Physics and Astronomy, Tel Aviv University, Tel Aviv
69978, Israel \\
}

\begin{document}

%\date{Accepted: Received ; in original form 2010 month day}

\pagerange{\pageref{firstpage}--\pageref{lastpage}} \pubyear{2009}

\maketitle

\label{firstpage}

\begin{abstract}

Continuum SED models  of gamma-ray burst (GRB) and  obscured GRB host galaxies  at moderately high redshifts
are presented and compared with those  of superluminous supernovae (SLSN), starburst (SB) and active galactic nuclei (AGN).
 We consider that  continuum radiation (bremsstrahlung) is emitted  from the same  clouds which emit
the line spectrum in each object. 
Therefore, we have selected from the samples of  the GRB host  continuum observations  those that
were  previously modelled on the basis of the line spectra, because modelling the continuum SED
is less constraining. 
The bremsstrahlung is generally recognised in the radio  and in the UV-X-ray frequency ranges, while
dust reradiation peaks in the IR.
We have found that GRB980703 host extended clouds  have  dust-to-gas ratio $d/g$=0.03, while for GRB980425 
  $d/g$ $<$ 0.0001.  
To reproduce the continuum SED of most  of the GRB, SLSN, SB and AGN  in the near-IR-optical range, 
 the  contribution  of an old star background  population is needed.   
This  radiation can be reproduced by a  black body (bb) corresponding to   temperatures  T$_{bb}$ $\sim$3000-8000 K.
The   best fit  of a few host SEDs
 includes also  the direct contribution of the bb  flux from the SB corresponding to \Ts $\sim$5$\times$10$^4$K. 
 $d/g$  calculated by modelling  the SEDs of obscured GRB hosts
roughly increases with z    resembling   the SFR trend.

\end{abstract}

\begin{keywords}
radiation mechanisms: general --- shock waves --- 
 galaxies: GRB host  --- SN host --- starburst--- AGN ---galaxies: high redshift

\end{keywords}

\section{Introduction}

Gamma-ray bursts (GRB) are explosions characterized by  huge energy release.  
GRB events  generally occur   at  redshifts higher than local.
  Their  characteristics are  analysed by the interpretation of phenomena 
connected with  the  explosion (light curves, ejecta, emitted and absorbed
spectra, etc)  and those  relative to the host galaxy, in particular the emitted spectra.
The  physical conditions and the element abundances of the host galaxies are investigated
from all points of view such as star formation, star formation rates
(SFR), star  chemical nucleosynthesis
and their evolution with time  (Kr\"{u}hler et al. 2015, Han et al. 2010,
Levesque et al. 2010, Savaglio et al. 2009, Contini 2016, 2017a,b etc).
In previous works (Contini 2017b and references therein) we have investigated  
  physical conditions and  element abundances of some basic
elements (H, N, O, S, etc) throughout the hosts  comparing model calculations
 with the line spectra emitted from galaxies hosting long and short period GRB,  supernovae
(SN) of different types, active galactic nuclei (AGN), starbursts (SB), 
HII regions. Some peculiarities were  revealed 
 e.g.  the interplay between oxygen and nitrogen   release  throughout the hosts  at different redshifts.

In this paper we refer to the continuum spectral energy distribution (SED) observed from GRB host galaxies
at moderately high redshifts, 
in particular those from the obscured ones and we compare them  with SLSN of different types.
The continua  are   generally investigated using
 population synthesis models of  more host galaxies to create a set of theoretical SEDs
 (e.g. Sokolov et  2001). 
We will start our study by the analysis of Hunt et al (2014) sample of obscured GRB hosts because  they present 
the continuum data on a large frequency range (from radio to NIR).
To reproduce the continuum SED Hunt et al (2014) applied the  fitting method introduced by Michalowski et al (2009, 2010) based on 
35000 templates in the library of Iglesias-Paramo et al (2007) plus templates of Silva et al (1998).
Kr\"{u}hler et al (2011)  reported the SEDs of  GRB dusty host galaxies.
The UV/optical/NIR photometry of the selected GRB hosts were analysed  in a standard way
using stellar population synthesis techniques to convert luminosities  into stellar masses.
 Perley et al (2016) analyzed the UV-optical-NIR SEDs of SLSN of different types to estimate host galaxy 
stellar parameters using a custom SED-fitting code (Perley et al 2013). 
To compare the GRB SEDs with those of SB and AGN hosts at relatively high redshifts,
we will refer to the 
Ramos Almeida et al (2013) sample presented  in the frame of AGN-SB feedback investigation. They
adopted the diagnostic SEDs  based on different types of galaxies (AGN type 1, type 2,  SB, etc., Polletta et al 2007).
The  diagnostics  result   from averages  on   hundreds of objects.

Nevertheless, star formation within galaxies at  various redshifts  depends also on
the interstellar medium (Spaans \& Carollo 1997) due to gas and dust heating  by   the
radiation flux  and  to compression by, e.g.    supernovae and stellar winds.
During  mergers of spiral galaxies at high redshifts (Springel  et al 2005)
the  collision and mixing of galaxy debris trigger nuclear gas inflow, which leads to
  energetic starbursts and  black hole accretion.
The gas photoionised and heated by   radiation from
 the SB,  by the AGN power-law flux and  by collisional process 
within the galaxies emits the line spectrum  as well as the continuum.
However, the continuum  emitted from the  gaseous and dusty clouds  within the host galaxy is
generally neglected in the  modelling of the  SED.

In this paper we  would like to demonstrate that gas and dust  contributions  from  clouds within the 
host galaxies have a leading role  to explain the observed continuum SEDs on a large frequency range.
We  focus on the continuum (by  free-free and free-bound radiation) emitted from the 
gas    inside the host clouds.  The physical conditions
and the element relative abundances   are revealed by the detailed modelling of the line ratios, while
the continuum SED permits to recognize  directly  the radiation sources dominating in the different
frequency ranges as well as the background stellar populations.
The  interpretation of the line ratios constrains the models, while
  the modelling of the  continuum SED  is   less straightforward, because
the   contributions from different radiation sources (e.g. background stars,
dust grains at different temperatures, synchrotron radiation, etc) overlap  within close
frequency ranges;  some of them  are  uncertain and    not well disentangled.
In this paper we   model the observed SED  consistently with  the line spectra
 by  the  code  {\sc suma} which has been successfully used to calculate  line and continuum  spectra 
from  clouds under the coupled effect of shock and photoionization in different cases.
Dust reprocessed radiation is consistently calculated by the code.
Shocks are significant because  most galaxies at high and low redshifts are the product of merging.
Therefore,   we will adopt  for the   analysis of the   continuum  SED  of each object
the same model as that  previously  calculated to reproduce  the line fluxes emitted from host galaxy.
Comparing the calculated continuum   with  the photometric data observed in the different wavelength domains,
(from radio to X-ray, when available)  the old star background contribution  is expected to emerge. 
The GRB host samples presented in the following are relatively poor because  few objects appear in both
line and continuum  samples. The SLSN host sample
presented by Perley et al (2016) is relatively abundant in number of objects because  the observations contain  both
spectroscopic and photometric data. The same occurs for the Ramos Almeida et al spectra. This sample  has been chosen
because in each galaxy a SB and/or an AGN  are disclosed by modelling
the line and continuum spectra.

To reproduce the Kr\"{u}hler et al (2011)  observed  SEDs we    select the objects  which appear
in  the Kr\"{u}hler et al (2015) sample of line
 spectra  that were analysed  in detail  by Contini (2016). For the Hunt et al (2014)  SEDs
we  use the models calculated by Contini (2016, 2017a)  for the  
Han et al (2010), Sollerman et al (2005), Kr\"{u}hler et al (2015), Levesque et al(2010),
Graham \& Fruchter( 2013) and Michalowski et al. (2014) line spectra.
In Sect. 2 we briefly describe the calculation process. In Sect. 3  the results of  modelling
 Hunt et al,  Kr\"{u}hler et al  and Sokolov et al  observed continuum SEDs are discussed
and compared with  those obtained for the  Perley et al SLSN host galaxy sample  and for the SB-AGN  survey 
presented by  Ramos Almeida et al.  Concluding remarks follow in Sect. 4.

\section{About the calculation code}

We use composite models which account consistently for   
photoionization and shocks.  The code {\sc suma}  is adopted.
The main input parameters are those which   are used for the  
calculations of the line and continuum fluxes.
They account for photoionization and heating by  primary and secondary  radiation and collisional  
process due to shocks.
The input parameters such as  the shock velocity \Vs, the atomic  
preshock density \n0 and the preshock
magnetic field \B0 (for  all models \B0=10$^{-4}$Gauss is adopted)   
define the hydrodynamical field.
They  are  used in the calculations  of the Rankine-Hugoniot equations  
  at the shock front and downstream
and  are combined in the compression equation which is resolved   
throughout each slab of the gas
in order to obtain the density profile downstream.
Primary radiation for SB in the GRB host galaxies is approximated by a black-body (bb).
The input parameters    are  the effective temperature \Ts and
the ionization parameter $U$.
 For an AGN, the primary radiation is the power-law radiation
flux  from the active centre $F$  in number of photons cm$^{-2}$ s$^{-1}$ eV$^{-1}$ at the Lyman limit
and  spectral indices  $\alpha_{UV}$=-1.5 and $\alpha_X$=-0.7. The primary radiation source
 does not depend on the host physical condition but it affects the surrounding gas.   This  region  is not considered
as a unique cloud, but as a  sequence of slabs with different thickness calculated automatically
following the temperature gradient. The secondary diffuse radiation is emitted
from the slabs of 
gas heated  by the radiation flux reaching the gas and by the shock.
Primary and secondary radiation are calculated by radiation transfer.

In our model the line and continuum emitting  region throughout the galaxy covers  an ensemble of fragmented clouds.
The geometrical thickness of the clouds is  an input parameter of the code ($D$) which is  calculated
consistently with the physical conditions and element abundances of the emitting gas.
The fractional abundances of the ions are calculated resolving the ionization equations
for each element (H, He, C, N, O, Ne, Mg, Si, S, Ar, Cl, Fe) in each ionization level.
Then, the calculated line ratios, integrated throughout the cloud thickness, are compared with the 
observed ones. The calculation process is repeated
changing  the input parameters until the observed data are reproduced by the model results, within 10 percent
for the strongest line ratios and within 50 percent for the weakest ones.

However,  some parameters regarding the continuum SED, such as the dust-to-gas  ratio $d/g$  and the dust grain radius
\agr ~ are not directly constrained by fitting the line ratios.
Dust grains are heated by the primary radiation and by mutual collision with  atoms.
The intensity of dust reprocessed radiation in the IR  depends on $d/g$ and \agr.
In this work we use $d/g$=10$^{-14}$ by number for all the models which corresponds to
4.1 10$^{-4}$ by mass for silicates (Draine \& Lee 1994).
The distribution of the grain size along the cloud starting from an initial radius 
is automatically  derived by {\sc suma},
 which calculates sputtering of the grains  in the different zones downstream of the shock. 
The sputtering rate depends on the gas temperature, which is $\propto$ V$_s^2$ in the immediate post-shock region. 
In the high-velocity case (\Vs$\geq$ 500 \kms) the sputtering rate is so high that the grains 
with \agr $\leq$0.1 \mum are rapidly destroyed downstream. 
So, only grains with large radius (\agr $\geq$0.1 \mum) will survive.
On the other hand, the grains survive downstream of low-velocity shocks ($<$200\kms). 
Graphite grains are more sputtered than silicate grains for T= 10$^6$ K (Draine \& Salpeter 1979). 
Small grains (e.g. PAH) survive in the extended galactic regions on scales of  hundred parsecs and lead to the 
characteristic features that appear in the SED.
In conclusion, cold dust or cirrus emission results from heating by the  interstellar radiation field, 
warm dust is associated with star formation regions and hot dust appears around  AGN (Helou 1986) and in high velocity shock regimes. 
Therefore, we will consider relatively large grains, e.g. silicate grains with an initial radius of 0.1 -1.0 \mum. 

In the radio range the power-law spectrum of synchrotron radiation
created by the Fermi mechanism at the shock front is seen in most galaxies.
It is calculated by {\sc suma} adopting a  spectral index of -0.75 (Bell 1977).

\section{Analysis of the SED}

   The  models constrained by the fit of the line spectra
   give a hint about the relative importance of the different  
ionization and heating mechanisms which  are recognised throughout the continuum SED
in each of the objects. In particular :

1) The black body radiation corresponding directly to the temperature  
dominating in the starburst  is seldom observed  in the
UV, because  absorption  is very strong in this frequency range
due  to  strong line formation.

2) The shock effect throughout the SED can be recognized from  the maximum frequency and  
intensity of the dust reprocessed radiation peak
in the infrared and of the bremsstrahlung   at high frequencies   
(see Contini, Viegas \& Prieto 2004, Contini \& Viegas 2000).

3) The gas ionized by the  SB (or AGN) radiation flux  emits  bremsstrahlung  
from  radio to  X-ray.
The black body emission from the background old star  population with  
T$_{bb}$$\sim$ 3000-8000 K  generally emerges
over  the bremsstrahlung  throughout the   SED in the near-IR(NIR) - optical range.

4) In the radio range synchrotron radiation created by the Fermi  
mechanism  is recognized by its spectral index.
   Thermal bremsstrahlung in the radio range has a steeper trend which becomes even steeper
by self-absorption at low $\nu$ (Sect. 3.5).
In the far-IR only comparison with the observation data  indicates the source of the continuum 
radiation flux, because  thermal bremsstrahlung, synchrotron radio and cold dust reradiation may be 
blended.

 Figs.  1-6 show the  calculated SEDs which best  fit  the  data. 
  In  all the diagrams two curves appear for each model.
The continuum calculated by the models (which  reproduce also  the observed line  ratios)   
in the radio-X-ray range
 refers to free-free and free-bound radiation (hereafter  addressed  
to as bremsstrahlung).
In the IR range  dust reprocessed radiation dominates. 
The bremsstrahlung  at $\nu$$<$ 10$^{14}$ Hz has a similar slope in  
all the diagrams.
In fact, the
bremsstrahlung continuum, emitted by free electrons accelerated in  
Coulomb collisions with positive ions
(mostly H$^+$, He$^+$ and He$^{++}$) in nebulae of charge Z has an  
emission coefficient (Osterbrock 1974):

J$_{\nu}$ $\propto$\Ne N$_+$Z$^2$($\pi h\nu$/3kT)$^{1/2}$ e$^{-(h\nu/kT)}$    (1)

\noindent
The  photoionization radiation flux can heat the gas to T$\sim$ 2-4$\times$ 10$^4$ K, while the  
gas is heated  collisionally by
the shock to a maximum of T= 1.5$\times$ 10$^5$ (\Vs /100 \kms)$^2$ K, where  
\Vs is the shock velocity.
The cooling rate downstream  depends on  \Ne N$_+$ (N$_+$ is the proton density).
The trend of the bremsstrahlung as function of $\nu$ depends from the  
interplay between T  and $\nu$.
High temperatures of the emitting gas  determine the maximum bremsstrahlung at high $\nu$.
At T$\sim$1-4$\times$ 10$^4$ K the exponential term is significant 
at frequencies between  10$^{14}$ and 10$^{15}$ Hz.
The  temperatures  are calculated by thermal balancing between the heating rates which depend on the
photoionizing flux and  the cooling rates by free-free, free-bound and line emission. 
Therefore,  the radiation effect is seen mainly in this frequency  range.
In the radio range,   the exponent  in eq (1) 
tends to 0 and  the continuum is $\propto \nu^{1/2}$  (see Figs 1-6). 
So the SEDs in all the  diagrams (Figs. 1-6) of all  galaxy types have  similar trends  at 
relatively low frequencies and  the dust reprocessed radiation bump in the IR is clearly recognizable.

\subsection{The Hunt et al (2014) GRB host sample. Herschel observations}

\begin{table}
\centering
\caption{Models of  line spectra for  selected Hunt et al (2014) GRB obscured host galaxies}
\begin{tabular}{lccccccccccc} \hline  \hline
\ GRB   & z    &   \Vs       & \n0      &$D$    &      \Ts   &    $U$        \\
\       &      & \kms        & \cm3     &pc     & 10$^4$K    &  -          \\ \hline
\ 980703$^1$& 0.966& 190     &  50      &10     & 34         &1.2           \\
\ 980425$^2$&0.0085&120      &150       &0.216  &6.5         &0.01           \\
\ 080207$^3$& 2.086& 324     & 80       &0.367  &9.7         &0.04           \\
\ 070306$^3$& 1.496& 280     &150       &0.367  &7.0         &0.05          \\
\ 051022$^4$& 0.807& 120     &150       &0.23   &6.5         &0.01          \\ \hline
%\ 020819B$^5$&0.411&120      &100       &0.367  &4.1         &0.0086        \\ \hline
\end{tabular}

$^1$  Han et al (2010);
$^2$  Sollerman et al (2005);
$^3$  Kr\"{u}hler (2015);
$^4$  Levesque et al (2010);

\centering
\caption{Modelling  Hunt  et al (2014) SEDs}
\begin{tabular}{lccccccccccc} \hline  \hline
\ GRB    &    d     & r     &$d/g$     &$\eta_g$ &  $\eta_d$   & T$_{bb}$   \\
\        &  Mpc     &  kpc  &0.0004    &      -    &     -          &  1000  K  \\  \hline
\ 980703 & 6419.4   & 0.906 &79.       &-13.7      & -11.8          &  -   \\
\ 980425 & 36.9     & 0.826 &$<$0.2    &-9.3       & -10.3          & 6    \\
\ 080207 &16687.1   &  1.3  &5.        &-14.2      &-13.5           & 1    \\
\ 070306 &11052.3   &  4.93 &25.       &-12.7      &-11.3           & 8      \\
\ 051022 &5137.6    & 0.13  &0.56      &-11.3      &-11.55          & -     \\ \hline
\end{tabular}

\end{table}

\begin{figure*}
\centering
\includegraphics[width=8.6cm]{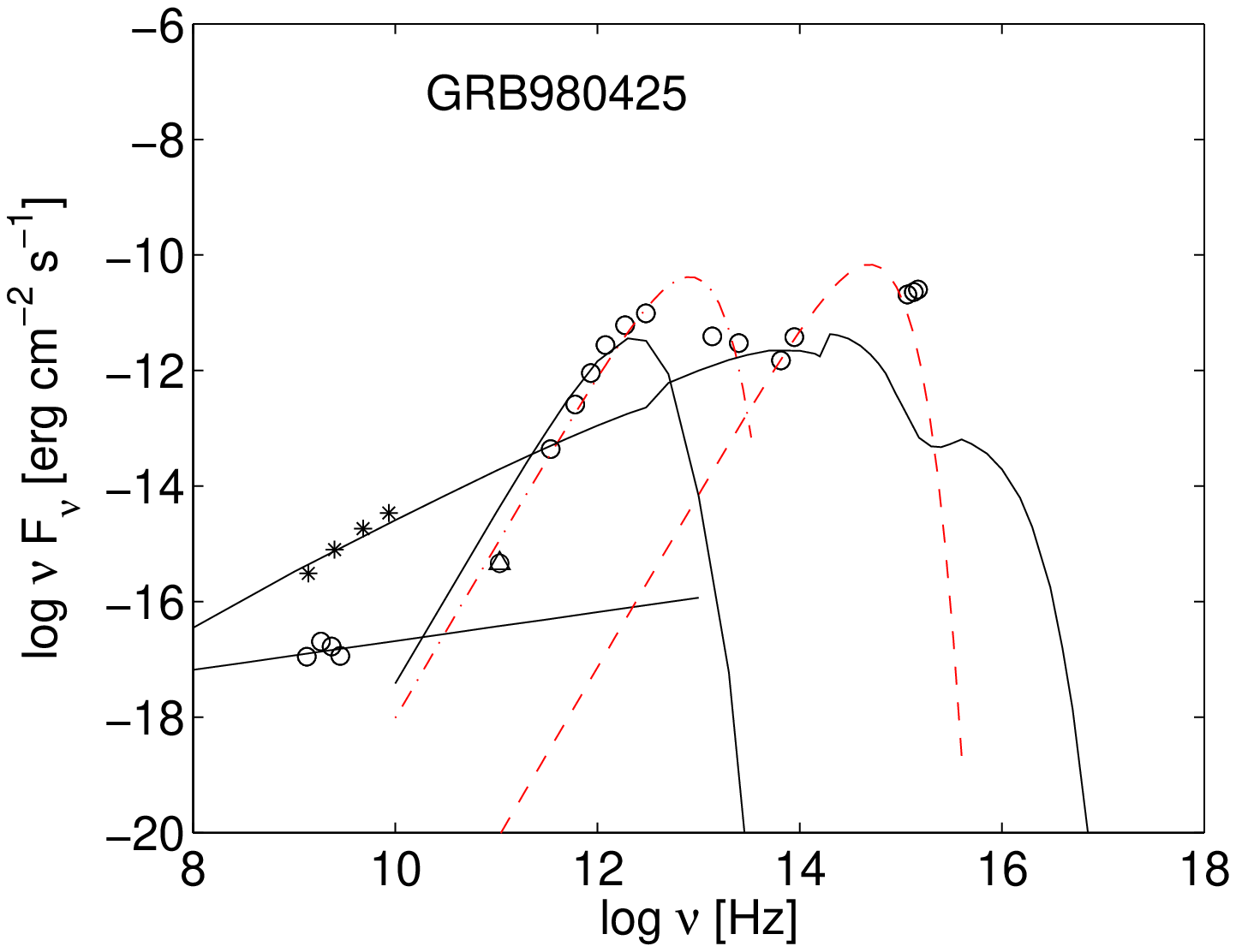}
\includegraphics[width=8.6cm]{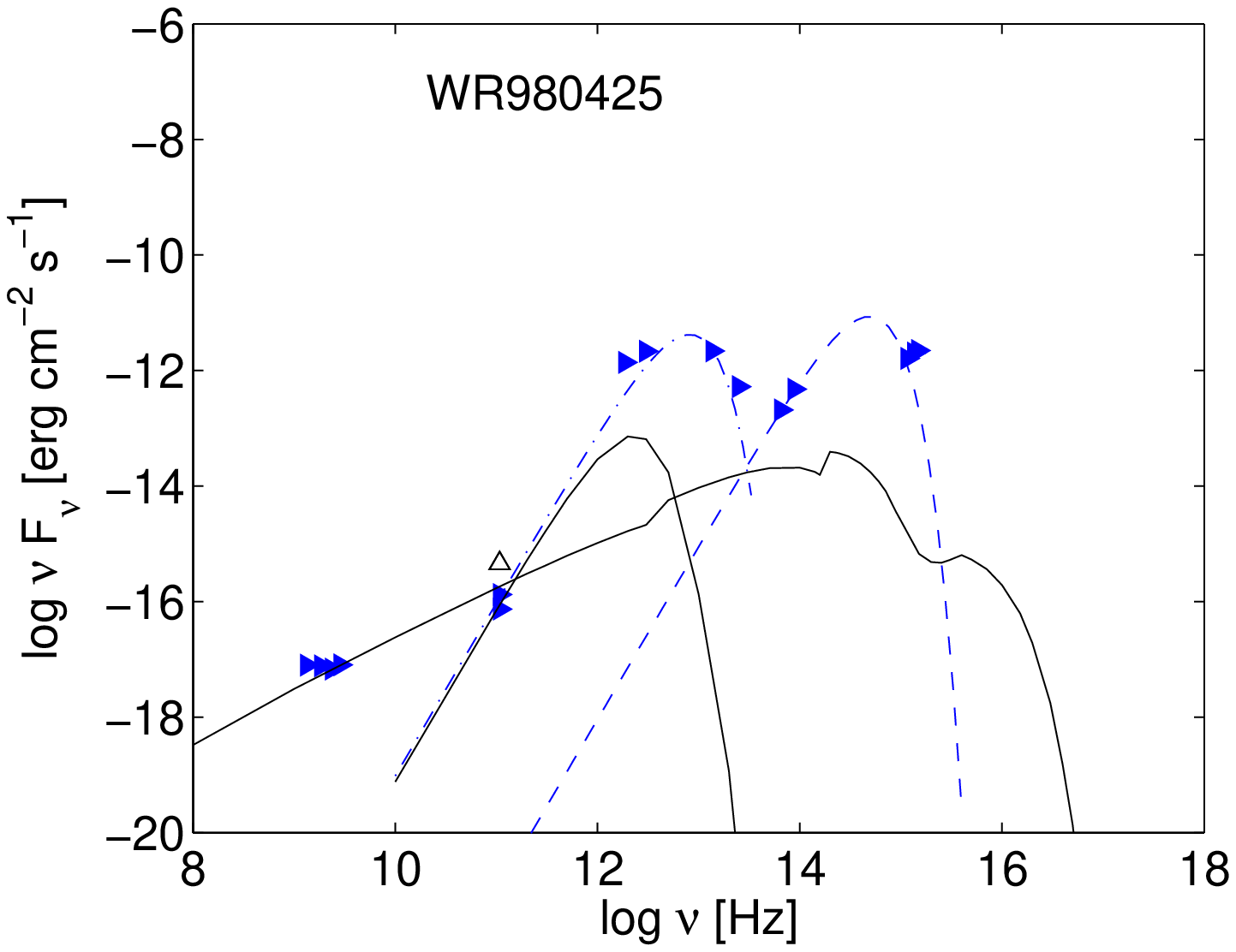}
\includegraphics[width=8.6cm]{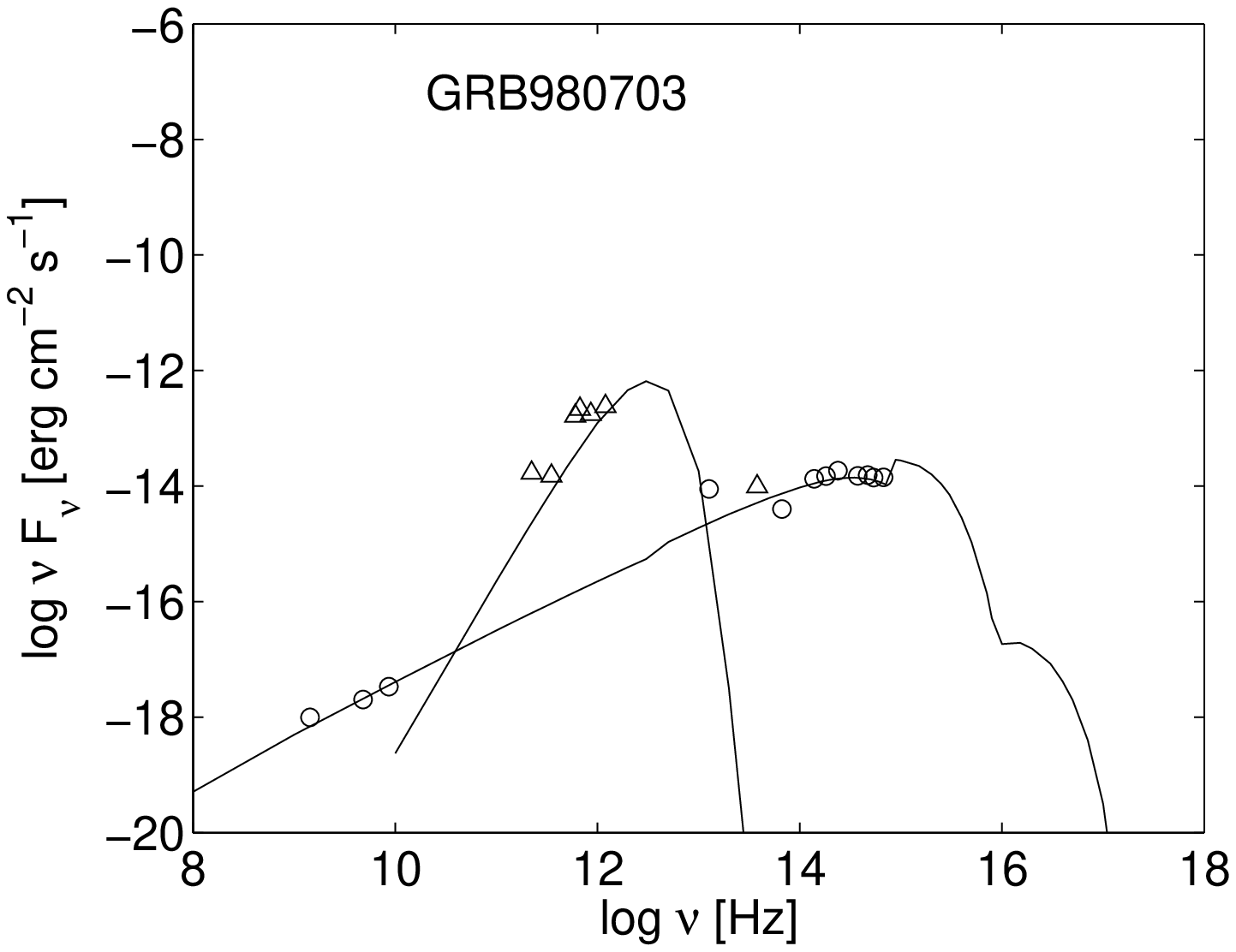}
\includegraphics[width=8.6cm]{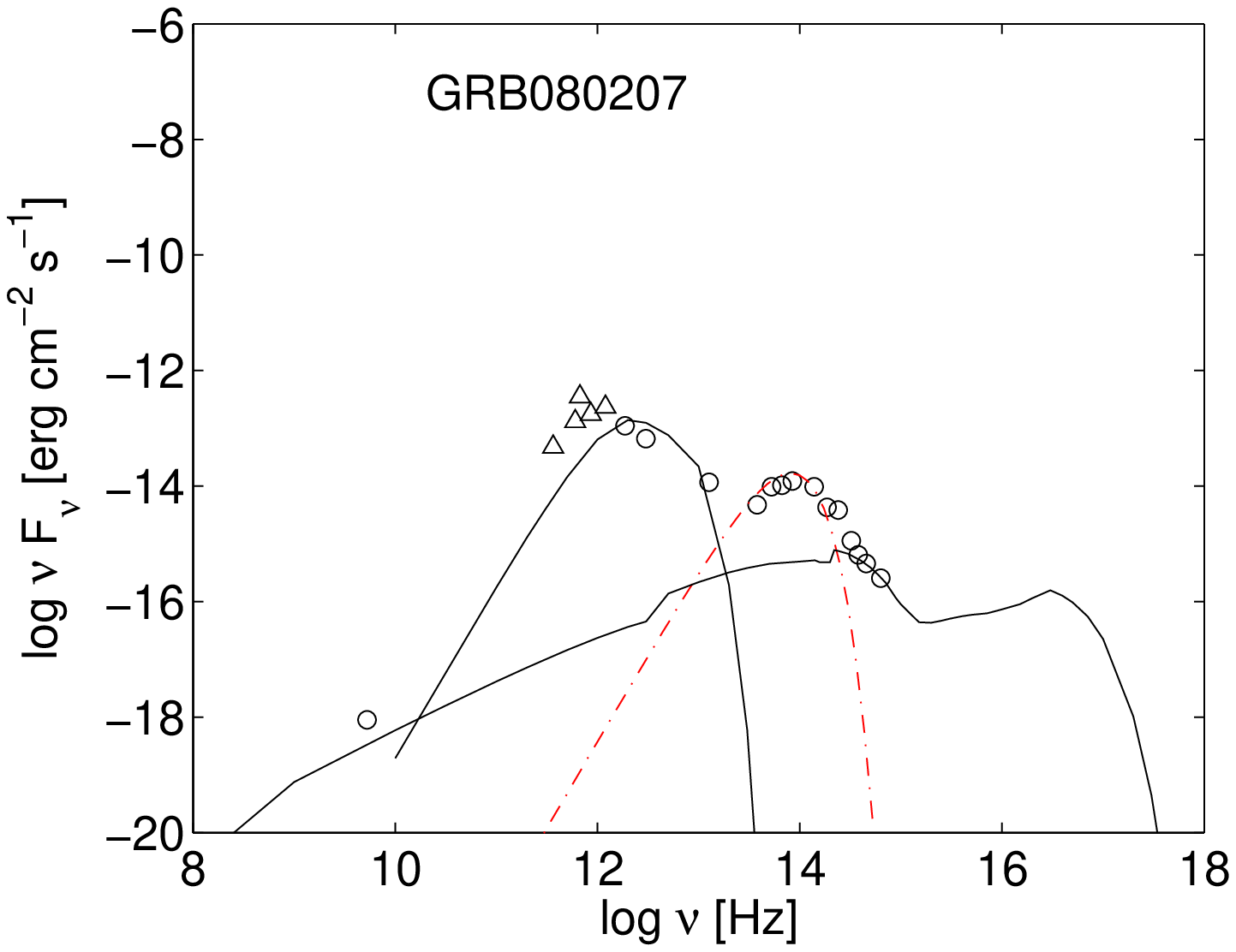}
\includegraphics[width=8.6cm]{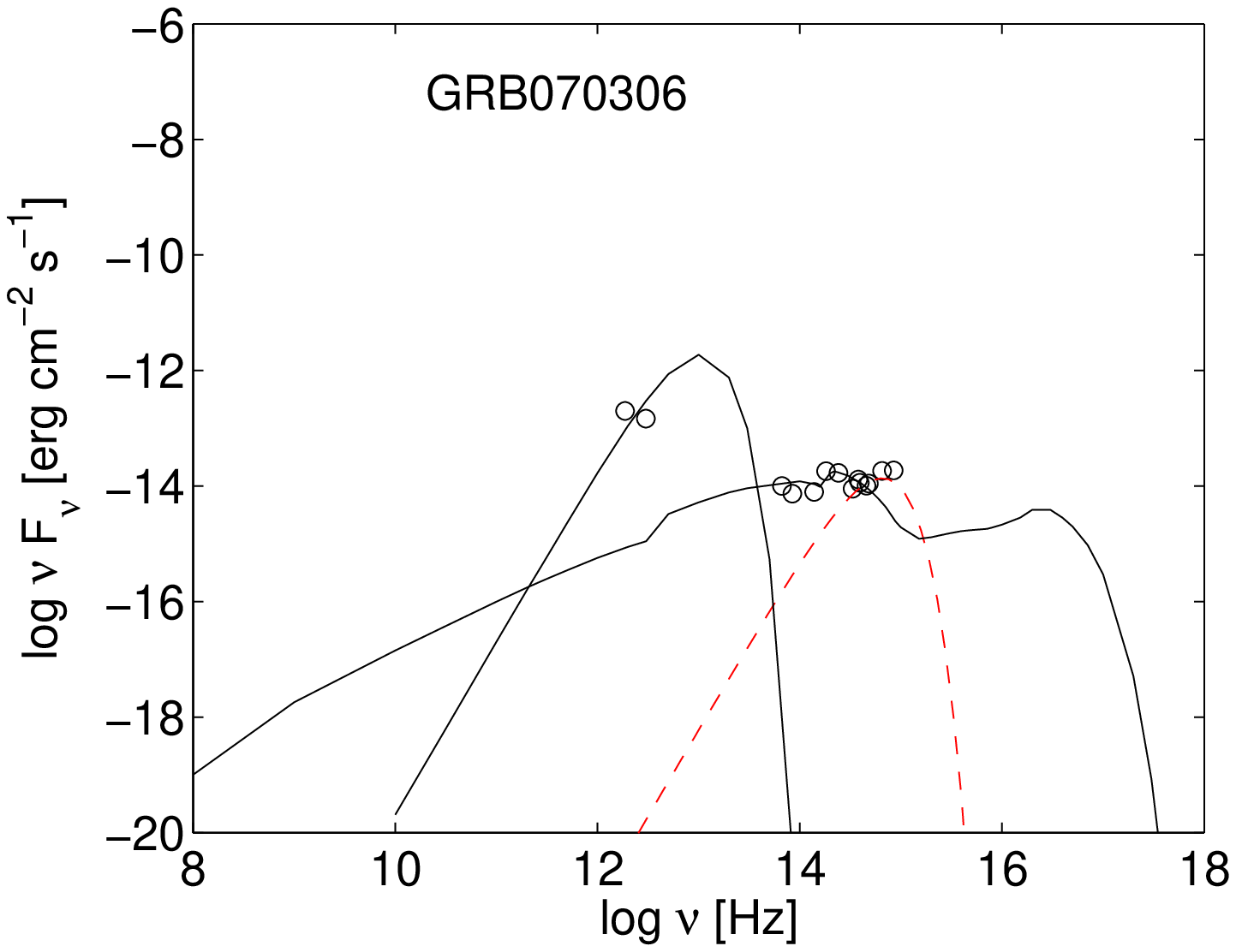}
\includegraphics[width=8.6cm]{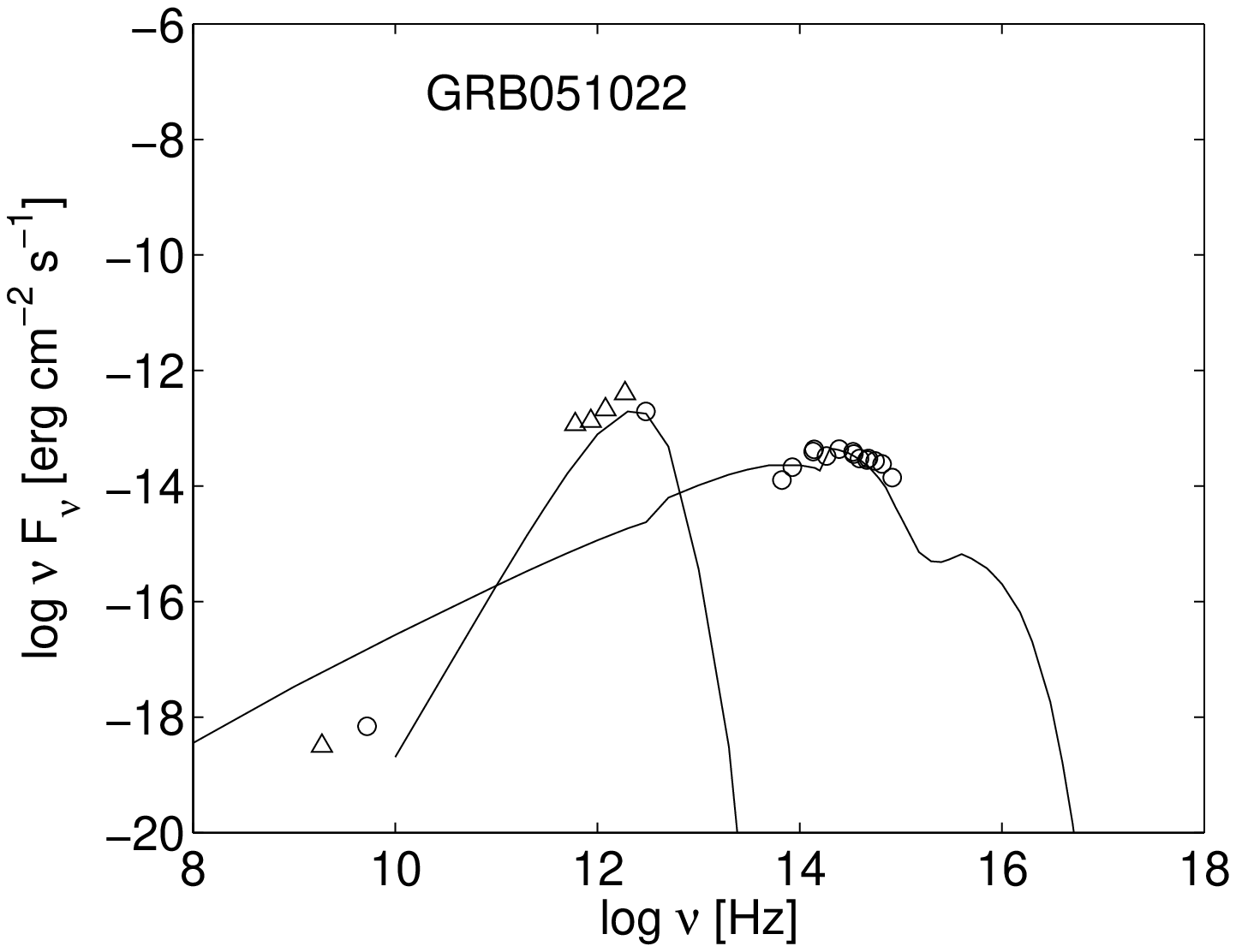}
\caption{The obscured GRB hosts from the Hunt et al sample. 
 Open triangles: upper limits. Filled triangles: data by Michalowski et al (2014) for the WR region}
\end{figure*}

Hunt et al. (2014) claim that   dark GRBs which are characterised by the  very faint observed  
optical afterglow  
relative to the extrapolation from the X-ray, can be found in massive,  
star forming galaxies with red colour, high extinction and large SFR.
Observations with Herschel (by the Photodetector Array Camera \& Spectrometer)
  up to redshift $\sim$ 3 have detected 7 out of 17 GRB.
Combining the IR data with optical, near IR and radio data from the  
literature,  Hunt et al
have successfully modelled  by GRASIL the SEDs on 6 orders of magnitude.
Hunt et al found that GRB host galaxies  are medium-seized   
with relatively high specific SFRs.
Hunt et al  concluded that the fraction of observed dark hosts   suggests that they are more likely to be detected
 at IR/submm wavelength than their optically bright counterparts.

We  compare the  models calculated for the Hunt et al GRB host sample in Fig. 1 on a large frequency range
because the data cover the radio-UV domain. The errorbars are absent for sake of clarity. The errors are 
relatively small ($<$ 10 percent) but could  confuse the dataset trends. 
  In Table 1 we report the  objects  selected from the  Hunt  et al (2014) sample that were
already analysed by modelling  the line ratios (Contini 2016)  emitted from the  host galaxies presented in 
different samples (described in the bottom of Table 1) and  the model results.  
(Model results are represented by the code input parameter set which leads to the best fit of the data).
For GRB980425, GRB980703 and GRB080207             
\agr=1. \mum has been adopted, for the other objects of the Hunt et al sample, \agr=0.1 \mum.
Dust reradiation is calculated by the models  adopting  $(d/g)_0$ =0.0004 by mass for all the hosts.
In Fig. 1 we present the  best fit of calculated to observed SEDs. The data are observed at Earth while 
bremsstrahlung and dust reprocessed radiation are calculated at the emitting nebulae. Therefore an adjusting factor $\eta_g$
is adopted to shift the calculated flux. $\eta_g$ depends on the galaxy  distance to Earth and on the distance r  
of the emitting cloud from the central  radiation source in the host galaxy (Table 2).
We   calculate  r by
$(\nu F)_o$ 4 $\pi$ d$^2$=$(\nu F)_c$ 4 $\pi$ r$^2$, where $(\nu F)_o$ (in \erg) is the  flux observed at Earth and
$(\nu F)_c$   is the bremsstrahlung  calculated at the nebula.
Considering that dust and gas coexist in the same cloud, to fit the observation data in the IR we can enhance or reduce 
the reradiation bump 
by multiplying $(d/g)_0$ by a factor f$_{d/g}$.  We obtain $d/g$ =f$_{d/g}\times (d/g)_0$ (Table 2).
 The average $d/g$  in the Milky Way is 0.007   (Dwek \& Cherchneff 2011).
If dust and gas coexist  in the nebula at the same radius, $d/g$   ranges between 8.2$\times$ 10$^{-5}$ and 0.032.
Adopting a patchy distribution of dust (\ff$<$ 1)  the results can change towards higher $d/g$.
In Table 2,  the galaxy distances to Earth are given  in column 2 followed by r in column 3,
 $d/g$, in terms of (d/g)$_0$ = 0.0004 in column 4, $\eta_g$ and $\eta_d$ in columns 5 and 6, respectively.  
Table 2 shows that $\eta_g$  $>$  $\eta_d$ for  GRB980425 and GRB051022.  
For a few GRB hosts (GRB980425, GRB080207 and GRB070306) the contribution of the old star population is 
shown throughout the SED.
In column 7  the temperatures corresponding to  the   background old stars   are given.

In particular, 
GRB980703  shows a very high \Ts (3.4 10$^5$K) similar to that of stars close to outburst, high $U$ and  extended clouds 
with a relatively low preshock density. 
The high \Ts is similar to those found in GRB990712 and GRB020903  (Contini 2016)
which are also included in the sample of Han et al. for hosts containing  W-R stars.
$d/g$ is  very high, reaching $d/g$=0.03. 
We have found that the dusty clouds are relatively far from the high temperature radiation source. 
Dust grains could evaporate at  temperatures reaching $>$ 1000 K. 
They  are also partly sputtered throughout shocks with \Vs =190 \kms. 
In this  case we suggest that  dust grains are  created in the outskirt of the host galaxy. 

GRB980425  was investigated in detail by Michalowski et al (2014) who presented data for both the entire GRB host and 
for the W-R star formation region located at 800 pc from the GRB position.
In a previous modelling of the spectra in the different regions of GRB980425 (Contini 2017a) we have found  
similar physical conditions throughout the host, in particular next to the SN and the GRB regions.
GRB980425 is associated with SN 1998bw (Galama et al 1998) which hosts the closest GRB.
By modelling  the SEDs  Michalowski et al found a low dust content in the GRB host. 
Moreover, the W-R region contribution 
is relatively high in the far-IR-radio range.
They claim that the presence of dust is connected with star formation.  
The modelling of GRB980425 SED by the physical parameters (see Table 1) constrained by the detailed 
modelling of the line ratios is shown in Fig. 1 top  left diagram. 
We added in this diagram the data from  Poonam \& Frail (2012) in the radio range which are attributed to the afterglow 
because they are well fitted by 
 the thermal bremsstrahlung  calculated by the  model which accounts for  photoionization and shocks.
On the other hand, the data in the radio range provided by Michalowski et al show a lower flux and a different slope,
which could suggest synchrotron radiation by the Fermi mechanism at the shock front. For these reasons we leave in
the diagram all the observation data throughout  the frequency range.
The bb flux from  background star  radiation corresponding to  6000 K and another bb flux at  100 K are 
used to reproduce the whole SED.
The 100 K black body could be easily explained by dust grains heated throughout the host.
Our modelling shows that reradiation by dust calculated consistently with gas should be  very low, because
constrained by the data in the far-IR. 

The modelling of the flux from the whole galaxy shows some ambiguities.
First, the fit of the afterglow radio data by the model referring to the host is rather suspicious.
Second, the suppression of dust reradiation  calculated from the emitting clouds. 
Therefore we  now model the SED presented by Michalowski et al for the W-R star formation region (Fig. 1 top right).
The data from this  region  contribute to the SED by a low  percent which increases
sensibly in the radio range.
We thus believe that the radiation source is identified with W-R stars rather than with other sources 
 in the extended GRB region.
The composite model reproduces the data in the radio, in the far-IR corresponding to both
dust reemission and thermal bremsstrahlung. The bb radiation from  6000K stars and  from 100K dust contribute  
to the SED also in the W-R region.
 Then, we can remove the bremsstrahlung from the GRB extended region and  confirm that the Poonam \& Frail (2012)
observed radio data  come from the afterglow.  
Concluding, from the extended host region we see the bb flux at 100K and 6000K, while
 from the W-R region we see the thermal bremsstrahlung from the clouds 
heated and photoionized by the SB at an effective temperature of 6.5$\times$ 10$^4$ K, and some  percent of the
bb fluxes at 100K and 6000K. It seems that the bb flux at 100K,  too low to  represent background stars, derives
from  matter  spread all over the host galaxy.

GRB070306 appears in the Hunt et al sample and in the Kr\"{u}hler et al (2011) one. The line ratios were modelled on the
basis of Kr\"{u}hler et al (2015) data for both, but the continuum observations come from different sources.
 The results are the same, but Hunt et al survey covers also some data in the IR. Therefore this GRB SED  is reported
in both Figs. 1 and 3.

\begin{figure}
\centering
\includegraphics[width=8.6cm]{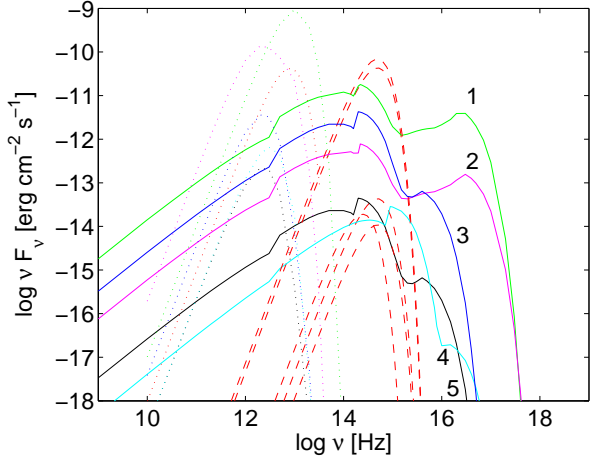}
\caption{Models for the Hunt et al (2014) host sample.   
1 : GRB070306; 2 : GRB080207: 3 : GRB980425; 4 : GRB980703; 
 5 : GRB051022.
Solid lines : bremsstrahlung; dashed lines : background old star bb flux ; dotted lines :  dust reradiation flux 
}
\end{figure}

In Fig. 2 we  report together the calculated continuum SEDs for the objects presented in Fig. 1
because they cover the radio-IR-optical and UV frequency range. Fig. 2 shows that
the bremsstrahlung, dust reradiation and radio emission domains  are well recognizable and  indicate that
at relatively high frequencies the shock velocity leads to  X-ray emission.
The background old stars are mostly at temperatures of $\sim$ 6000 K.
The dust reradiation bump does not  invade  the radio emission range at $\nu \leq$ 10$^{10}$ Hz.

\subsection{Continuum SEDs   for  selected objects from the   
Kr\"{u}hler et al (2011) GRB host sample}

\begin{table}
\centering
\caption{Models of line spectra for  selected  Kr\"{u}hler et al (2011) GRB host galaxies}
\begin{tabular}{lcccccccccccccccc} \hline  \hline
\ GRB     &z     & \Vs  & \n0  & $D$       & \Ts & $U$       \\
\  &             &  \kms& \cm3 & pc        &10$^4$K & -    \\ \hline
\ 070306  & 1.496&  280 & 150  & 0.33       & 7.    & 0.05   \\
\ 070802  & 2.45 &  140 & 100  & 0.33       &5.2    &0.03      \\
\ 080605  & 1.64 &  200 & 80   & 0.33       & 5.    & 0.05  \\
\ 080805  & 1.5  &  130 & 100  & 0.4        & 7.8   & 0.01    \\
\ 090926B &1.24  & 160  & 110  & 0.27       &  9.   & 0.01   \\
\ 100621A& 0.543 & 200  & 150  & 0.67       & 5.    & 0.05  \\ \hline
\end{tabular}

\centering
\caption{Modelling  the   Kr\"{u}hler et al (2011) SEDs}
\begin{tabular}{lccccccccccc} \hline  \hline
\ GRB    &   d    &    r    & $\eta_g$    & T$_{bb}$         \\
\        &  Mpc   &  kpc    & -           & 1000  K               \\  \hline
\ 070306 & 11052.3& 4.93    & -12.7       & 6              \\
\ 070802 & 20278.3&38.19   & -11.45       &  -             \\
\ 080605 & 12384.5&13.11   & -11.95       & 6           \\
\ 080805 & 11088.9&5.25    & -12.65       &  -             \\
\ 090926B& 8754.9 & 14.7   & -11.55       & 8        \\
\ 100621A& 3162.9 & 2.23   & -12.19       & 6       \\  \hline

\end{tabular}
\end{table}

\begin{table}
\centering
\caption{Approximated modelling of   Sokolov et al (2001) SEDs}
\begin{tabular}{lccccccccccc} \hline  \hline
\ GRB    &   z    &   $\eta_g$    & T$_{bb}$         \\
\        &        &             & 1000  K               \\  \hline
\ 970508 & 0.8349 &   -12.70      &   -         &                \\
\ 980613 & 1.0994 &    -12.12      &   -            \\
\ 990123 & 1.6    &    -12.70      &  8          \\
\ 991208 & 0.7063 &    -12.50      &  6             \\ \hline
\end{tabular}
\end{table}

\begin{figure*}
\centering
\includegraphics[width=8.8cm]{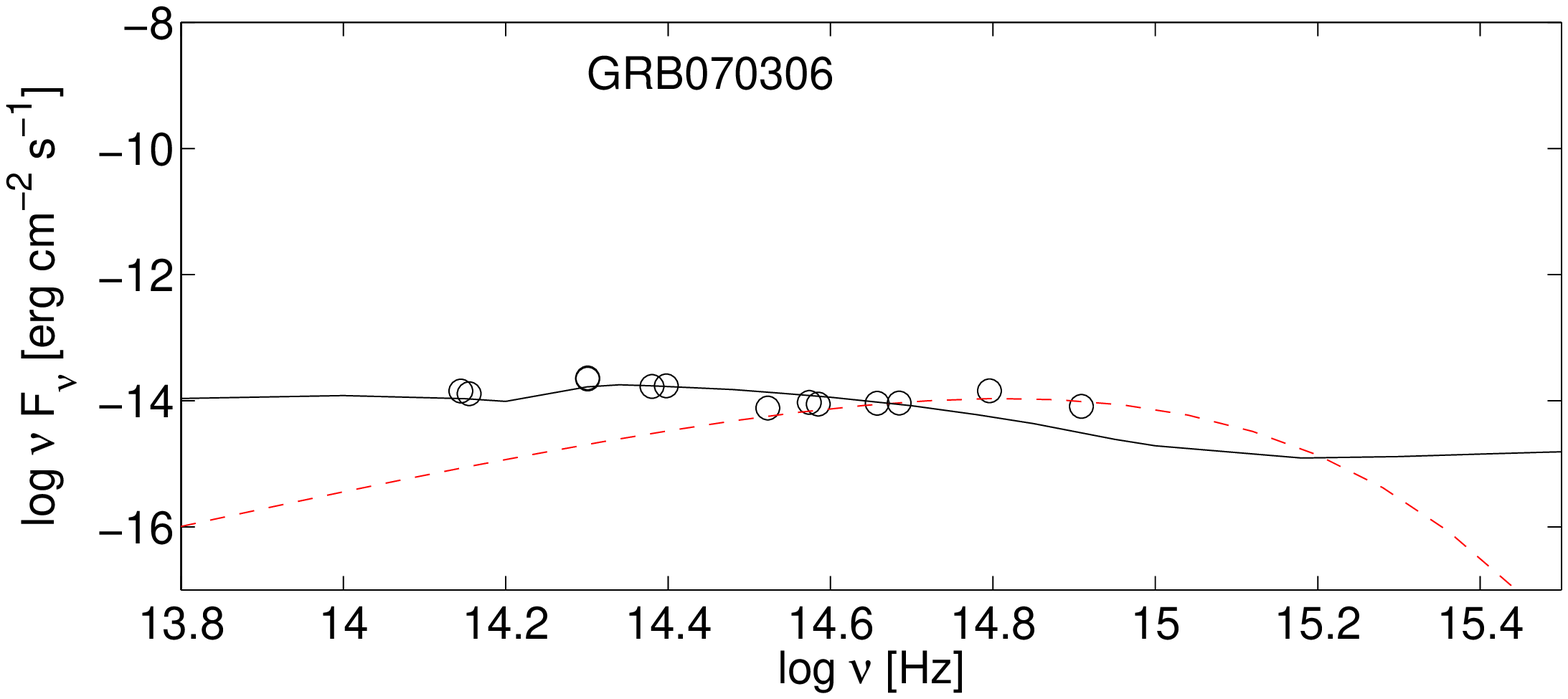}
\includegraphics[width=8.8cm]{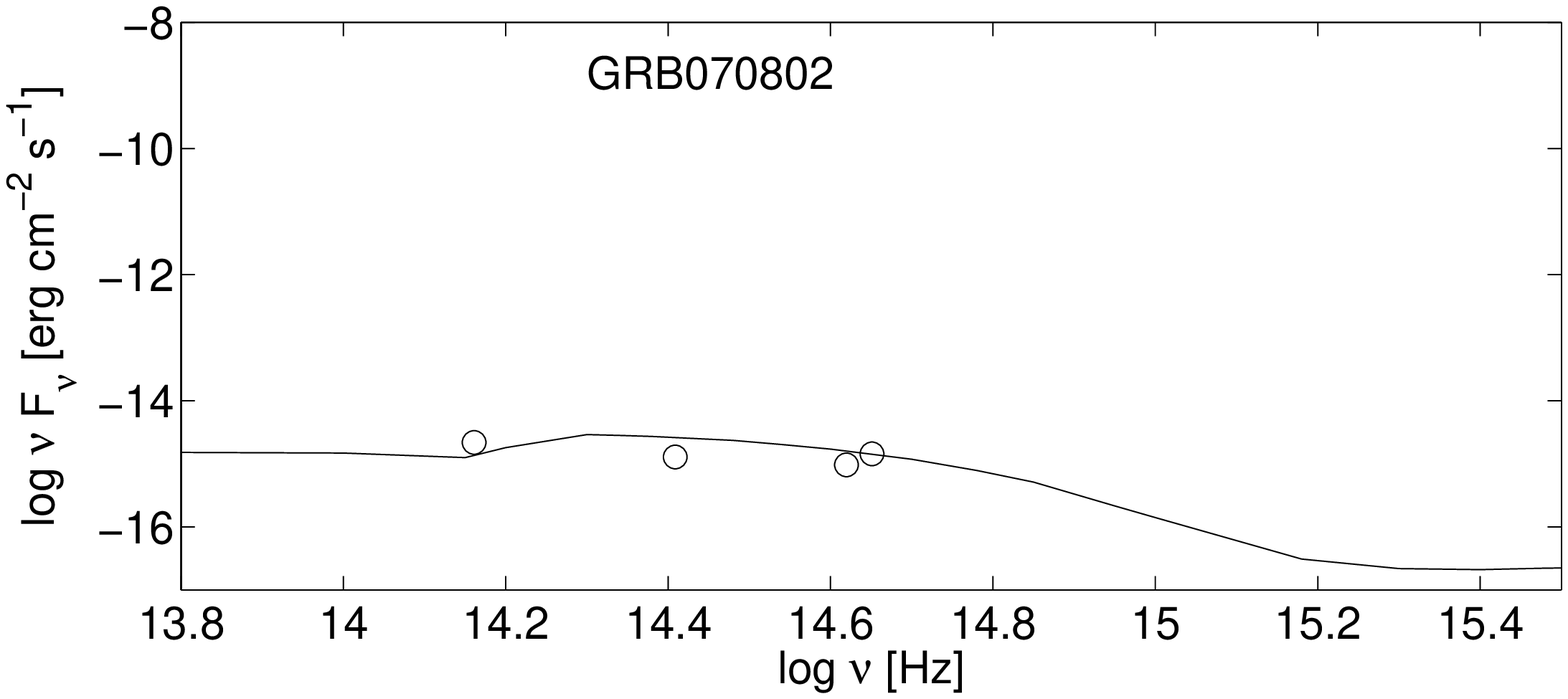}
\includegraphics[width=8.8cm]{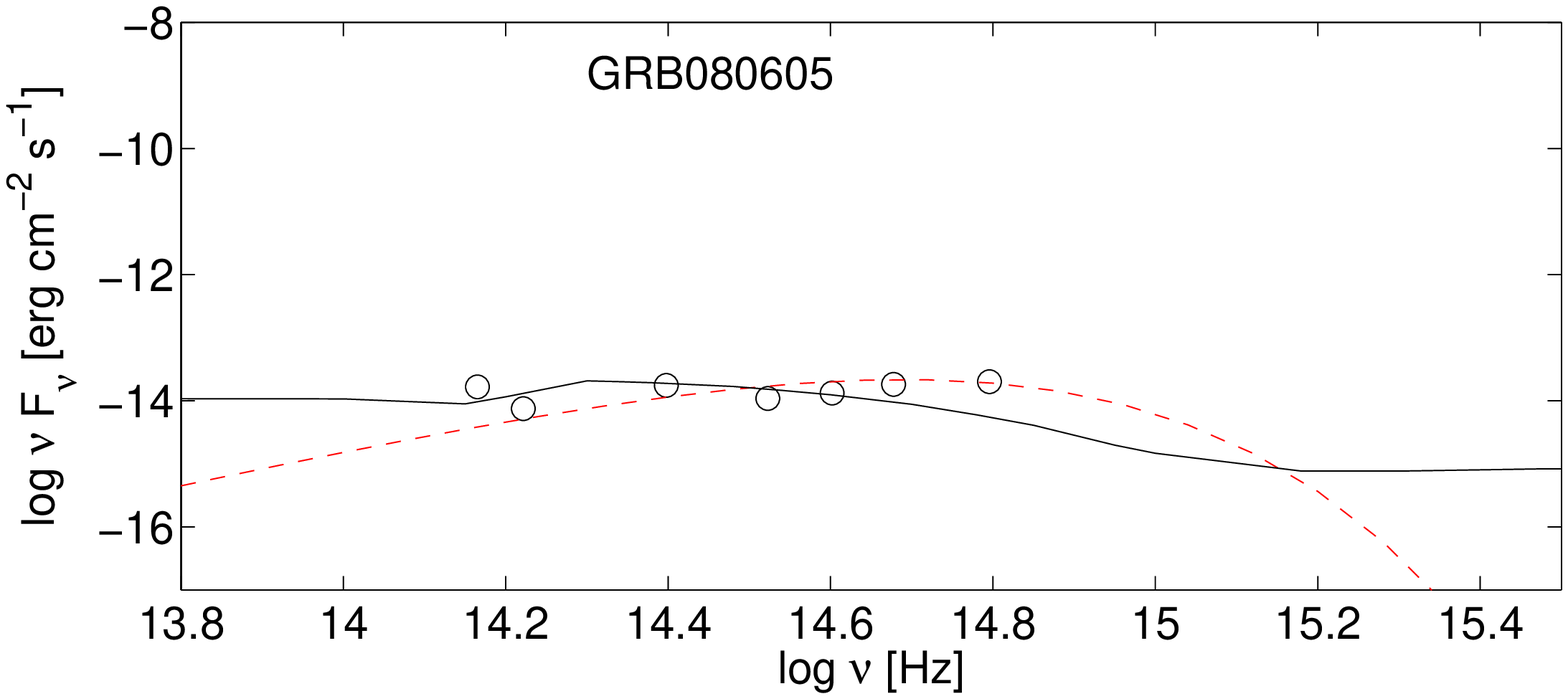}
\includegraphics[width=8.8cm]{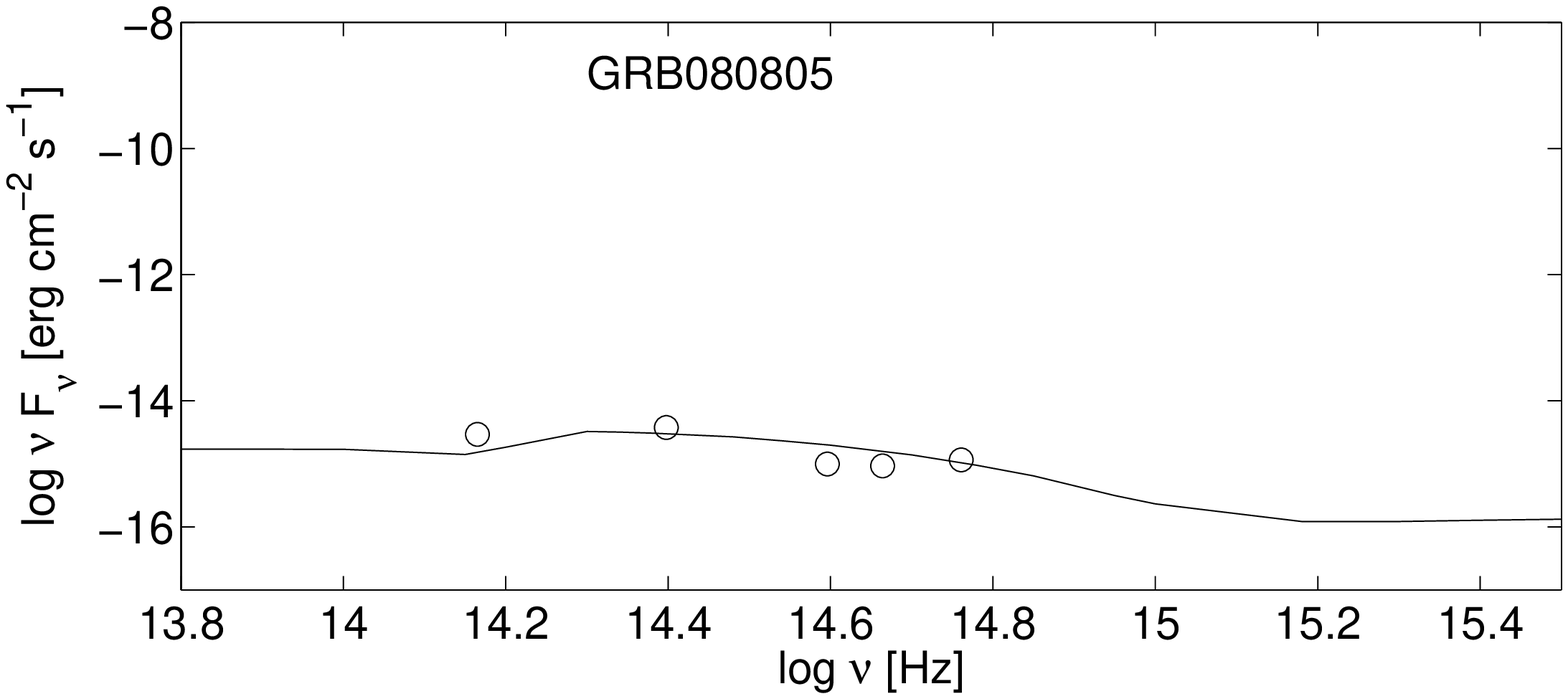}
\includegraphics[width=8.8cm]{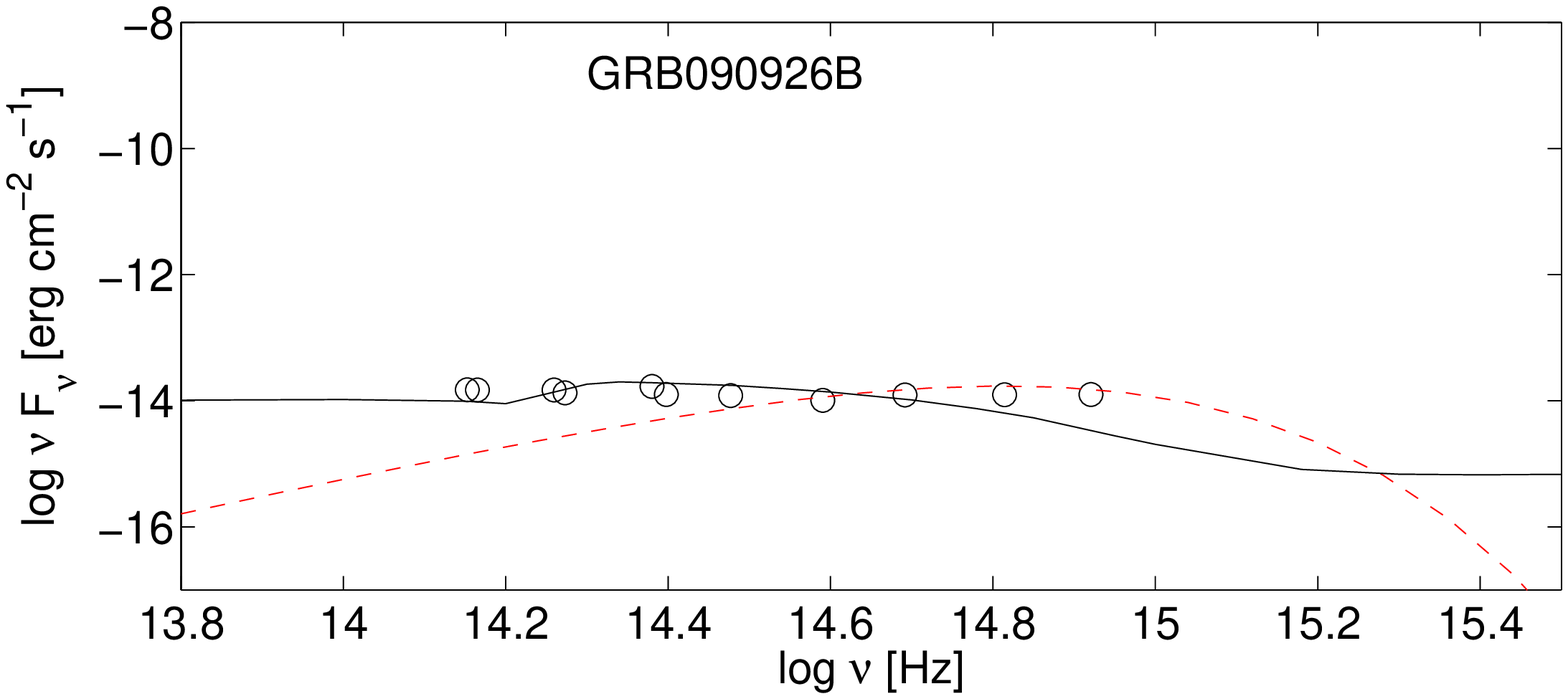}
\includegraphics[width=8.8cm]{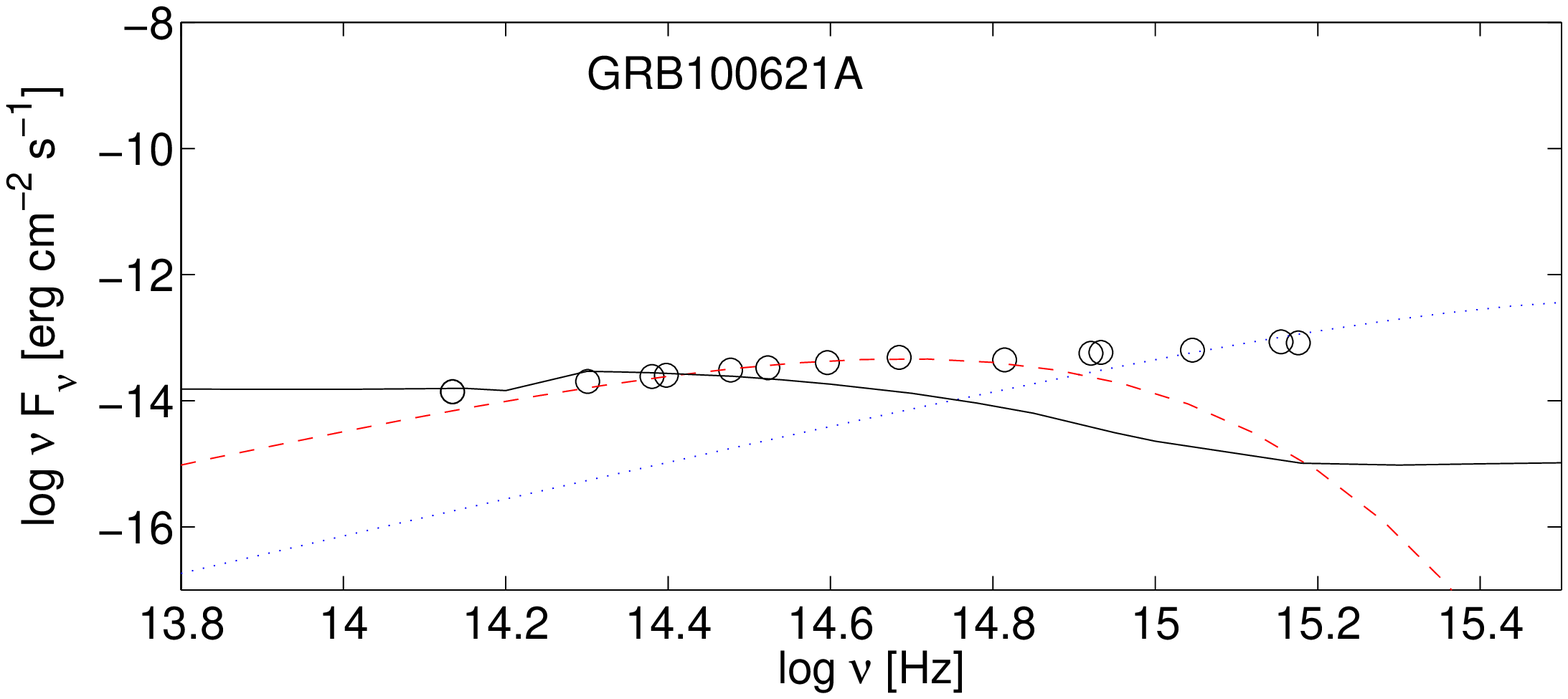}
\caption{The continuum SED of selected galaxies from  Kr\"{u}hler et al. (2011) sample. 
Asterisks : data from Punha \& Freil (2012);
open circles : the data from Kr\"{u}hler et al (2011); solid lines : the
result of  models (Contini 2016, table 6b) obtained  by fitting  line ratio observations of
Kr\"{u}ler et al (2015);
dashed line : bb radiation  corresponding to T  $\geq$6000K ;
dotted line : bb radiation corresponding to T= 5$\times$10$^4$K;
}
\end{figure*}

\begin{figure*}
\centering
\includegraphics[width=8.8cm]{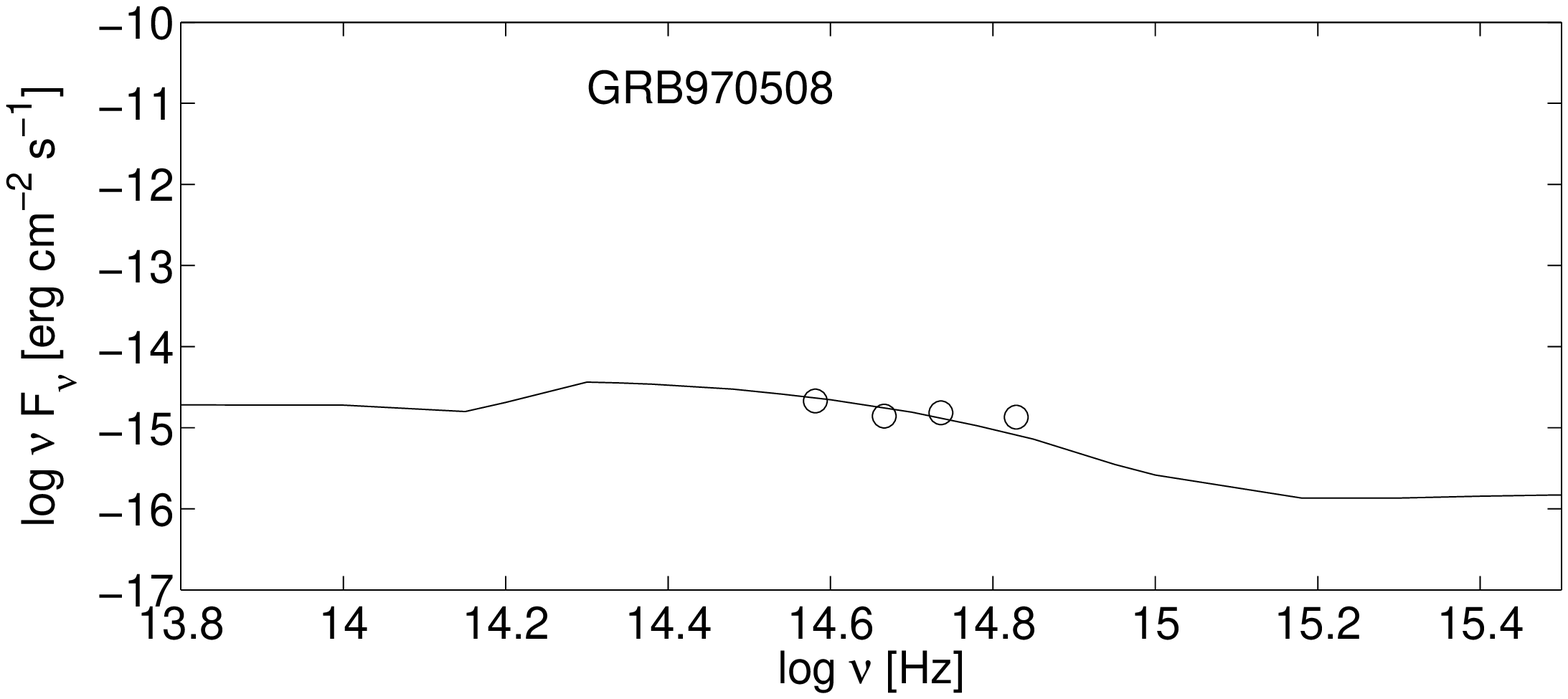}
\includegraphics[width=8.8cm]{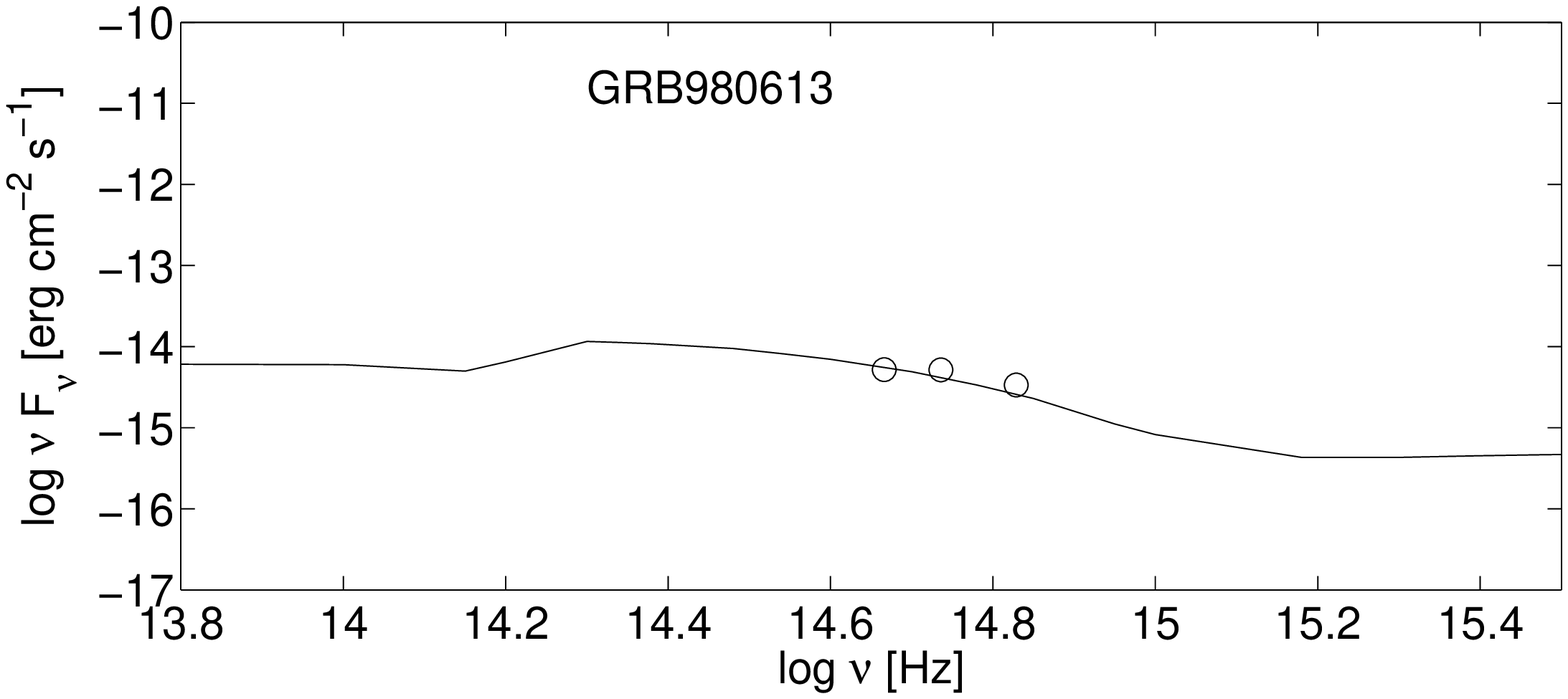}
\includegraphics[width=8.8cm]{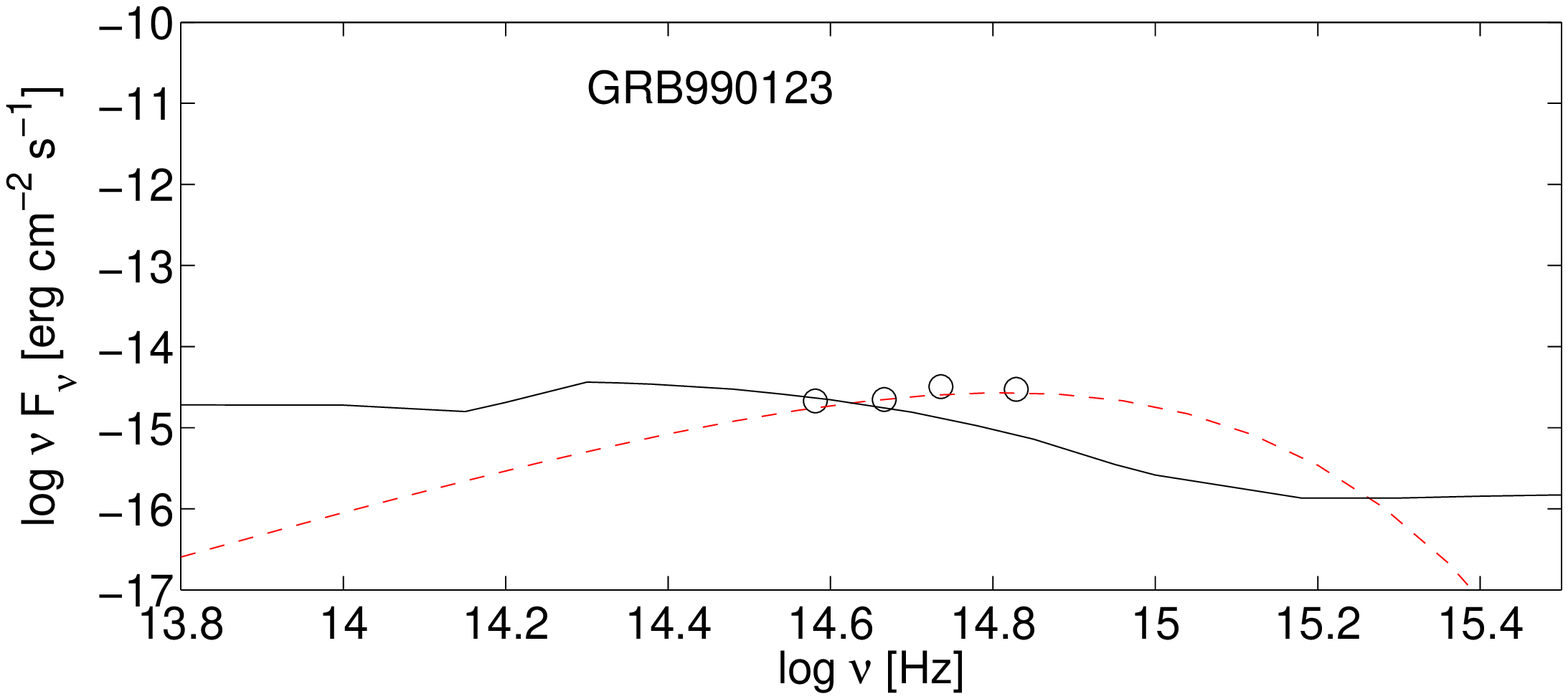}
\includegraphics[width=8.8cm]{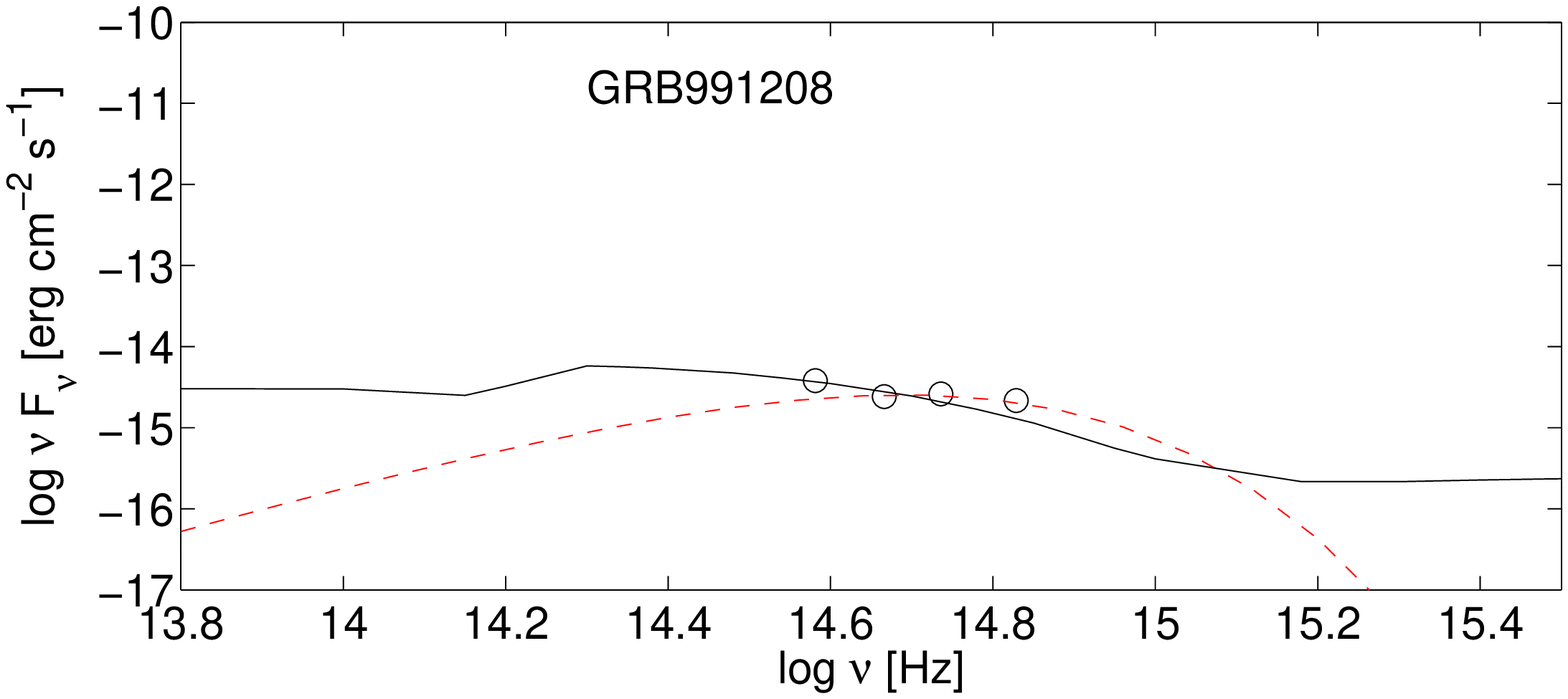}
\caption{The best fit of GRB hosts from the Sokolov et al (2001) sample. Symbols as in Fig. 3}
\end{figure*}

The observations were  initiated by GROND and in the case of non-detections
in individual filters they were continued  by EFOSC/SOFI at the NTT(4m  
class) and FORS2/HAWKI at the VLT(8m class).
The UV/optical/NIR photometry of the selected GRB hosts were analysed  
in a standard way
using stellar population synthesis techniques to convert luminosities  
into stellar masses.
We selected the objects that were modelled on the basis of the line  
spectra presented by
Kr\"{u}hler et al (2015). The models are described in Table 3   (see Contini 2016, table 6).

In  Fig. 3  the  observed data from the GRB host (Kr\"{u}hler 2011)  correspond to the  
10$^{14}<\nu<10^{15}$ Hz frequency  domain, therefore
they do not contain any contribution from   dust. 
  The modelling of the SEDs for the  selected GRB hosts does not   
constrain the dust-to-gas ratios, but the line ratios were   reddening
corrected indicating that dust could be present.

In GRB070306 optical-near IR data are  well fitted by the bremsstrahlung calculated by the
 model presented in  Table 3 which is characterized by
a relatively high \Ts (7$\times$ 10$^4$K). In this host galaxy as well as in   
GRB080605 and GRB090926B
the contribution of the bb flux from background stars with T$\sim$ 6000K is evident.
GRB080805 continuum SED is well reproduced by the bremsstrahlung from the clouds. 
 El\'{i}asd\'{o}ttir et al (2009) claim that the dust-to-gas ratio for GRB070802 host is significantly
lower than that of other GRB hosts. There are no data available to confirm it.
The bremsstrahlung calculated by the model fitting the line spectrum of  GRB100621A 
(described in Table 3) reproduces  only a few  data at $\leq$3$\times$10$^{14}$ Hz  of the SED. 
The data reported by Kr\"{u}ler et al  cover a large  frequency range (see Kr\"{u}hler et al 2011, fig. 8).
To reproduce the entire observation trend the bb fluxes  referring to 6000K and 5$\times$10$^4$K are added
in the diagram.
The bb flux at T=6000K from the background star population  is  common also  to the other sample galaxies. 
 The   bb flux at T=5$\times$10$^4$K 
which fits the  continuum data at the highest observed frequencies can be   considered as the 
direct flux (not reprocessed by the gasous clouds)
from the photoionizing source. Actually, a SB temperature  \Ts=5$\times$10$^4$K 
was found phenomenologically by  modelling the line ratios (Contini 2016, Table 6).

Factors $\eta_g$  calculated adopting a filling factor \ff=1 for each object appear in Table 4.
 They are the same for gas and dust for the Kr\"{u}hler (2011) sample hosts, considering that dust and gas coexist.
However,  there are no data in the IR 
to constrain $d/g$.

\subsection{Approximated models to the SEDs of the GRB hosts  from  Sokolov et al (2001) } 

Sokolov et al (2001) presented B V R$_c$ I$_c$ broad band flux spectra for a few  GRB host  galaxies obtained by the
6m telescope of SAO RAS and compared them with templates of local  galaxies of different types.
They claim that the  SEDs are best reproduced by SB galaxies. GRB hosts show internal extinction.
We have tried to reproduce the SEDs by using a model calculated by the input parameters
similar to those used for GRB hosts and obscured GRB hosts as presented in previous sections.
We have selected model GRB080605 (Table 3) which shows intermediate \Vs=200 \kms, \n0=80 \cm3 and 
\Ts=5$\times$10$^4$K.
The results are shown in Fig. 4. We have selected  GRB970508, GRB980613, GRB990123 and GRB991208
from the Sokolov et al sample, and we have neglected  GRB971214 which 
is defined by only the V and R$_c$ fluxes. The detailed analysis results of GRB980703 appear in Table 1.
Fig. 4 shows that  the bremsstrahlung emitted from the clouds within the GRB hosts
reproduces satisfactorily the SEDs of GRB970508, GRB980613 and GRB991208 but
not the data in the B and V$_c$ bands of GRB990123. For it and perhaps also for GRB991208 the contribution of the bb flux 
corresponding to T$_{bb}$ $\geq$6000K is needed.

\subsection{Line spectra and  continuum SEDs for SLSN host galaxies from the Perley et al (2016) SLSN survey}

\begin{table*}
\centering
\caption{Modelling  Perley et al (2016) long GRB host galaxy line ratios to \Hb}
\begin{tabular}{lcccccccccccccccccc} \hline  \hline
\ PTF ID  &Class   &z      & \Ha     & \Hg       & [OII]  &[OIII]    &[OIII] & [NII]  &[SII]  &[SII]    \\
\         &        &       & 6563    &4340       &3727+    & 4363    & 5007+  & 6548+   & 6716  & 6731   \\\hline
\ 09as    & I      &0.1867 &  3.38   & 0.39      &1.49     &0.11     &7.2     &0.145    &0.23   &0.15     \\
\ mp1     &        &       &3.       & 0.46      &1.38     &0.28     &7.12    &0.1      &0.1    &0.09    \\
\ 09cnd   & I      &0.2584 & 2.5     & 0.3       &2.12     &$<$0.25  &3.5     &$<$0.55  &0.38   &0.3      \\
\ mp2     &        &       &2.95     & 0.46      &2.27     &0.02     &3.25    &0.33     &0.3    &0.28     \\
\ 10bfz   &I       & 0.1701&3.07     & 0.37      &1.53     &0.1      &5.52    &$<$0.17  &0.27   &0.16     \\
\ mp3     &        &       &3.       & 0.46      &1.6      &0.12     &5.5     &0.06     &0.72   &0.25     \\
\ 10bjp   &I       &0.3584 &3.13     &$<$0.65    &2.27     &$<$0.35  &5.1     &$<$0.39  &0.38   &0.226    \\
\ mp4     &        &       &2.95     &0.46       &2.16     &0.064    &5.      &0.12     &0.25   &0.22     \\
\ 10cwr   &I       &0.2297 &2.59     &0.44       &0.93     &0.22     &6.63    &$<$0.27  &0.13   &0.23     \\
\ mp5     &        &       &2.97     &0.46       &1.1      &0.05     &6.67    &0.06     &0.07   &0.11     \\
\ 10felb  &II      &0.2356 &3.58     &0.47       &2.23     &$<$0.13  &1.      &1.11     &0.95   &0.49     \\
\ mp6     &        &       &3.5      &0.45       &2.29     &0.08     &1.      &0.91     &0.5    &0.41     \\
\ 10heh   &II      &0.3379 &3.56     &0.44       &3.03     &$<$0.28  &3.2     &0.66     &0.72   &0.25     \\
\ mp7     &        &       &3.       &0.46       &3.2      &0.14     &3.26    &0.63     &0.4    &0.35     \\
\ 10jwd   &II      &0.477  &5.5      &0.65       &3.2      &0.16     &1.3     &$<$2.6   &1.36   &0.76     \\
\ mp8     &        &       &3.       &0.47       &3.4      &0.1      &1.3     &1.75     &1.3    &1.1      \\
\ 10nmn   &I-R     &0.1237 &4.       &0.46       &1.66     &$<$0.12  &6.96    &$<$0.34  &0.28   &0.18     \\
\ mp9     &        &       &3.       &0.46       &1.61     &0.3      &7.08    &0.1      &0.12   &0.11    \\
\ 10qaf   &II      &0.2836 &4.09     &$<$0.75    &3.52     &0.9      &2.33    &$<$0.78  &1.08   &1.03    \\
\ mp10    &        &       &3.1      &0.46       &3.43     &0.17     &2.5     &0.55     &0.7    &1.1     \\
\ 10tpz   &II      &0.0395 &5.5      &0.48       &0.74     &$<0.046$ &0.35    &3.78     &0.92   &0.68    \\
\ mp11    &        &       &3.3      &0.44       &0.77     &0.018    &0.23    &1.1      &0.7    &0.5     \\
\ 10uhf   &I       &0.2882 &4.67     &0.55       &1.44     &$<$0.57  &0.81    &2.38     &0.55   &0.4      \\
\ mp12    &        &       &3.4      &0.44       &2.0      &0.05     &0.7     &2.4      &0.33   &0.3     \\
\ 10vqv   &I       &0.4518 &3.06     &0.43       &1.3      &$<$0.28  &6.88    &$<$0.58  &0.15   &0.09   \\
\ mp13    &        &       &3.       &0.46       &1.3      &0.18     &6.      &0.4      &0.1    &0.08   \\
\ 10aagc  &I       &0.206  &3.2      &0.62       &2.77     &$<$0.25  &3.79    &0.32     &0.67   &0.41   \\
\ mp14    &        &       &3.2      &0.45       &2.8      &0.22     &3.8     &0.43     &0.11   &0.1    \\
\ 11dsf   &II      &0.3848 &2.83     &0.35       &2.04     &0.127    &4.54    &0.27     &0.25   &0.48   \\
\ mp15    &        &       &3.       &0.45       &2.2      &0.24     &4.5     &0.26     &0.07   &0.06   \\
\ 11hrq   &I       &0.057  &3.05     &0.46       &2.48     &0.04     &4.96    &0.22     &0.34   &0.23   \\
\ mp16    &        &       &3.03     &0.46       &2.21     &0.26     &5.      &0.29     &0.08   &0.075  \\
\ 12dam   &I-R     &0.1073 &3.       &0.52       &1.88     &0.1      &7.95    &0.125    &0.19   &0.138  \\
\ mp17    &        &       &3.       &0.45       &1.96     &0.4      &7.97    &0.1      &0.1    &0.1    \\ 
\ 12epg   &II      &0.3422 &3.34     &0.45       &2.98     &$<$0.12  &2.07    &0.93     &0.57   &0.41   \\
\ mp18    &        &       &3.       &0.45       &2.85     &0.07     &2.07    &0.7      &0.47   &0.41   \\
\ 12mue   &II      &0.2787 &4.18     &0.63       &2.76     &1.6      &5.16    &$<$0.93  &0.65   &0.63   \\ 
\ mp19    &        &       &3.04     &0.46       &2.9      &0.28     &5.16    &0.4      &0.25   &0.25   \\ \hline 
\end{tabular}

\end{table*}

\begin{table*}
\centering
\caption{Models used to reproduce  Perley et al (2016) spectral line ratios and SEDs}
\begin{tabular}{lcccccccccccccccccc} \hline  \hline
\  model     & \Vs  & \n0   &  $D$     &log(N/H)+12&log(O/H)+12&log(S/H)+12&  \Ts   &   $U$   & \Hb       & T$_{bb}$ &log(O/H)+12    \\
\            & (1) & (2)  & (3)        & (4)        & (5)       &(6)        & (7)     &(8)     & (9)       & (10)     &(11)          \\  \hline
\  mp1       &120   & 65    & 5.8       &7.54       &8.74       &7.7        &5.2      &   24    & 3.5         &  6        &     7.98             \\
\  mp2       &120   &65     &52.5       &7.3        &8.82       &7.3        &6.5      &   8.5    & 74.         &  5        &    8.21   \\
\  mp3       &120   &55     &26.        &7.0        &8.82       &7.48       &4.9      &   15    & 8.2         &  6        &   8.09  \\
\  mp4       &100   &63     &27.        &7.0        &8.82       &7.48       &6.1      &   6.5    & 18.3        &  6        &  7.41   \\
\  mp5*      &100   &260    &1.         &7.0        &8.82       &7.48       &5.6      &   20    & 100.        &  6        &  8.78   \\
\  mp6*      &100   &60     &5.         &7.95       &8.82       &7.48       &3.       &    3    & 8.2         &5          &  8.30   \\
\  mp7       &120   &55     &6.         &7.6        &8.82       &7.52       &5.5      &    3.5  & 6.5         &8          &   8.30   \\
\  mp8       &120   &55     &13.5       &7.9        &8.82       &7.52       &5.5      &    1.5  & 9.2         &3          & 8.87   \\
\  mp9       &140   &65     &5.8        &7.48       &8.73       &7.7        &5.2      &   24    &3.8          &8          &   8.14   \\
\  mp10      &200   &200    &0.55       &7.3        &8.76       &7.7        &5.5      &    3.5  &27.          &5          &  8.36   \\
\  mp11*     &100   &60     &500.       &9.         &8.82       &7.78       &1.       &    0.6  &17.          &3,5          & 9.22 \\
\  mp12*     &120   &90     &90.        &9.         &8.88       &7.78       &1.       &    0.6  &13.          &3,5           &  9.0 \\
\  mp13*     &100   &80     &2.         &8.         &8.82       &7.48       &5.6      &   13    &6.5          &8           &8.38 \\
\  mp14*     &100   &70     &1.6        &7.84       &8.82       &7.52       &4.       &   12    &3.8          &5           &8.19 \\
\  mp15*     &100   &70     &1.6        &7.84       &8.82       &7.52       &4        &   22    &3.5          &8           &8.20  \\
\  mp16*     &100   &70     &1.6        &7.84       &8.82       &7.52       &4.4      &   16    &3.4          &6            & 8.15 \\
\  mp17*     &140   &65     &1.2        &7.48       &8.82       &7.7        &5.2      &   24    &3.2          &8           &7.97   \\
\  mp18      &120   &55     &10.        &7.6        &8.82       &7.52       &5.5      &    3.5  &85.          &6           &8.64   \\
\  mp19      &120   &63     &10.        &7.7        &8.82       &7.7        &4.9      &    8.5  &3.6          &6           & 8.07 \\  \hline

\end{tabular}

(1): \kms; (2):  \cm3; (3): 10$^{16}$cm ; (4): solar = 8.0; (5): solar = 8.82; (6): solar =7.48; (7): 10$^4$K; (8): 0.001; (9): 10$^{-4}$ \erg ; (10):  10$^{3}$K; (11): Perley et al ('best');
*  models adopting outflowing clouds   

\end{table*}

Perley et al (2016) presented UV  - IR photometry and spectroscopy of SLSN hosts
 discovered by the Palomar Transient Factory (PTF) prior to 2013. They derive
the luminosities, star formation rates, stellar masses and gas-phase metallicities
by ground-based Imaging (SDSS, Palomar P60 telescope imaging camera, the WIRC on the 
Palomar 5-m Hale telescope, or the LRIS on the MOSFIRE on the KeckI 10 m telescope) and
by ground based  HST Observations, Spitzer and  WISE observations described in 
Perley et al (2016, sect. 3).
Regarding the spectral analysis, all the lines  were corrected
 for foreground extinction (Schlafly \& Finkbeiner 2011). The gas-phase oxygen abundance are calculated using 
direct methods, e.g. the empirical derivation of Nagao et al (2006).
Perley et al (2016) concluded that the host galaxies of SLSN I have  different properties from
the general star-forming galaxy distribution, in particular, lower masses and metallicities
and unusually high specific SFRs.

We focus on the spectra emitted from the gaseous clouds in each galaxy  and we calculate the physical conditions 
and the element abundances by the detailed modelling of the observed line ratios. 
In Table 6 we present the host galaxies selected for our analysis. They   include at least the data
(not only upper limits)  of \Ha, \Hb, [OII]3727+, [OIII]5007+, [NII]6548+  and possibly [OIII]4363. 
In Table 6 we compare  calculation results (models mp1-mp19) 
with Perley et al (2016, table 5) observed line ratios to \Hb.
The results of our analysis  will be used to constrain the models suitable to represent the emitting gas
SEDs for each galaxy. The models are given in Table 7. 
In Table 7 last column we report the metallicities in term of O/H determined by Perley et al (2016).
Metallicities calculated by detailed modelling (Table 6, column 6) are nearly all solar (O/H=6.6 10$^{-4}$, 
Grevesse \& Sauval 1998) while those  evaluated by Perley et al are mostly lower than solar (except PTF 10tpz and 10uhf).
Calculated  N/H are lower than solar (solar N/H=10$^{-4}$). 
In previous papers (Contini 2017a and references therein) we have demonstrated that the
 detailed modelling of the line ratios
leads generally to higher metallicities (in terms of O/H) than those calculated by the strong line methods. 

The  observed [SII] 6717/6731 line ratios are  $>$1,
indicating that the preshock density is $<$ 100 \cm3 and shock velocities \Vs cannot
be $>$ 140 \kms otherwise compression  would enhance the density downstream. Most of the lines are emitted
from the gas compressed downstream of the shock front  and photoionised by the SB radiation. 
In fact, the observed line ratios  show the characteristic values of  radiation dominated spectra e.g.
[OIII]5007+/[OII]3727+ $>$ 1. 
PTF 10qaf has  outstanding \Vs =200 \kms and \n0=200 \cm3. They were adopted to fit the oxygen line ratios.
In this case the calculated  [SII]6717/6731 line ratios are $<$ 1, while the observed  ratio is $>$ 1.
We suggest that the gas  emitting the [SII] lines   is merging  with  ISM matter or  with  another galaxy
as suggested by Perley et al (2016).

Regarding the continuum SED, to estimate stellar masses $M_*$,  SFR
 and interstellar extinction A$_V$,
  Perley et al (and generally the  author community) analyzed the UV-optical-NIR SEDs by
 a custom SED-fitting code (e.g. Perley et al 2013) using the population synthesis templates 
(Bruzual \& Charlot 2003)  summed according to a parametric star formation history
(see Perley et al  2016). Indeed, in their fig. 2  they  reproduced  the observed SEDs  with high precision.
However, we suggest that the clouds emitting the line spectra should also  contribute to  the continuum 
radiation (bremsstrahlung).
To  calculate the continuum SED  we adopted  the physical parameters and the element abundances
 which result by the  best fit of all the line ratios in each spectrum (Table 7).
In Fig. 5 we compare the calculated SED with Perley et al data.
 The fluxes (in \erg) are given as function of the frequency $\nu$ (in Herz) in a reduced range (log $\nu$=13.5-15.5) 
similar to that adopted in  Fig. 3 diagrams.

 Fig. 5 shows that  calculated  models  mostly fit the data at the lowest $\nu$ ($\leq 3\times10^{14}$ 
Hz), while to  reproduce the data at higher $\nu$ the flux from the old star population background  should be added.
This flux is generally approximated by a black body to  fit the SED of different galaxy types such as SB and AGN in
the 10$^{14}$-10$^{15}$ Hz range, even for objects at redshift higher than local.
Towards lower frequencies our models   diverge from   the bb flux. More data will
confirm the calculated trend.
It was  shown by  Contini \& Contini (2007, figs 5 and 6),  Contini (2013, fig. 8),
etc that the  old star population  data are nested within the Planck function corresponding to the effective
star temperature. The bremsstrahlung from the gas defines the lower bound.
 To fit the observed SEDs in Fig. 5 sometimes two bb fluxes are needed. The  temperatures  (T$_{bb}$)  are
reported in Table 7,  column 11. In the host galaxies PTF 11hrq and 12dam the bb fluxes corresponding 
to T=5$\times$ 10$^4$K are added to
 reproduce a few data in the SED  at relatively high frequencies. 
Those temperatures may represent the bb flux from the SB in the host galaxy.

\begin{figure*}
\centering
\includegraphics[width=4.6cm]{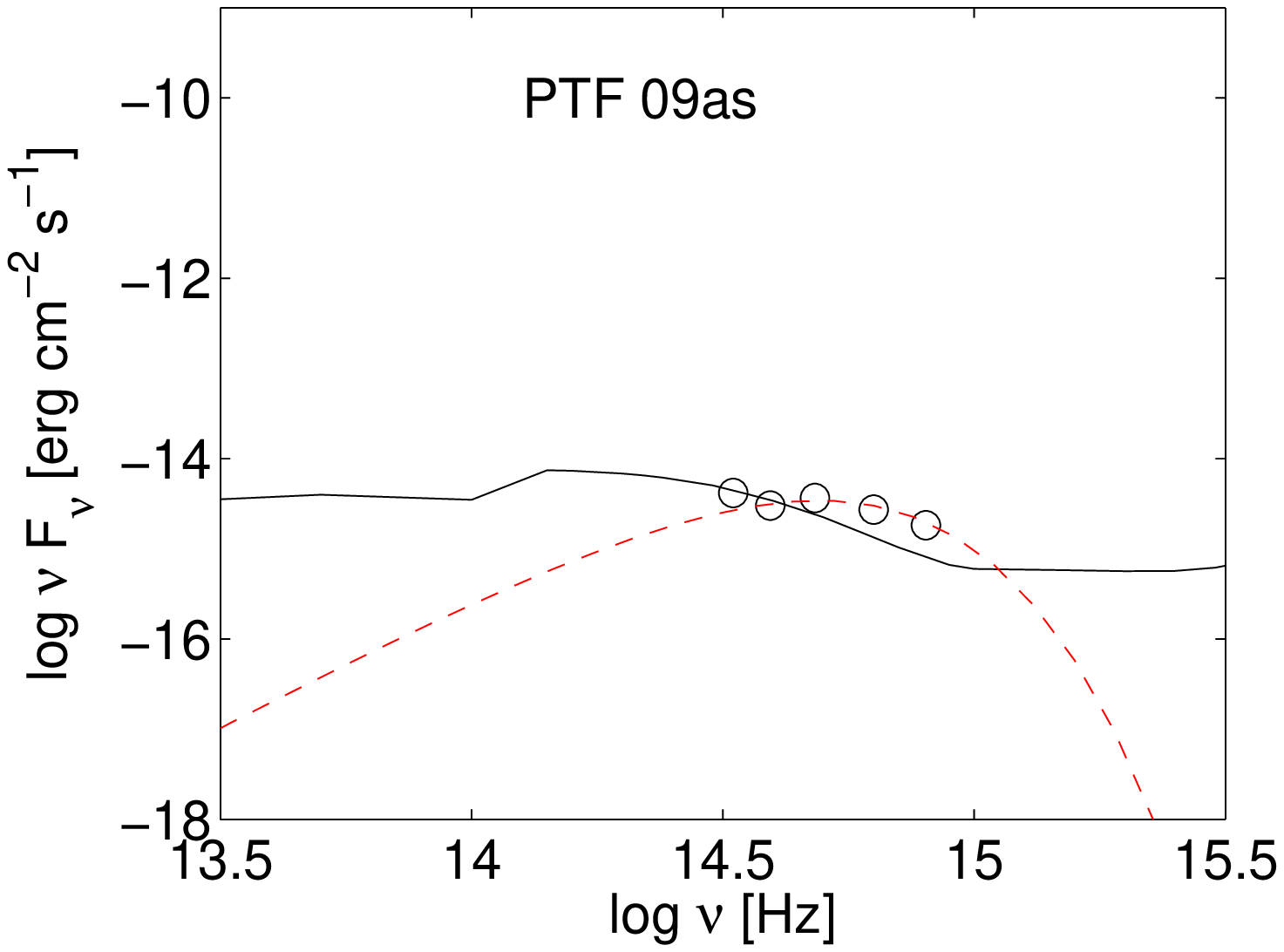}
\includegraphics[width=4.6cm]{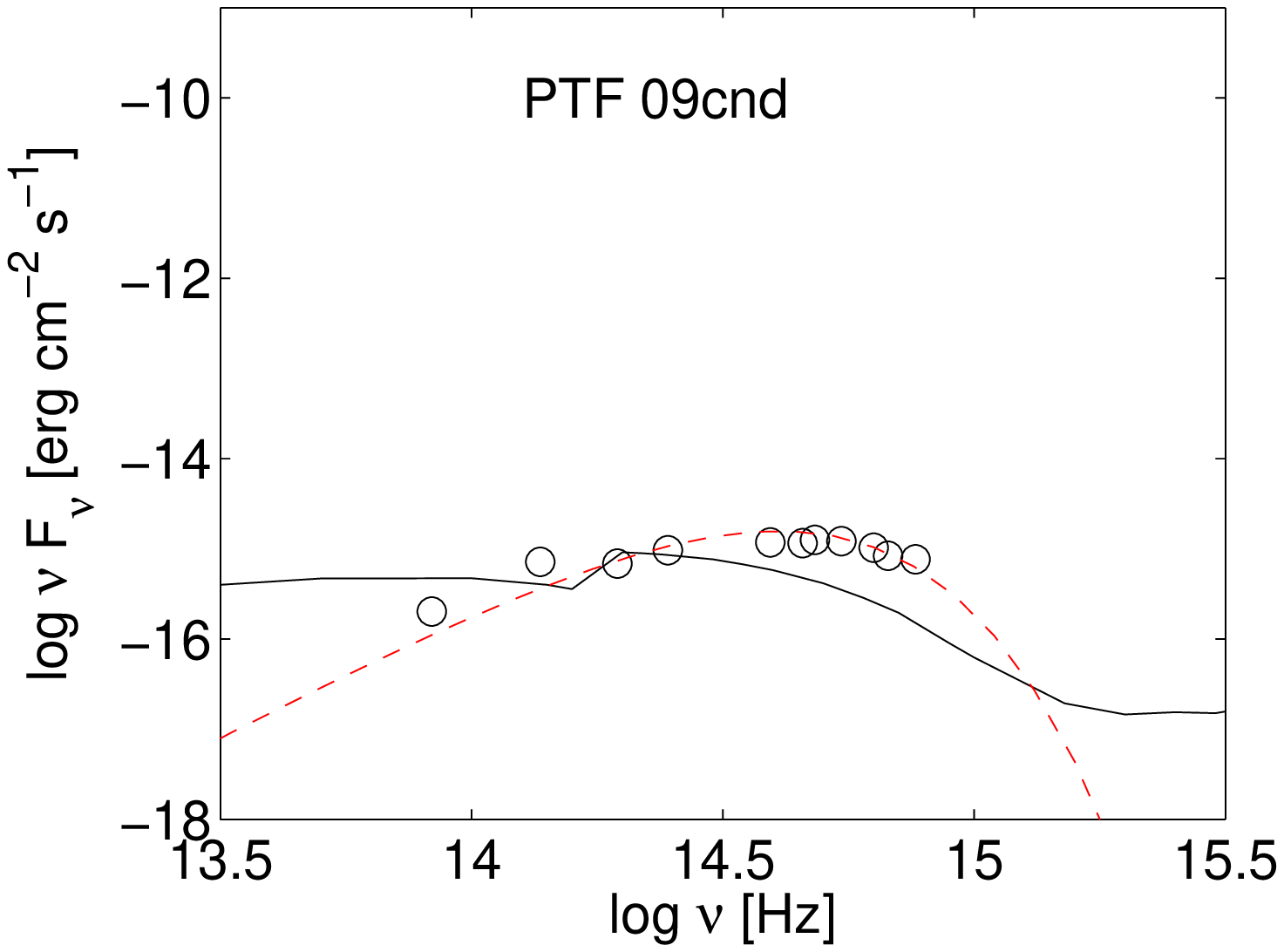}
\includegraphics[width=4.6cm]{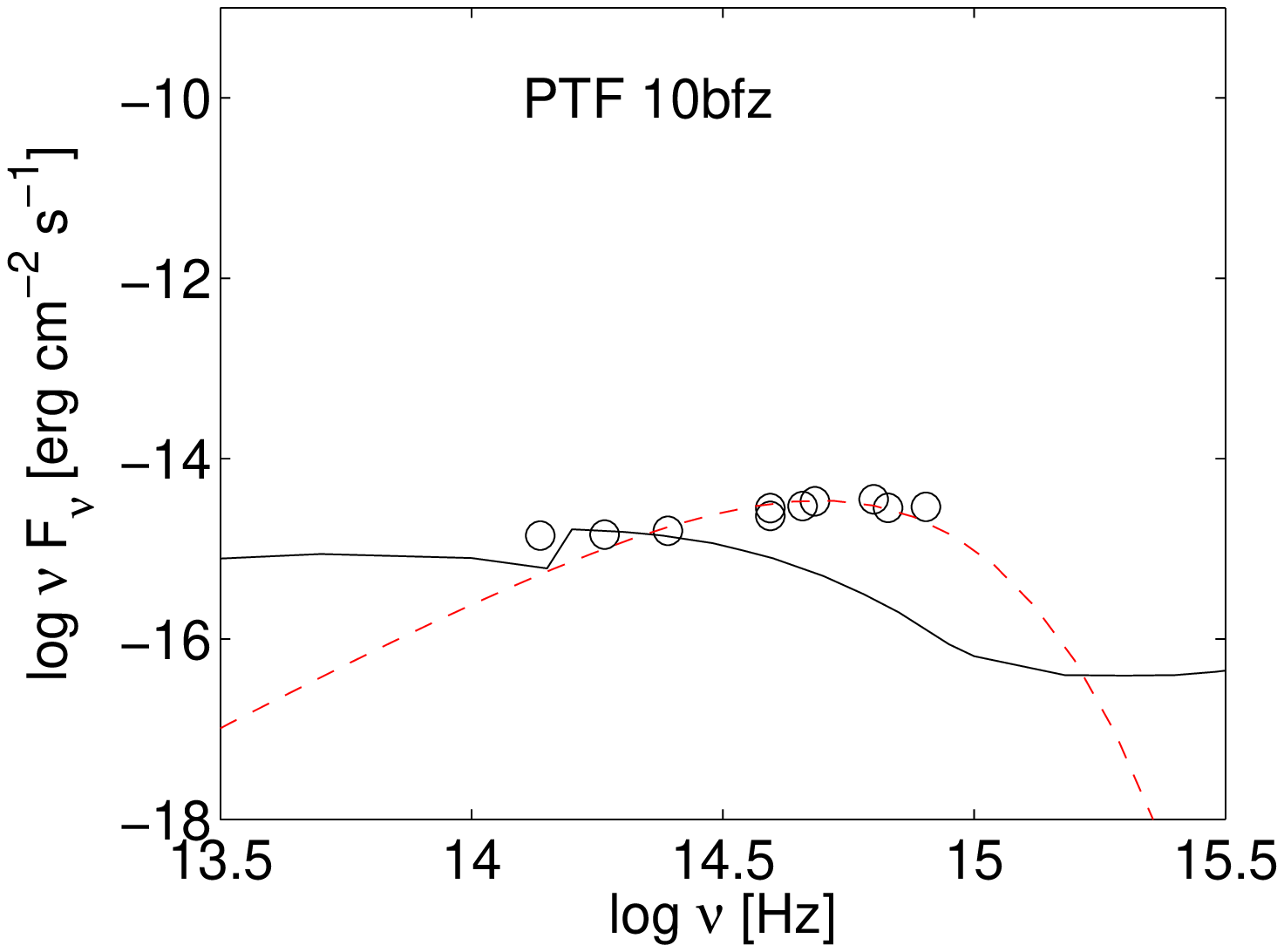}
\includegraphics[width=4.6cm]{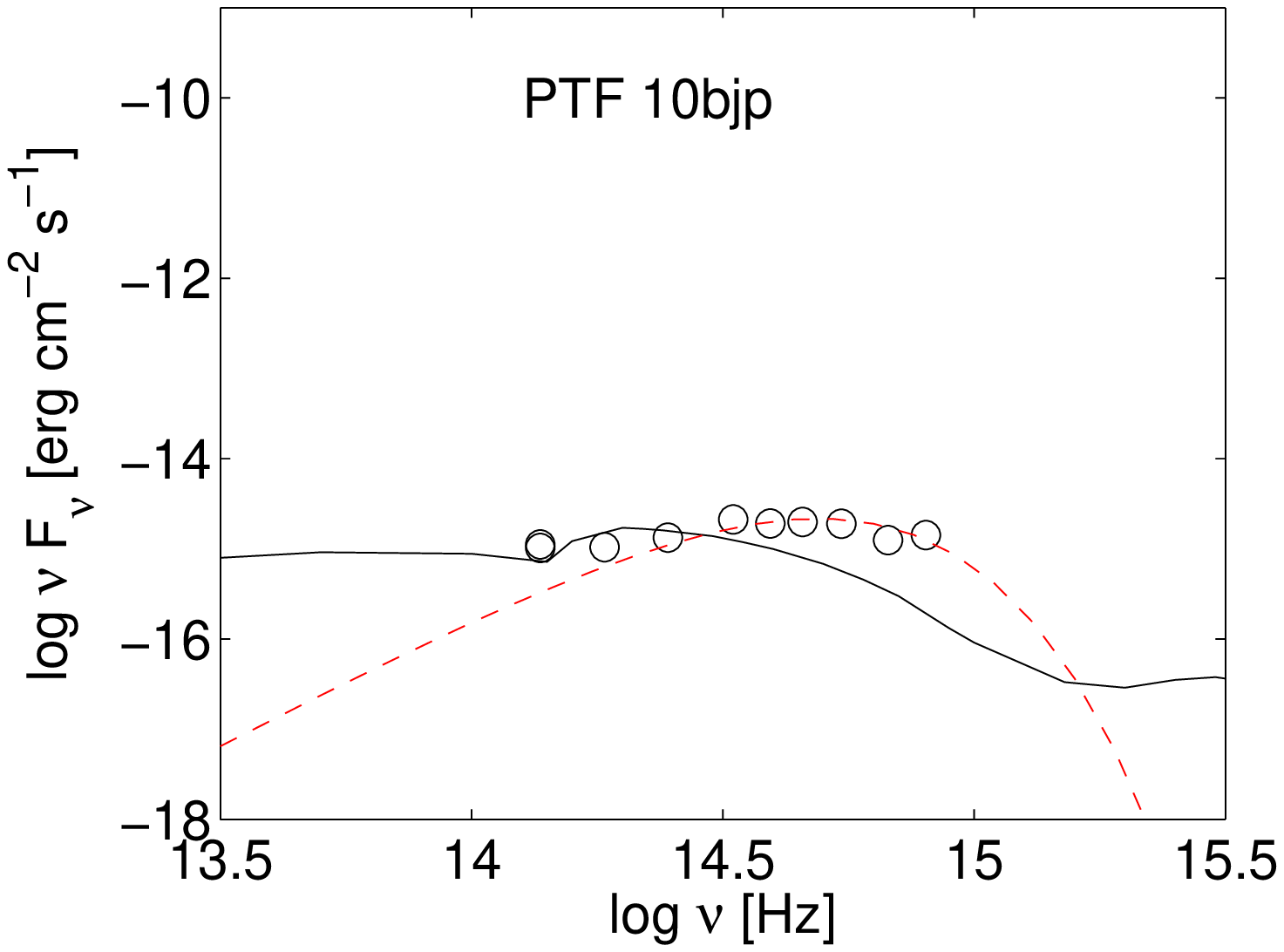}
\includegraphics[width=4.6cm]{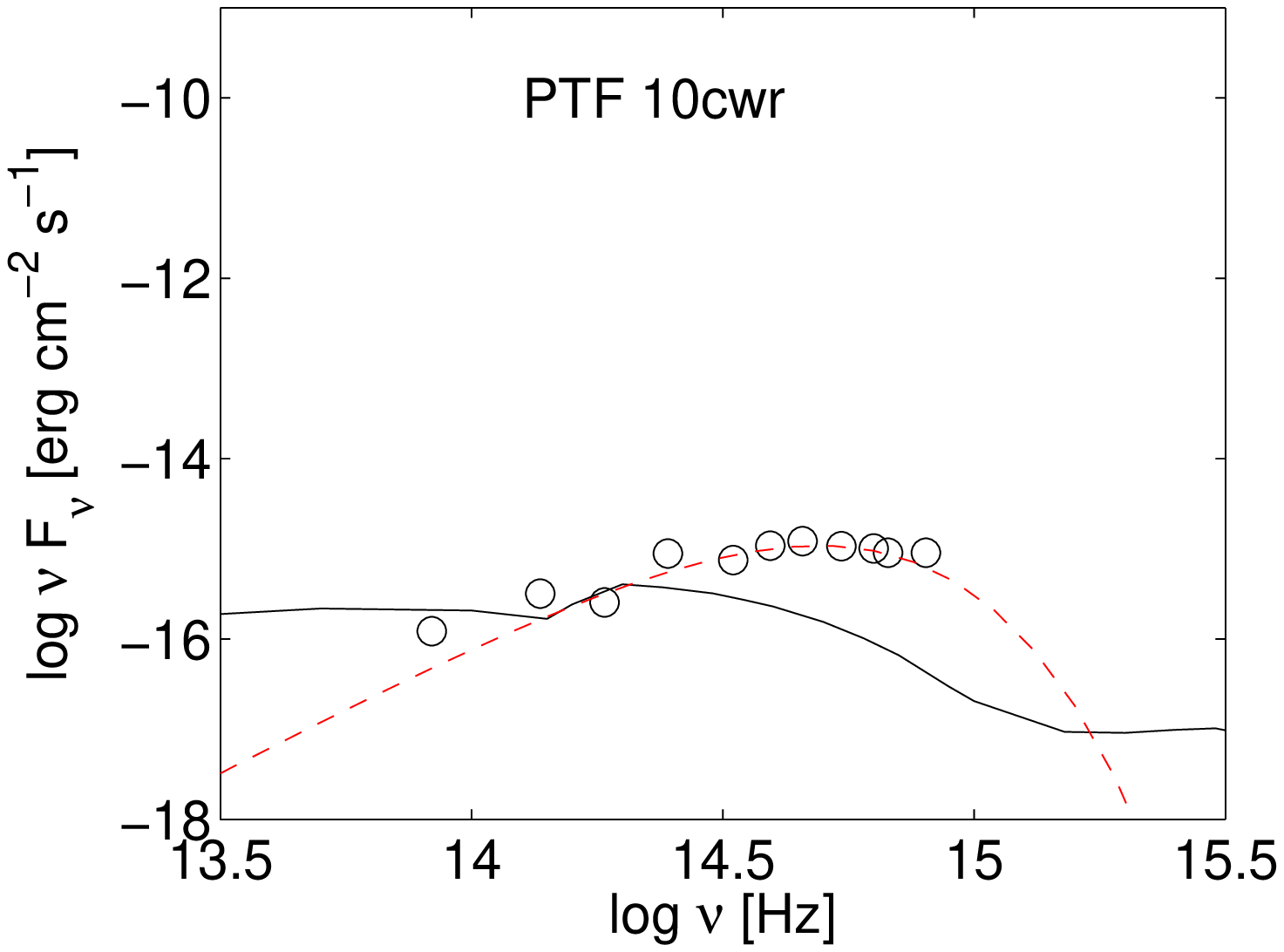}
\includegraphics[width=4.6cm]{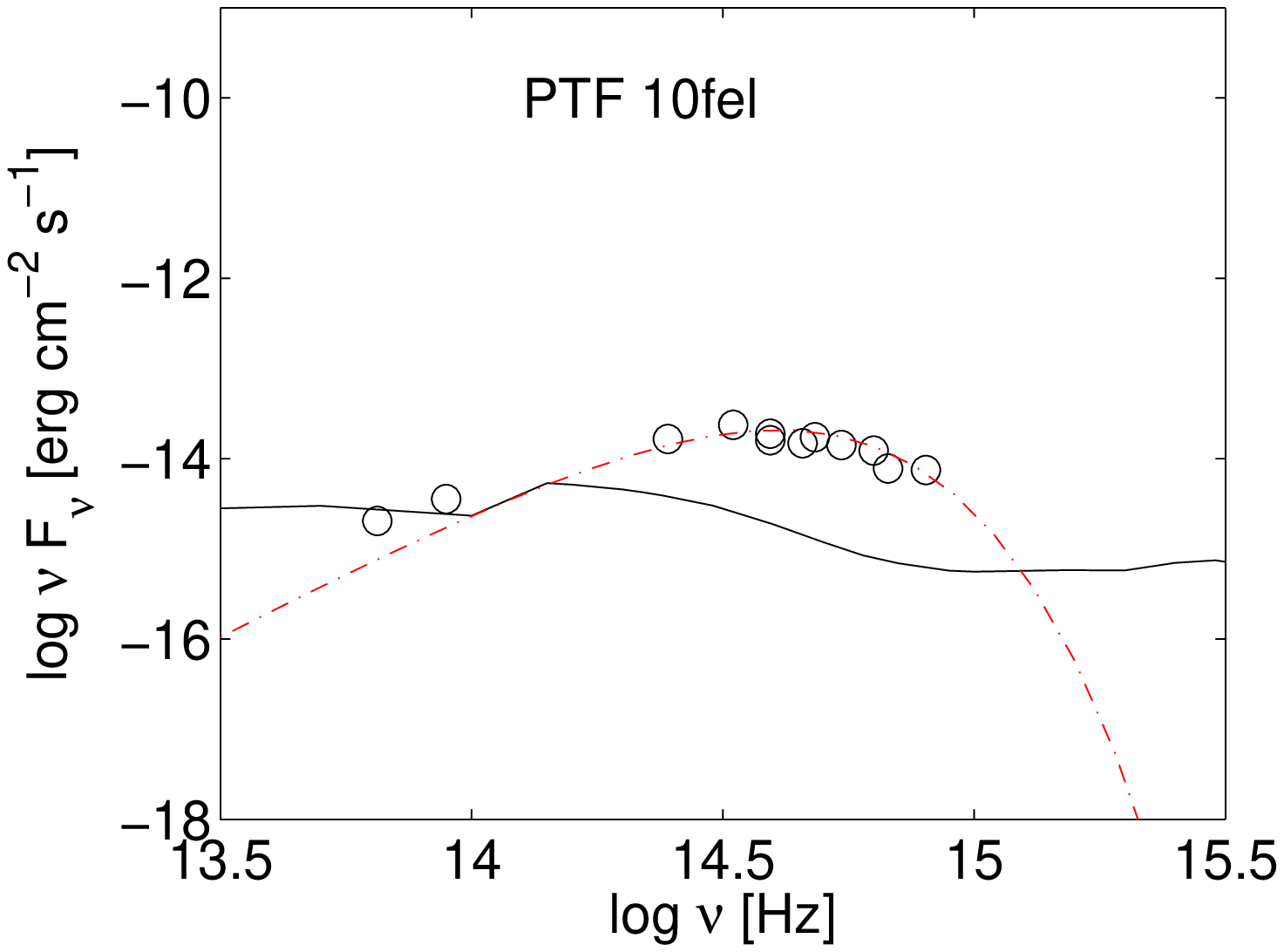}
\includegraphics[width=4.6cm]{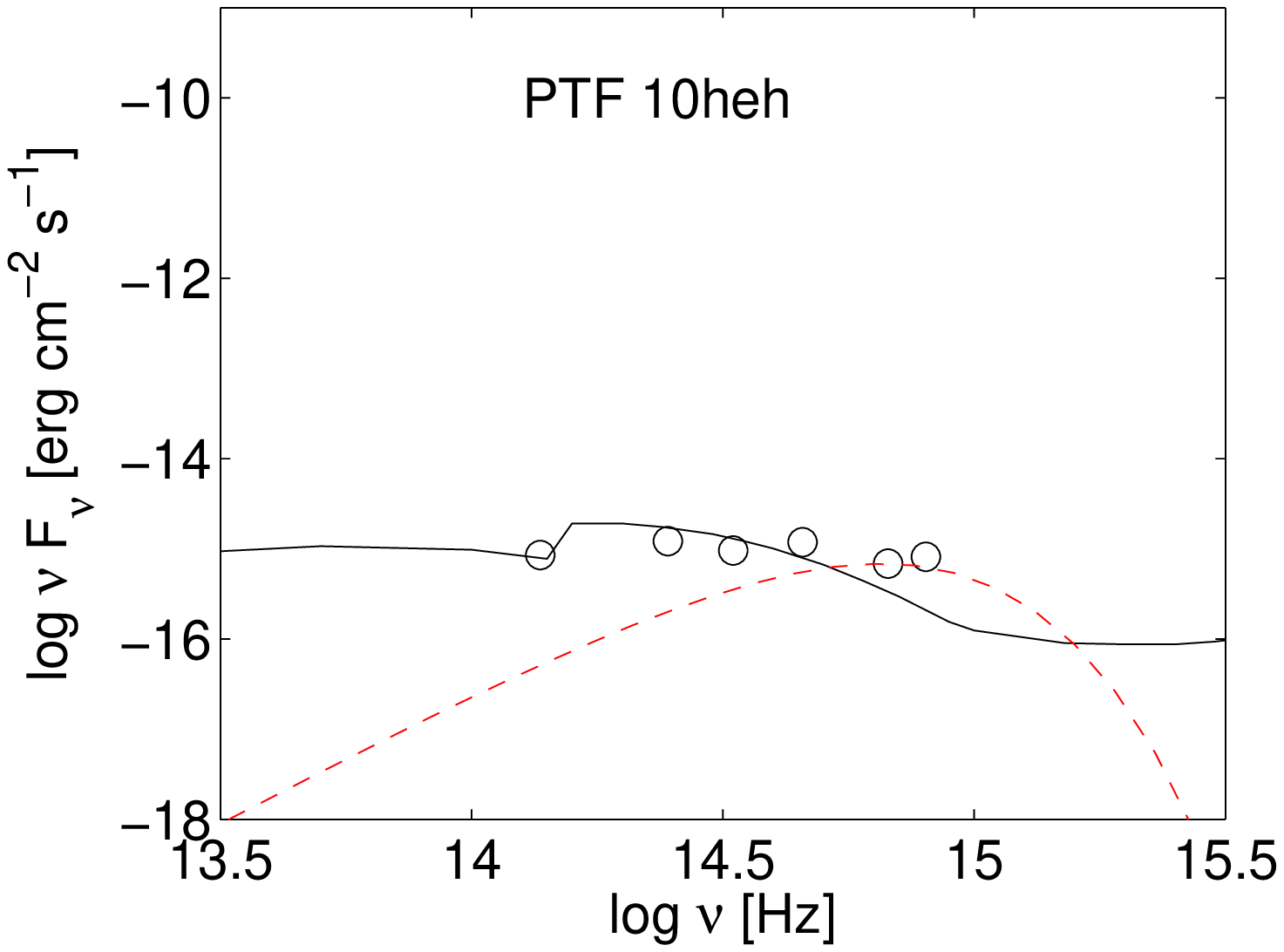}
\includegraphics[width=4.6cm]{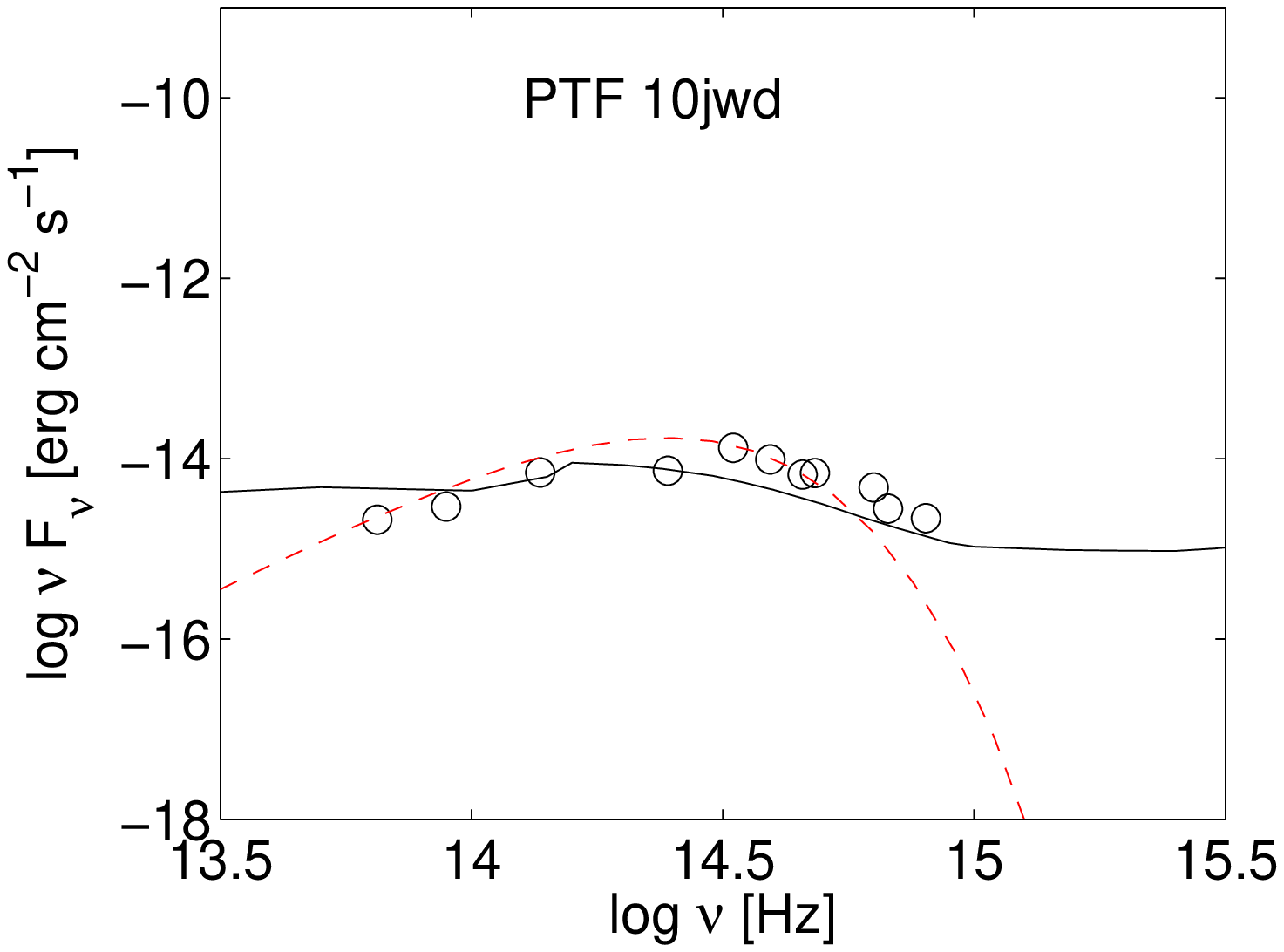}
\includegraphics[width=4.6cm]{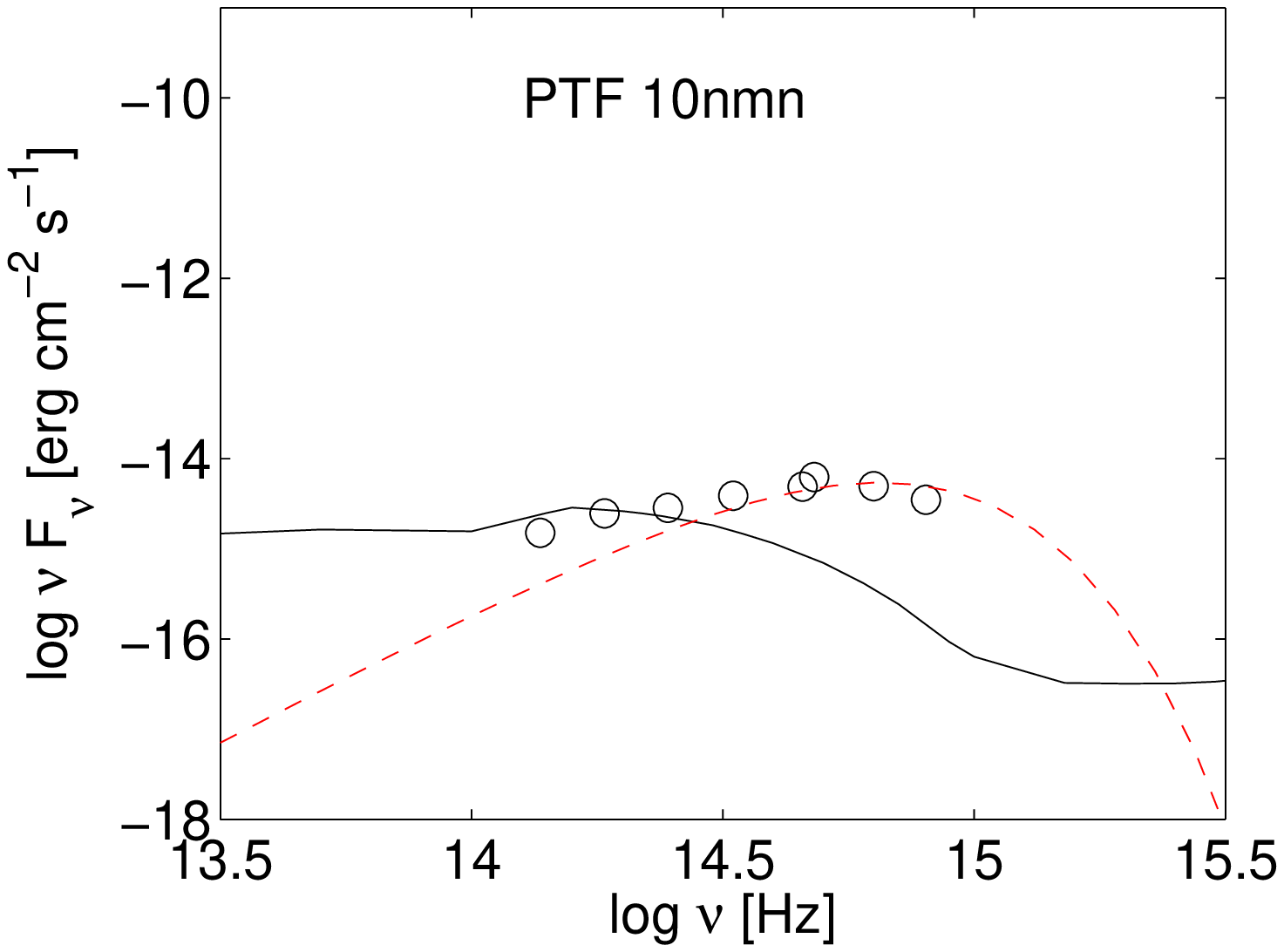}
\includegraphics[width=4.6cm]{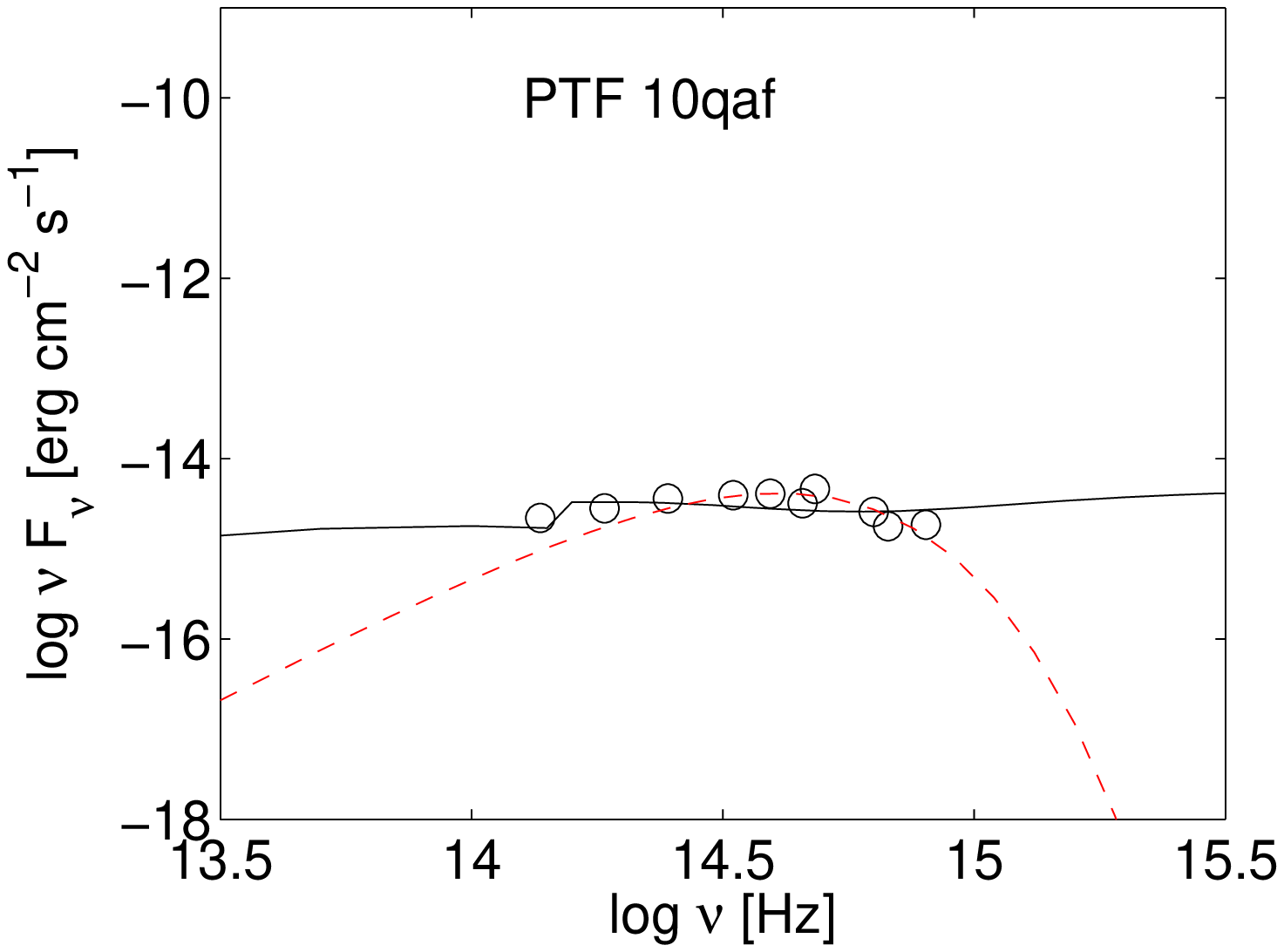}
\includegraphics[width=4.6cm]{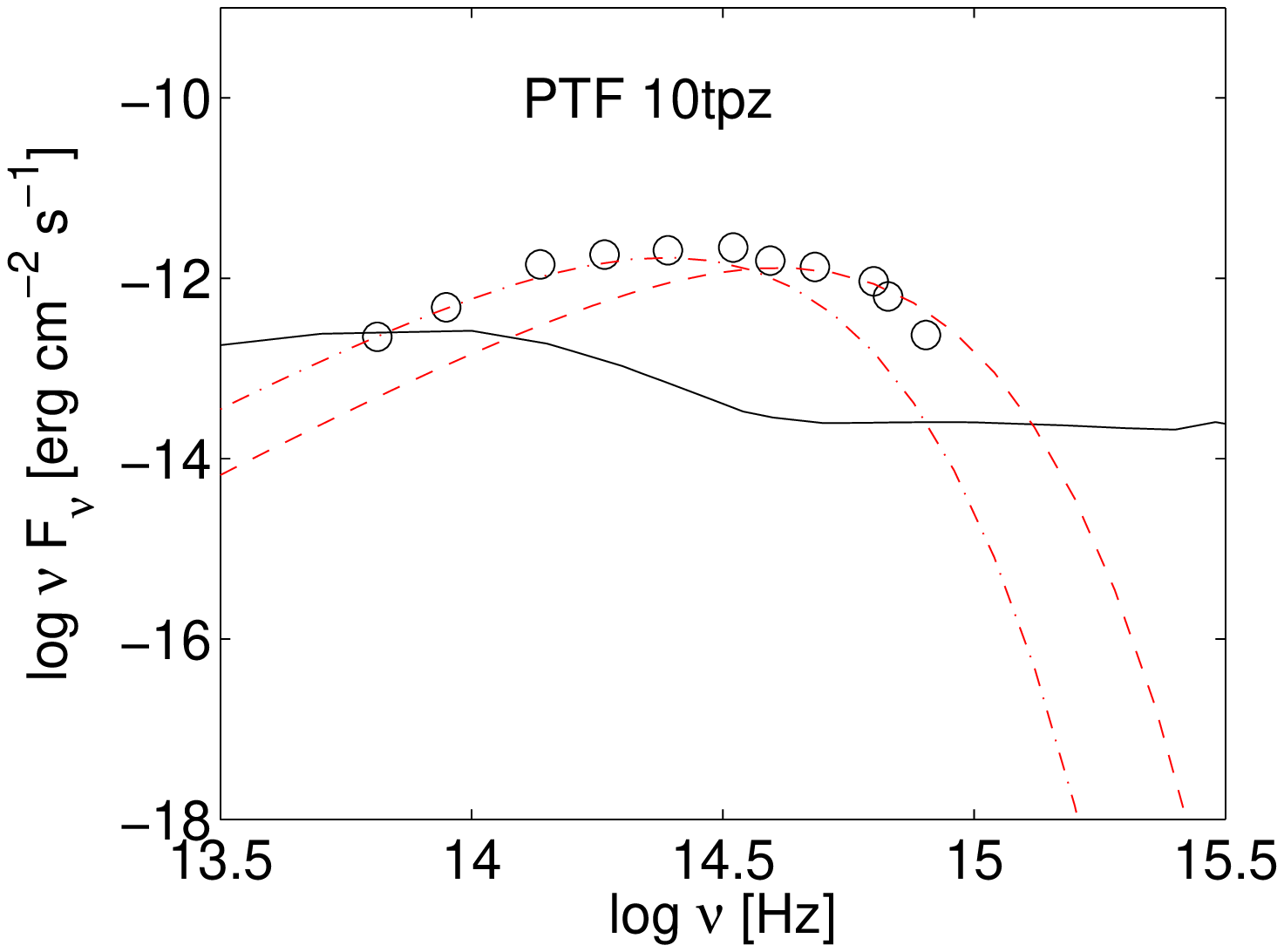}
\includegraphics[width=4.6cm]{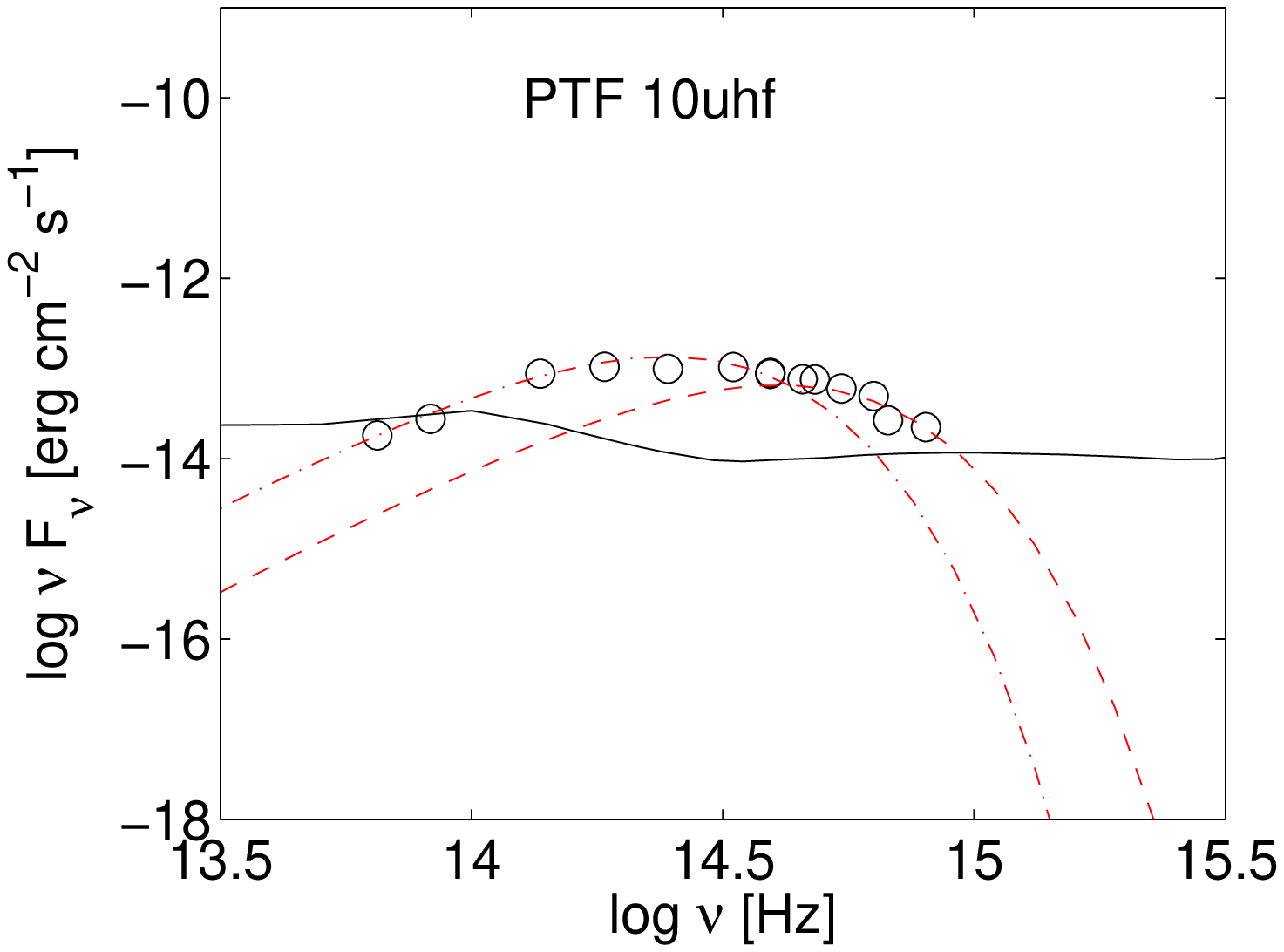}
\includegraphics[width=4.6cm]{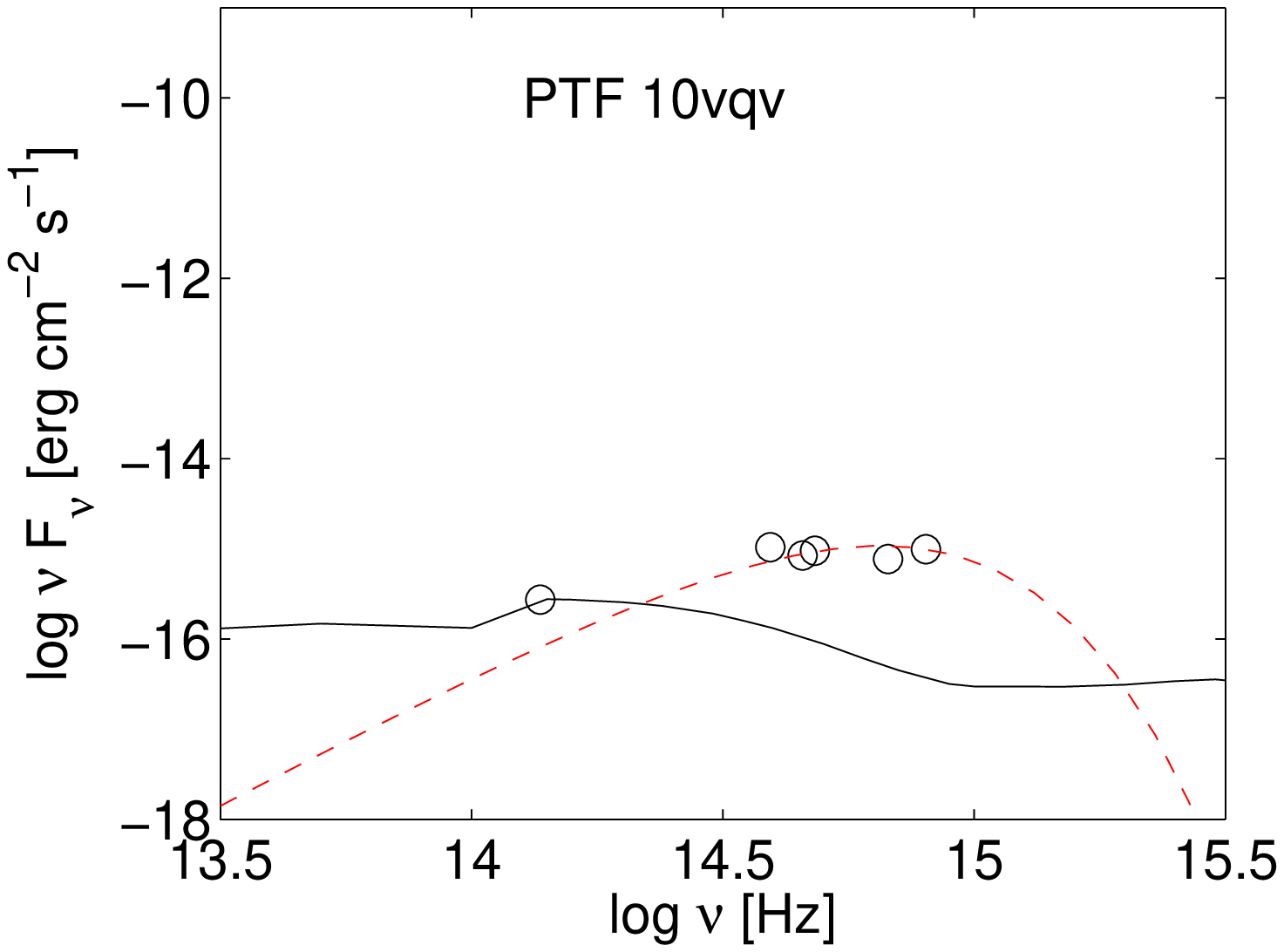}
\includegraphics[width=4.6cm]{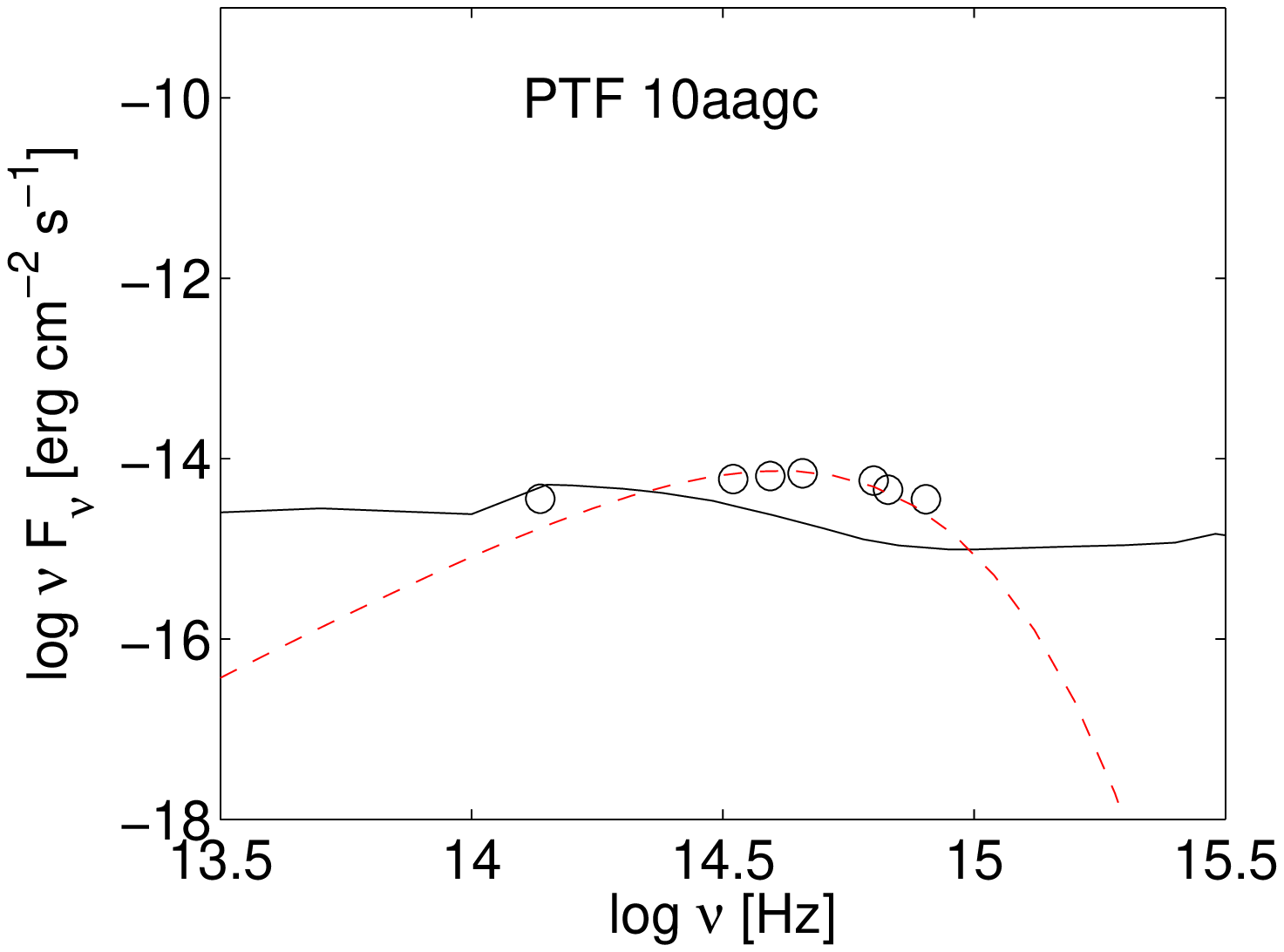}
\includegraphics[width=4.6cm]{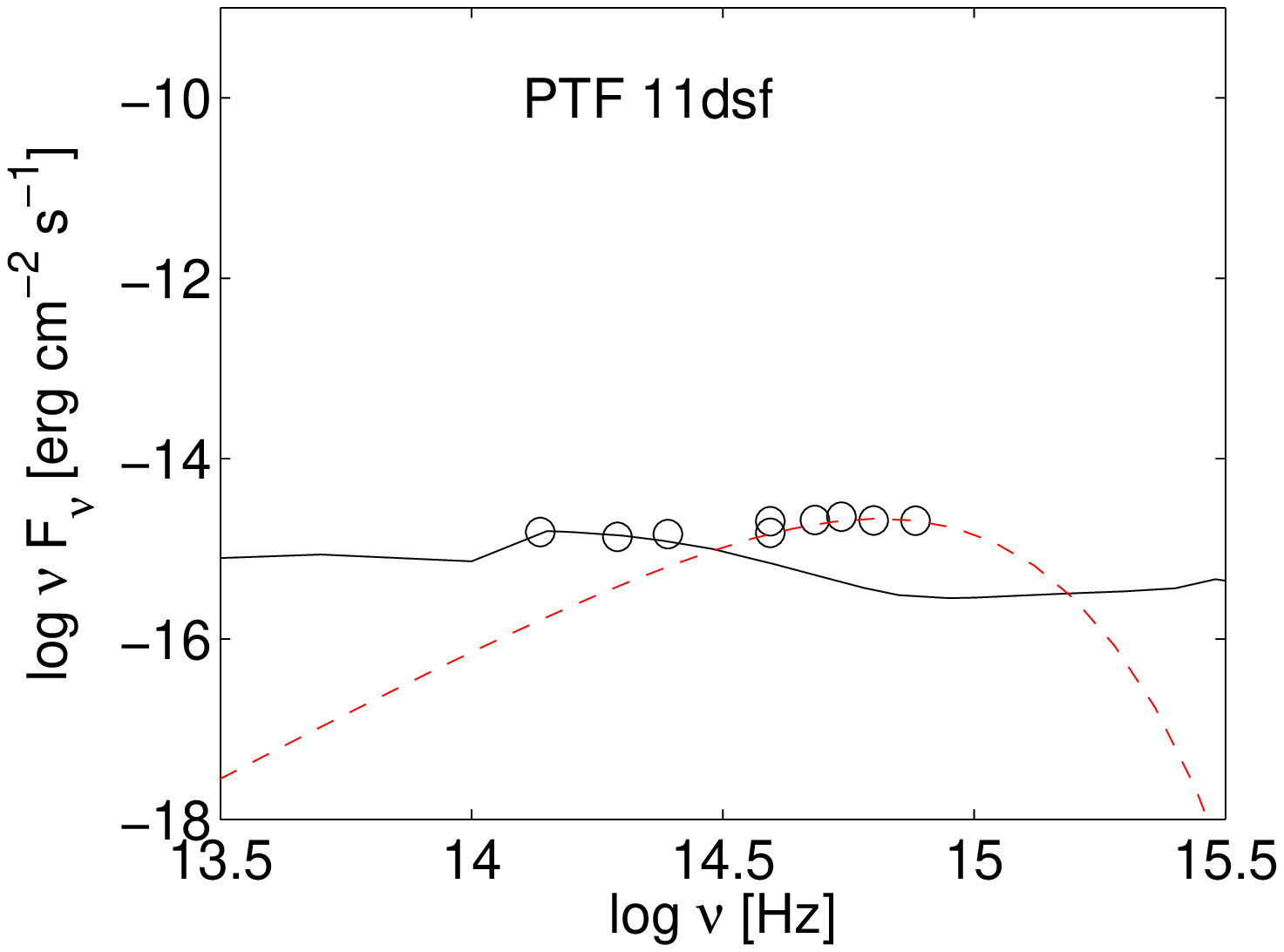}
\includegraphics[width=4.6cm]{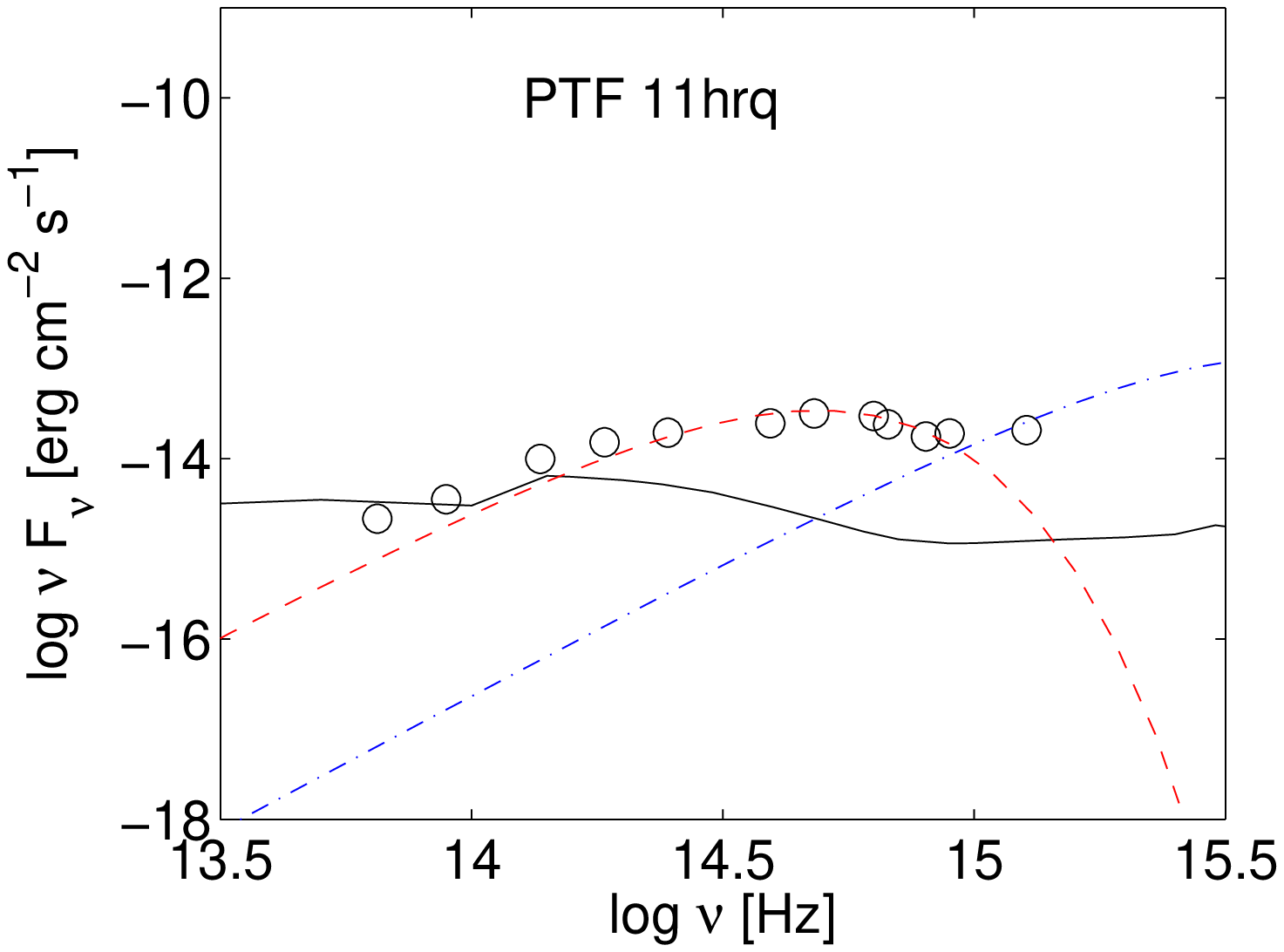}
\includegraphics[width=4.6cm]{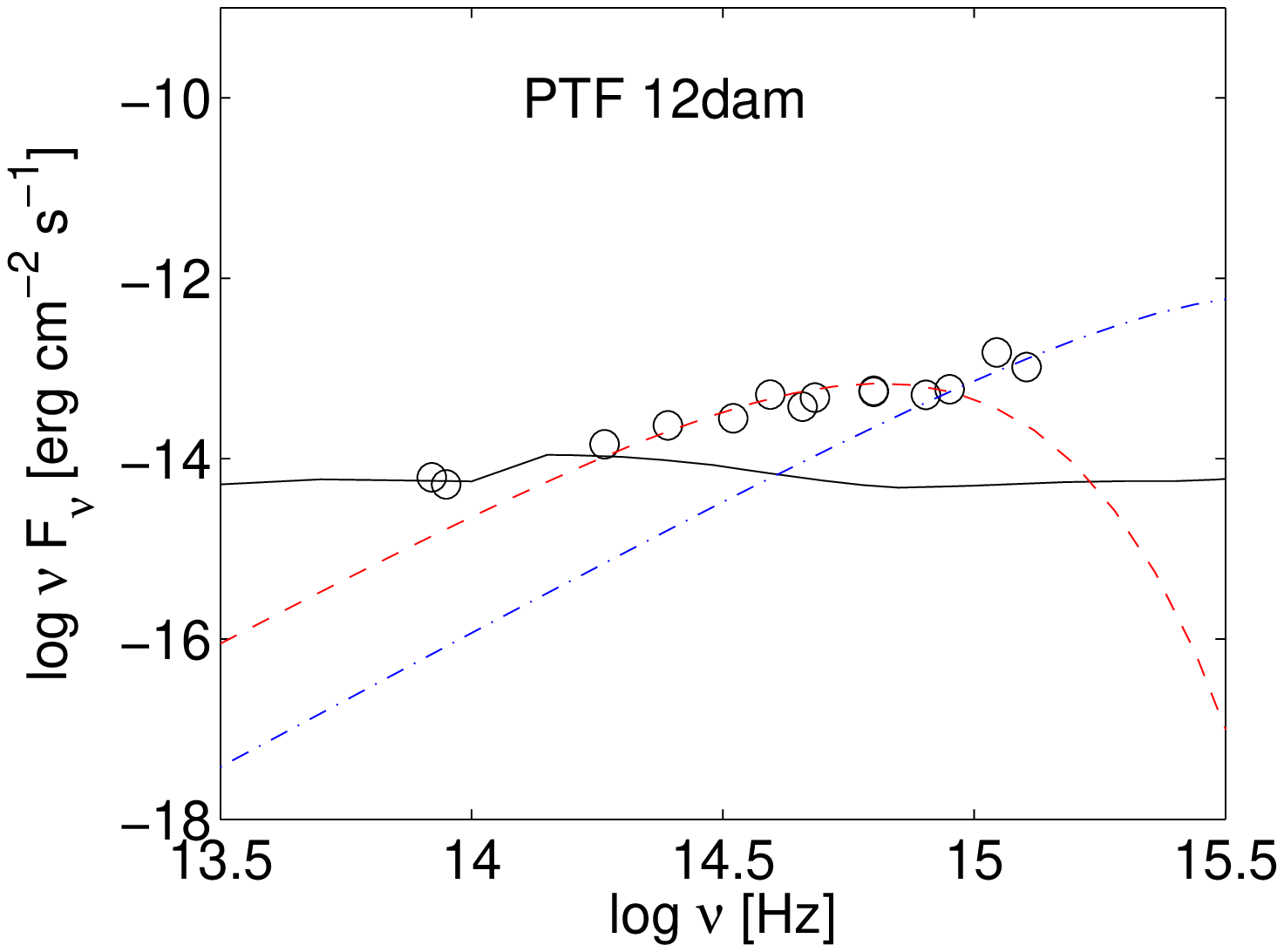}
\includegraphics[width=4.6cm]{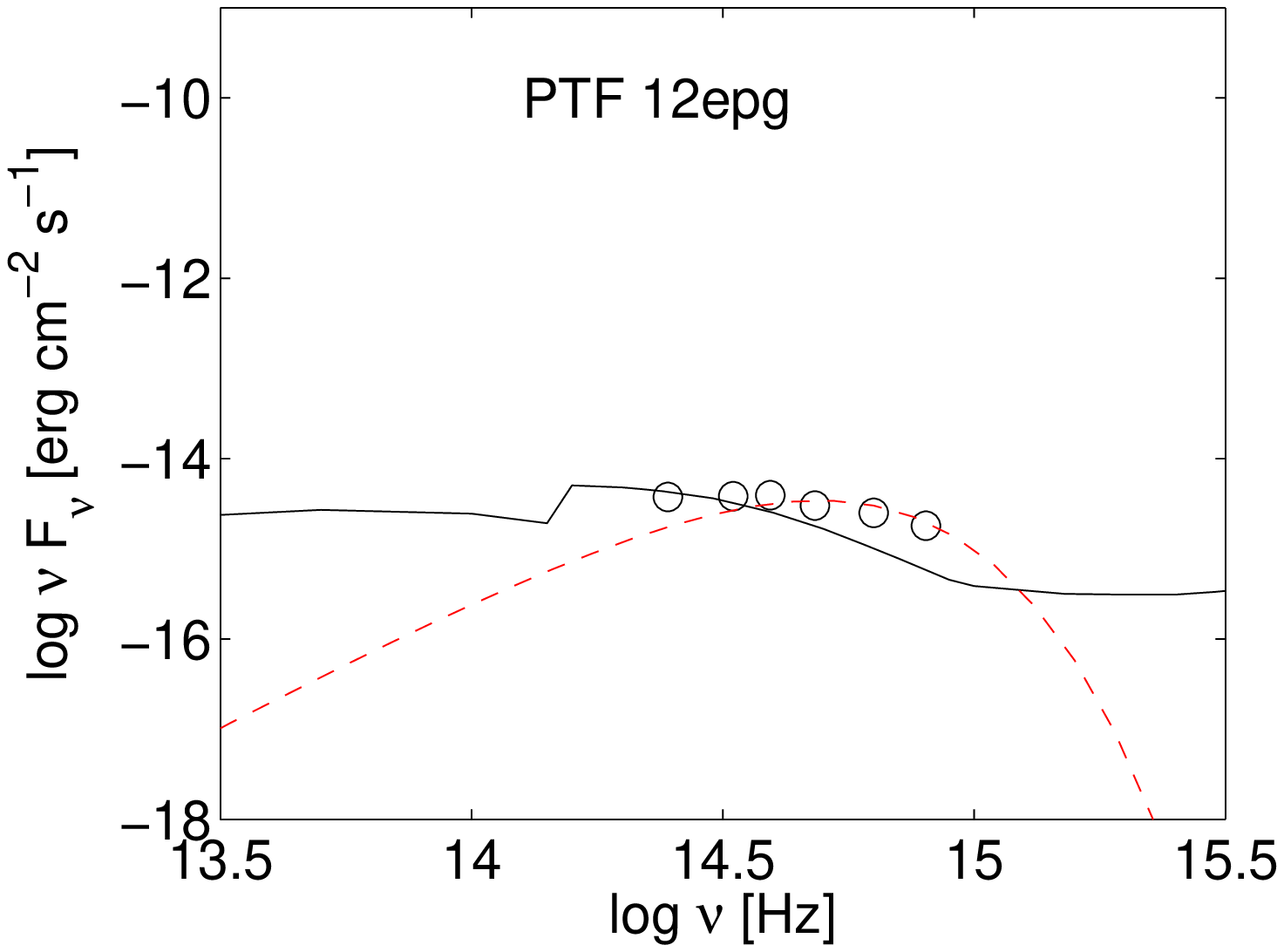}
\includegraphics[width=4.6cm]{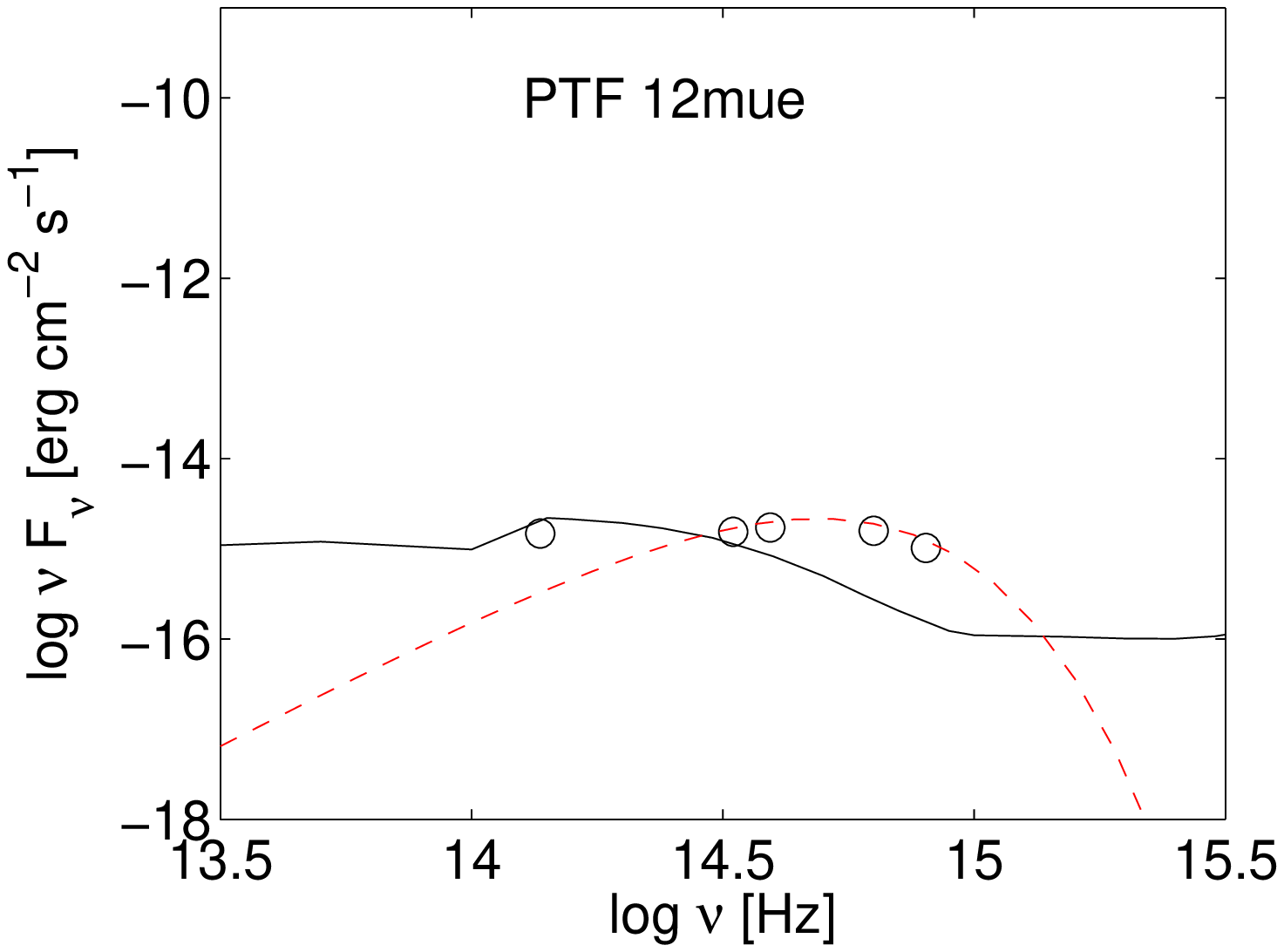}
\caption{SEDs of the Perley et al (2016) survey. Open circles : the data; solid lines : results obtained
by the detailed modelling of the line spectra; dashed lines : bb fluxes referring  to  temperatures T$_{bb}$.
}
\end{figure*}

\subsection{The SB-AGN sample by Ramos Almeida et al (2013)}

Ramos Almeida et al (2013)   examined  star formation    quencing in AGN  
galaxies at redshift 0.27$<$z$<$1.28
  on the basis of  AGN activity  and star formation delay.
To investigate
  the interactivity between AGN and star formation in galaxies
they  presented near-IR (NIR) spectroscopic observations of 28 X-ray  
and mid-infrared selected sources at median redshift
z$\sim$ 0.8 in the Extended Groth Strip :
  13 AGN dominated and 15  host-galaxy dominated. New NIR spectra  
referring to the continuum SEDs
of  objects at  z between $\sim$ 0.4 and $\sim$1.25 were  reported. Moreover,
they  showed the spectra of an AGN subset at 0.28$<$z$<$1.28
  including   \Ha~ and other key optical lines for each of them.
The data   reported  by  Ramos Almeida et al
in their table 3 and completed  by the data presented in their table 4  
  were observed by  the Long- Slit
Intermediate Resolution Infrared Spectrograph (LIRIS) at the 4.2m  
William Herschel Telescope (WHT) from the
Deep Extragalactic Evolutionary Probe 2 (DEEP2).
  SFRs  were   calculated  from the \Ha luminosities   and
the line fluxes  were obtained   subtracting the continuum SED from each  
observed spectrum.
The interpretation of the SED was done by the diagnostic SEDs of  
Polletta et al. (2007), that
  are based on different types of galaxies (AGN type 1, type 2,  SB, etc.).
The  diagnostics  result   from averages  on   hundreds of objects.

We have  selected   9 out of the 28 galaxies  from the Ramos Almeida et al sample which
  show  enough line ratios for a  suitable modelling  without any   
risk of degeneracy (see Contini 2013).
The modelling of the observed line ratios  from the selected objects
adopting the code {\sc suma}  by Contini (2013, table 2) is shown
in Fig. 6.

\begin{figure*}
\centering
\includegraphics[width=5.8cm]{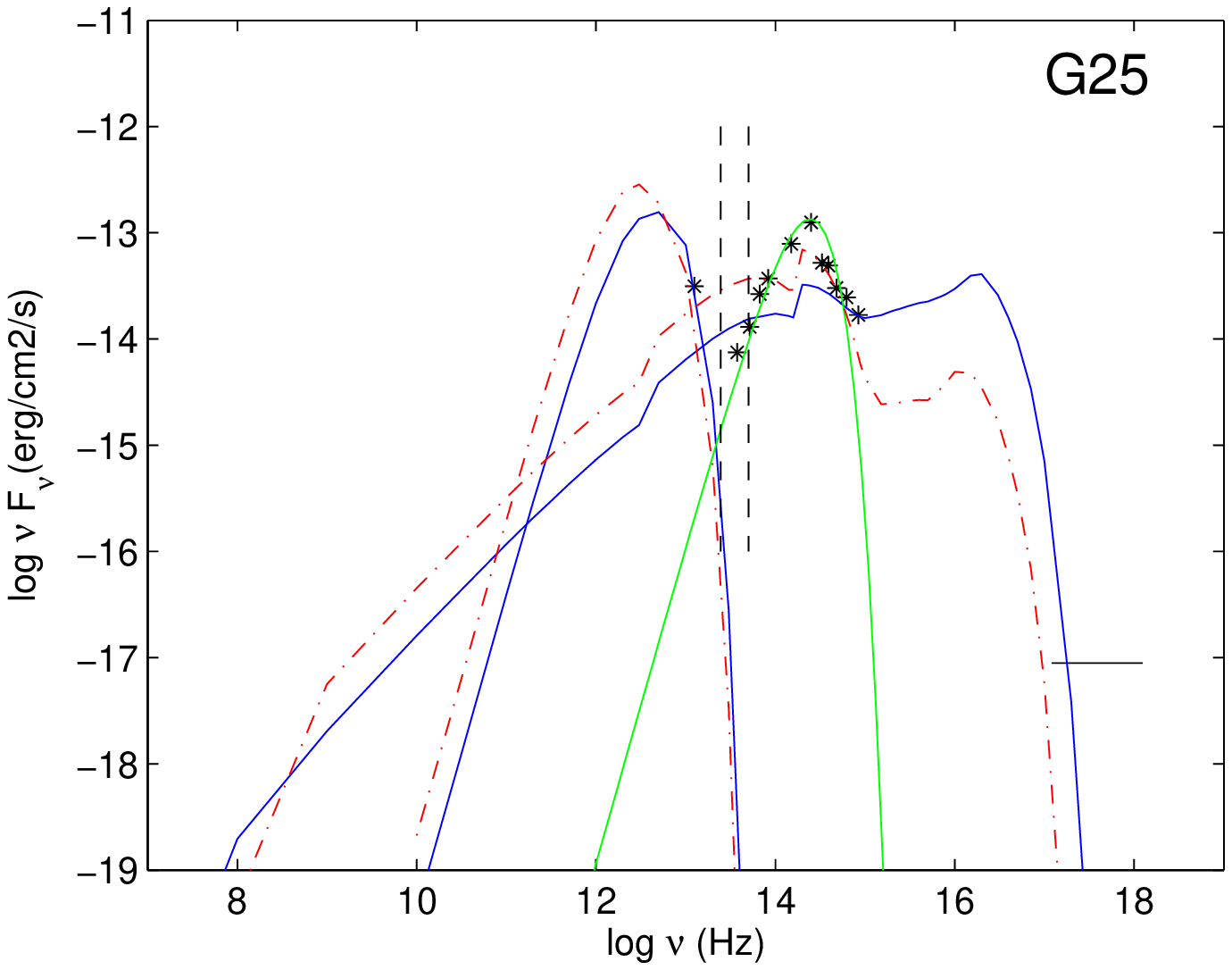}
\includegraphics[width=5.8cm]{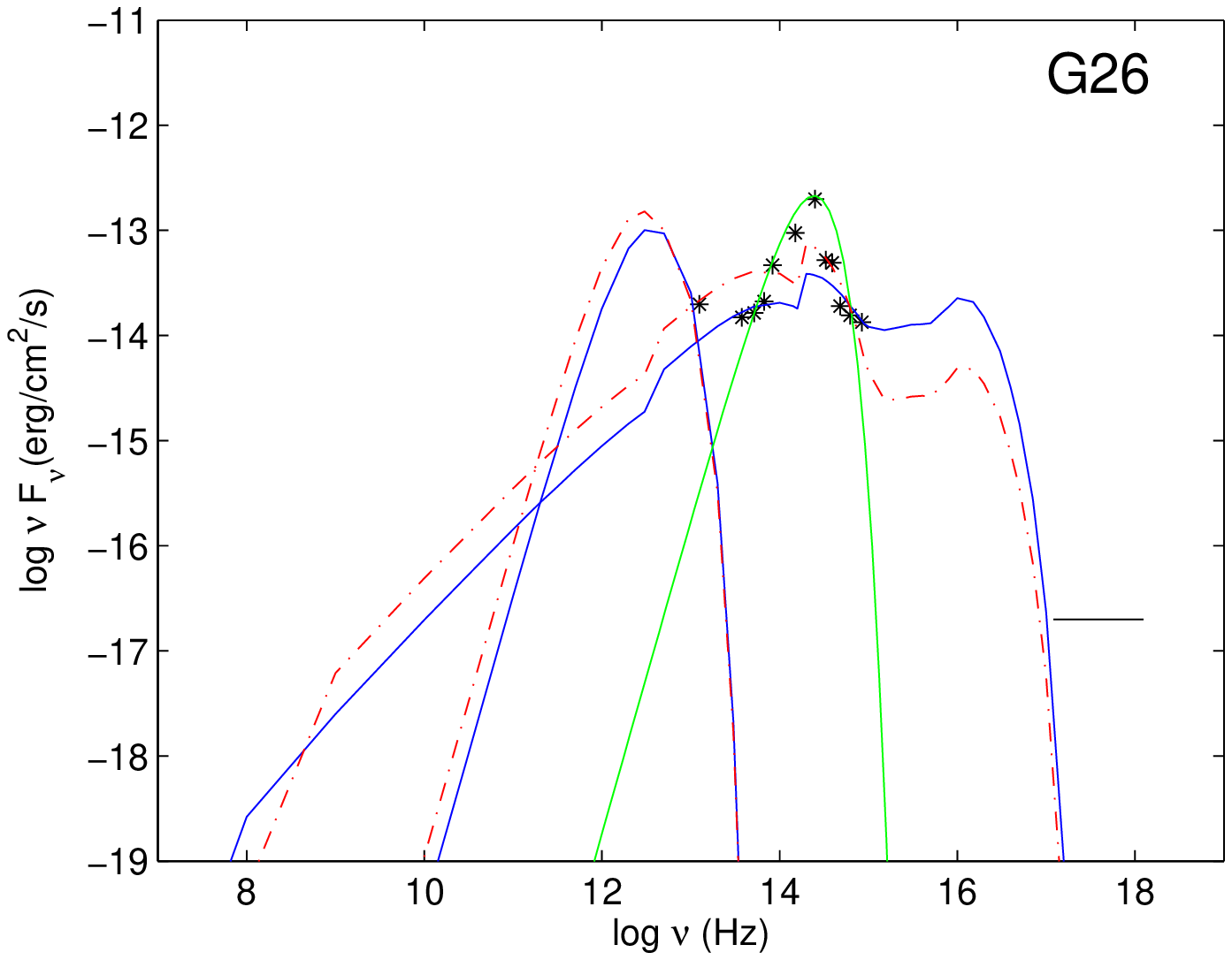}
\includegraphics[width=5.8cm]{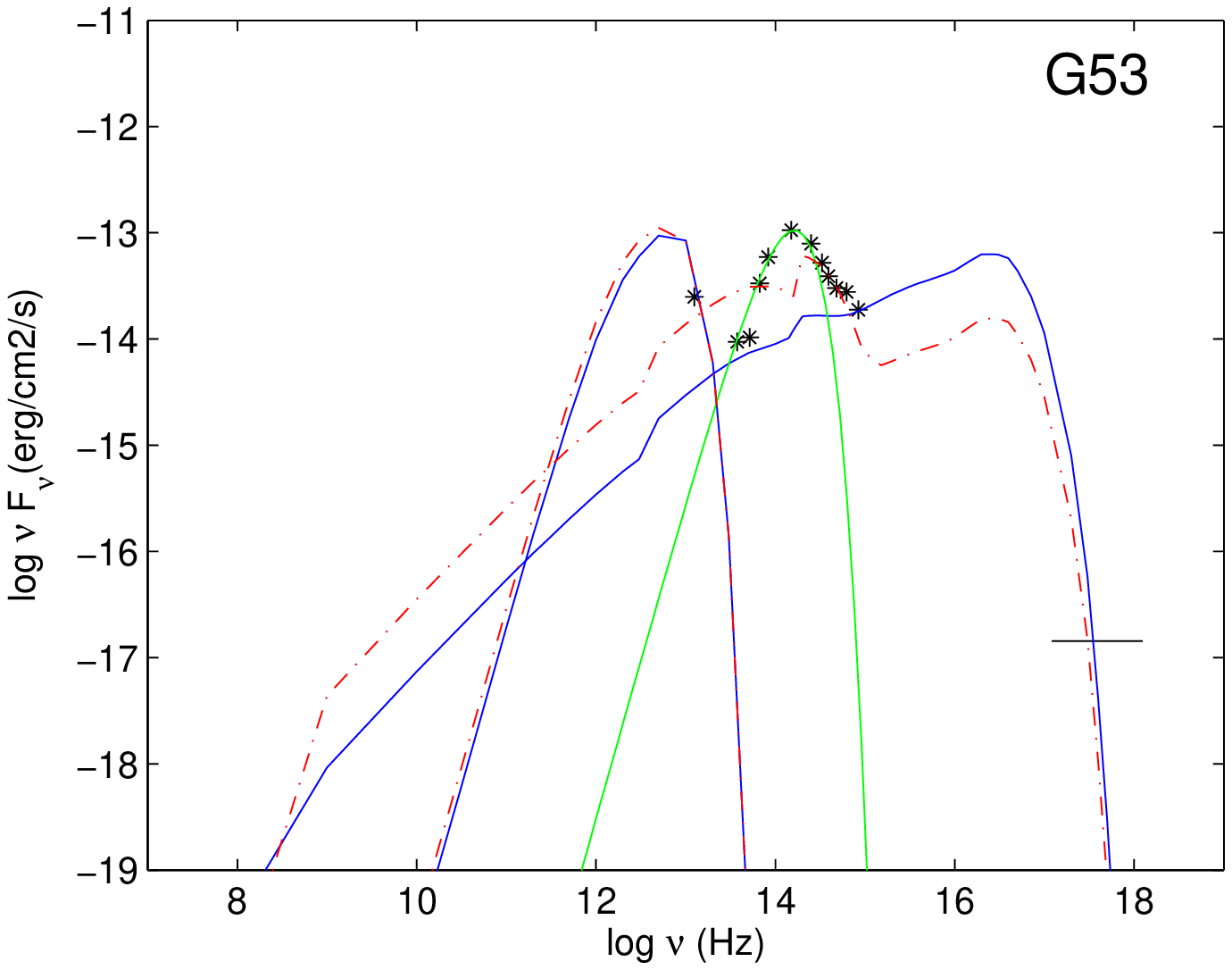}
\includegraphics[width=5.8cm]{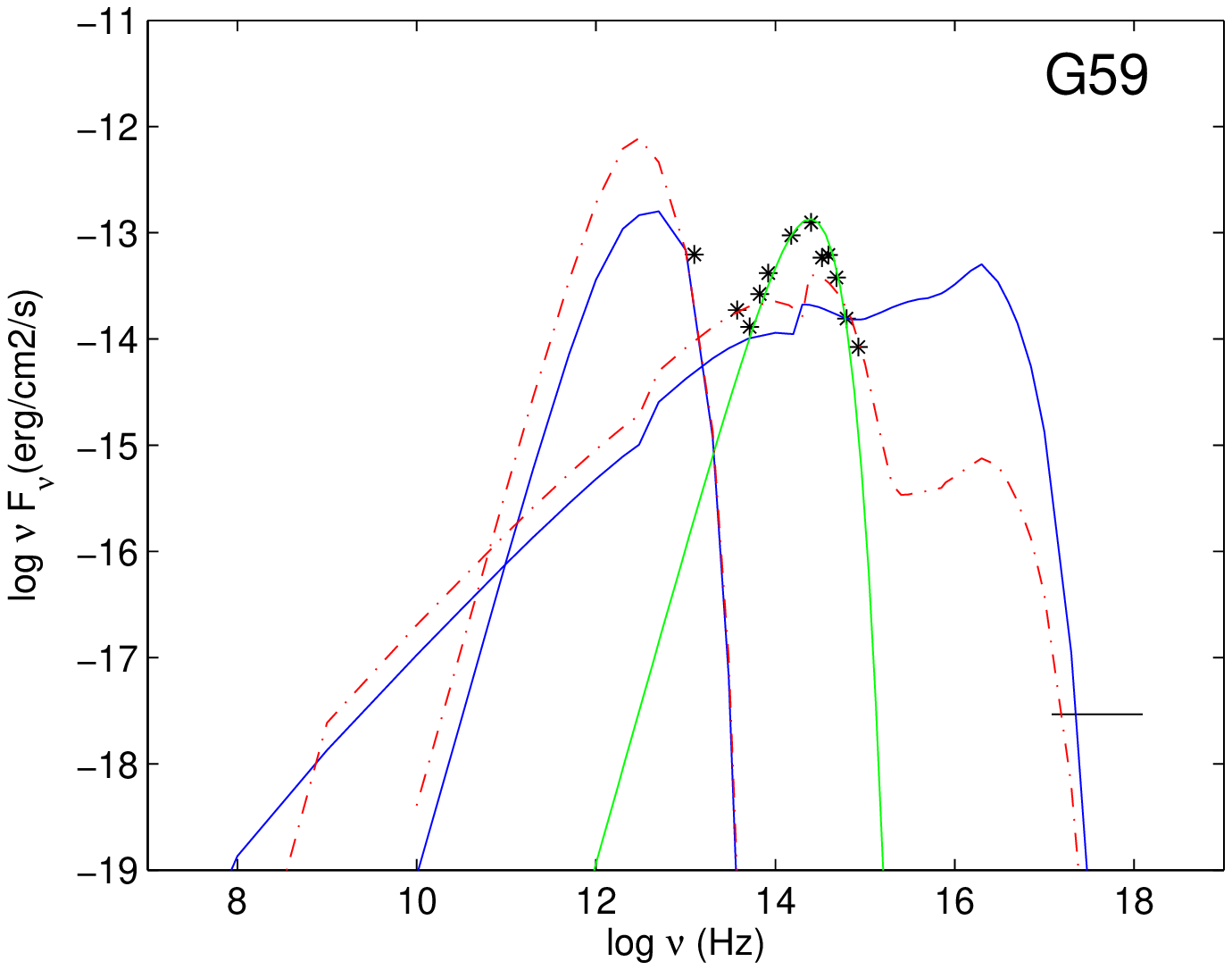}
\includegraphics[width=5.8cm]{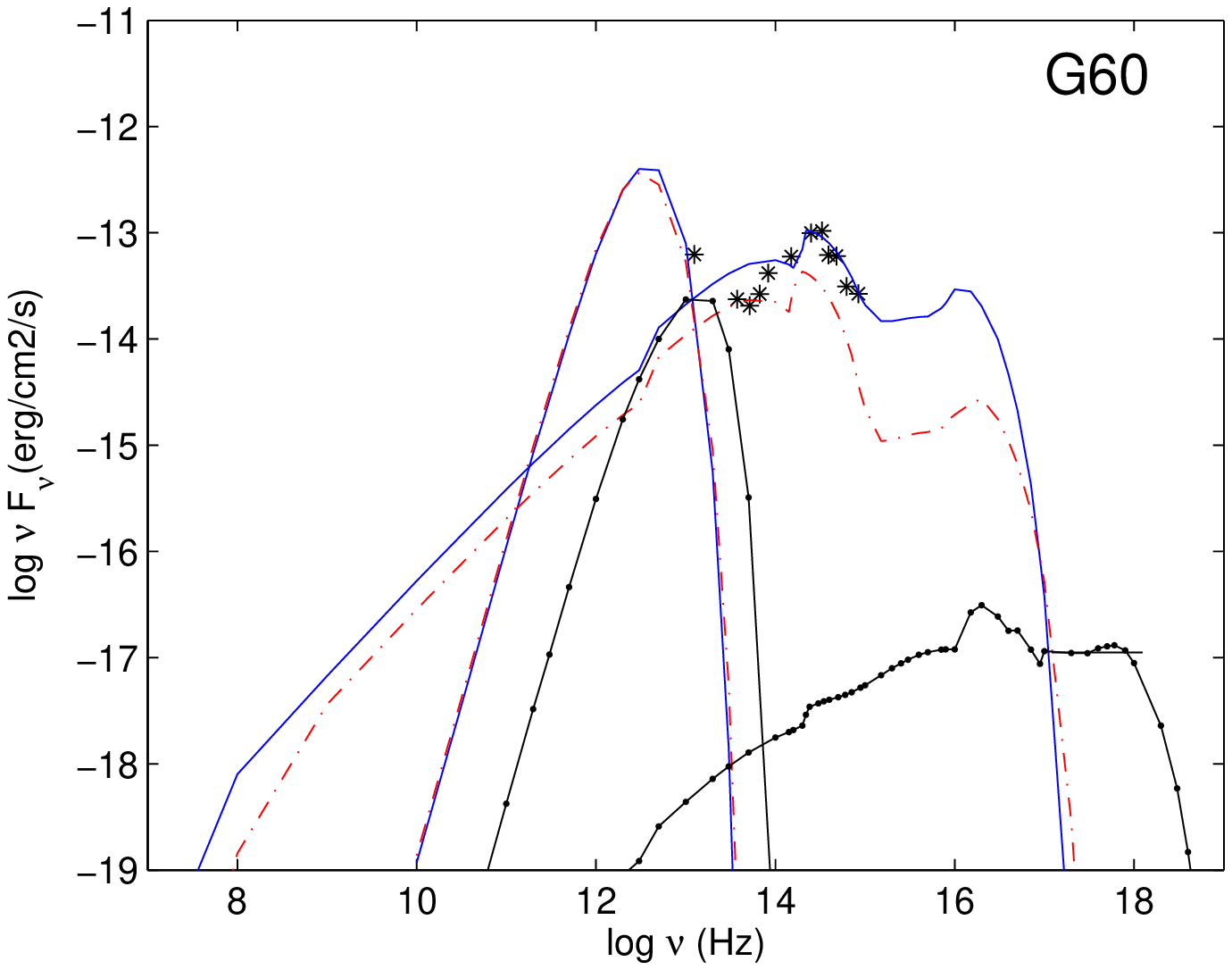}
\includegraphics[width=5.8cm]{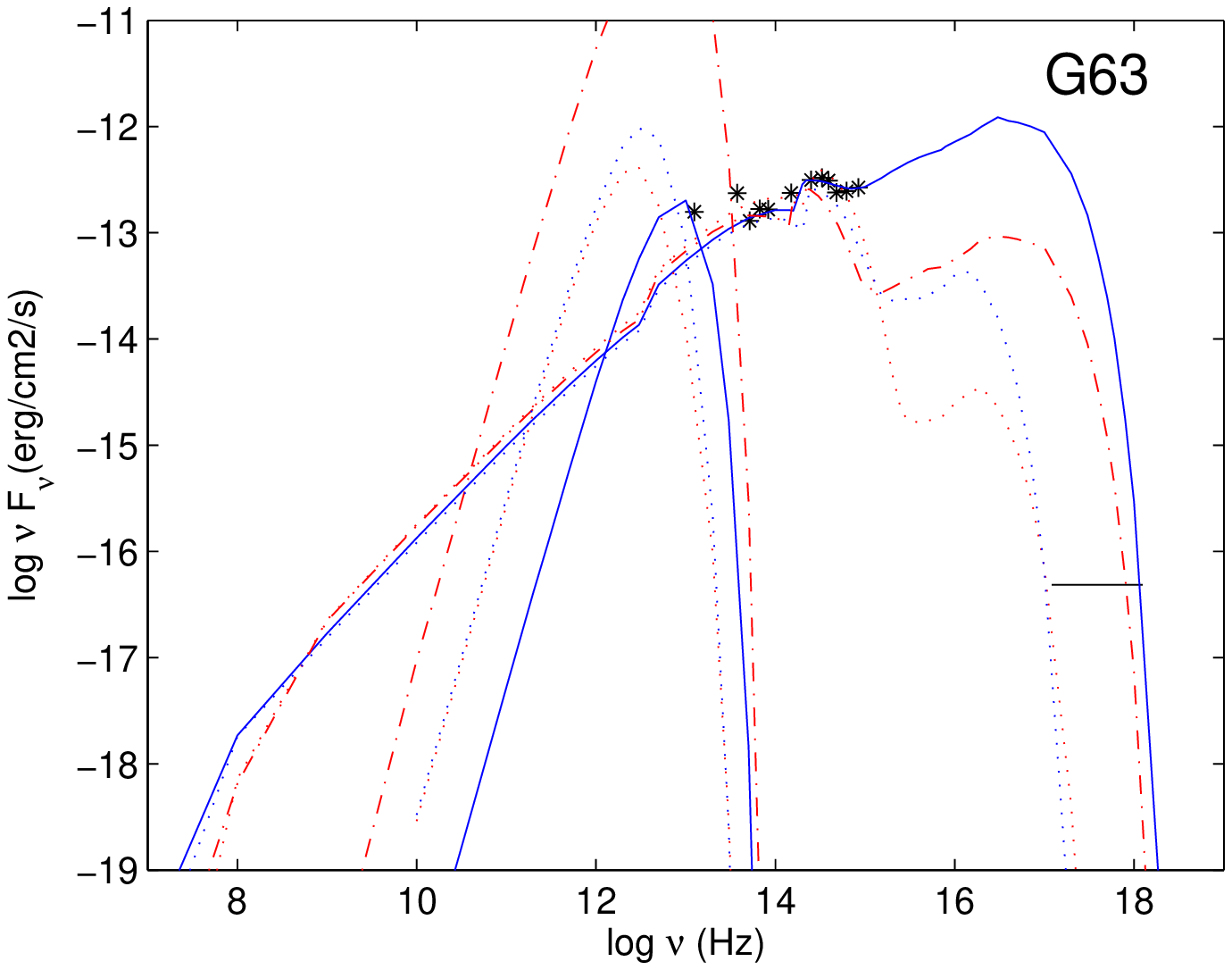}
\includegraphics[width=5.8cm]{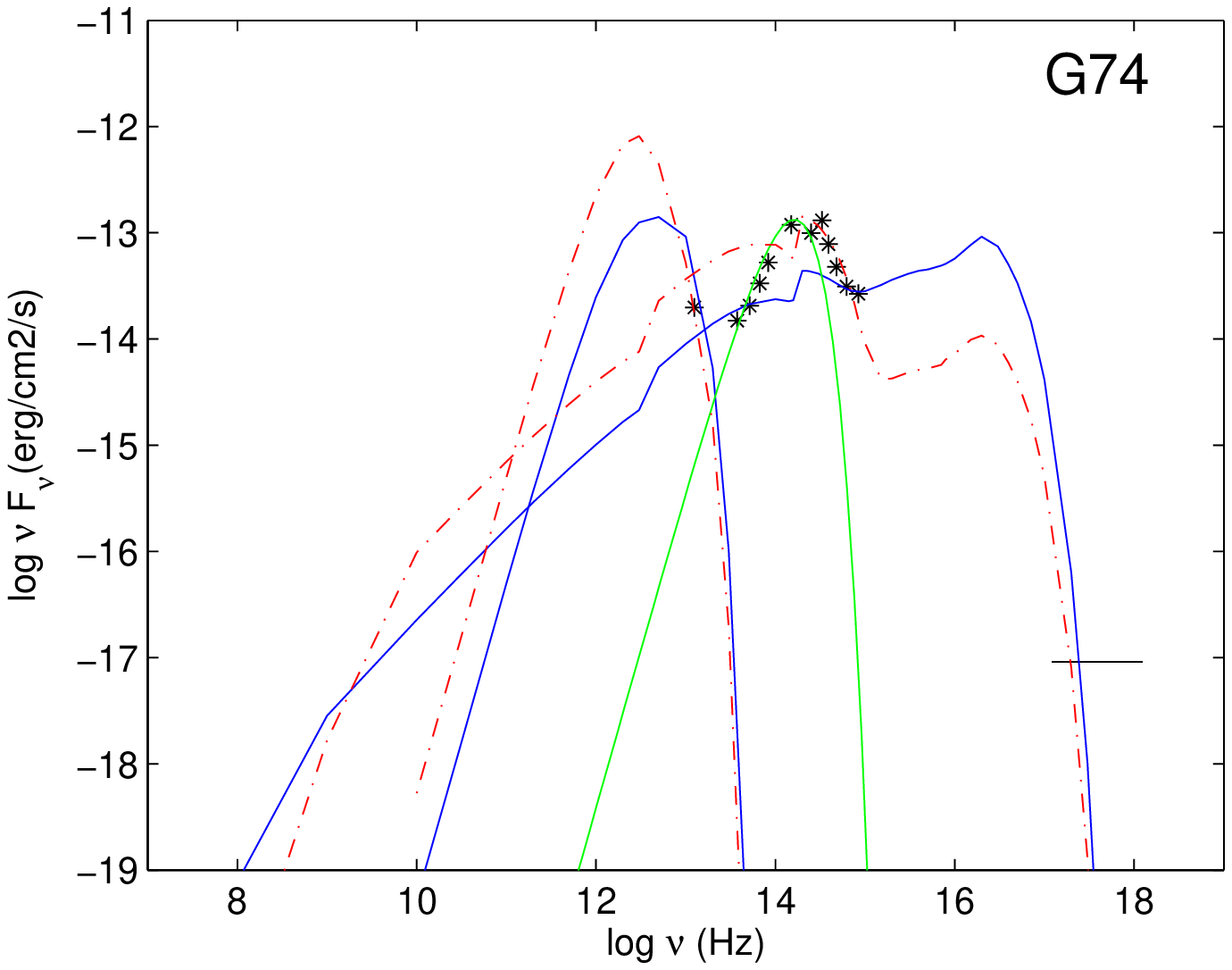}
\includegraphics[width=5.8cm]{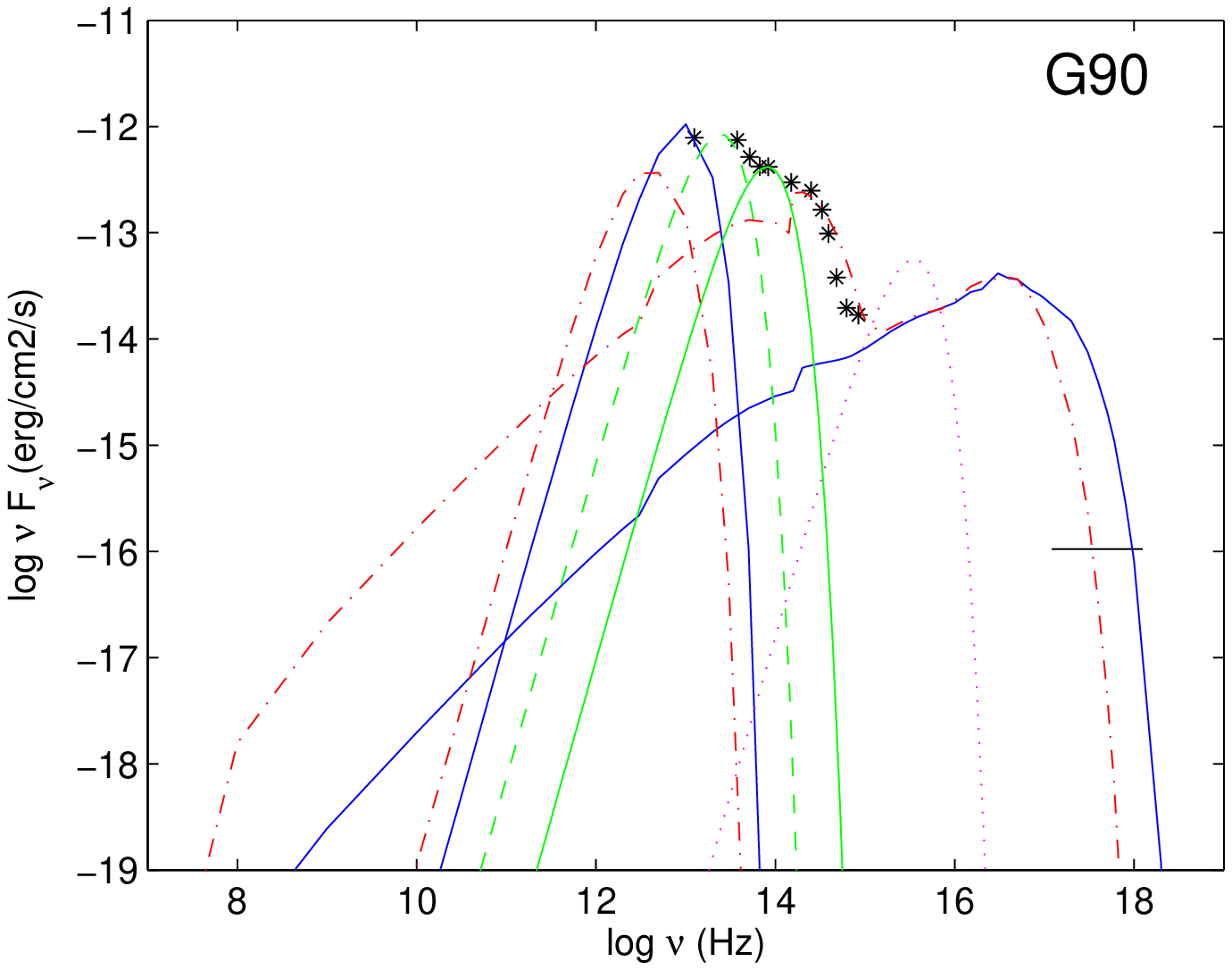}
\includegraphics[width=5.8cm]{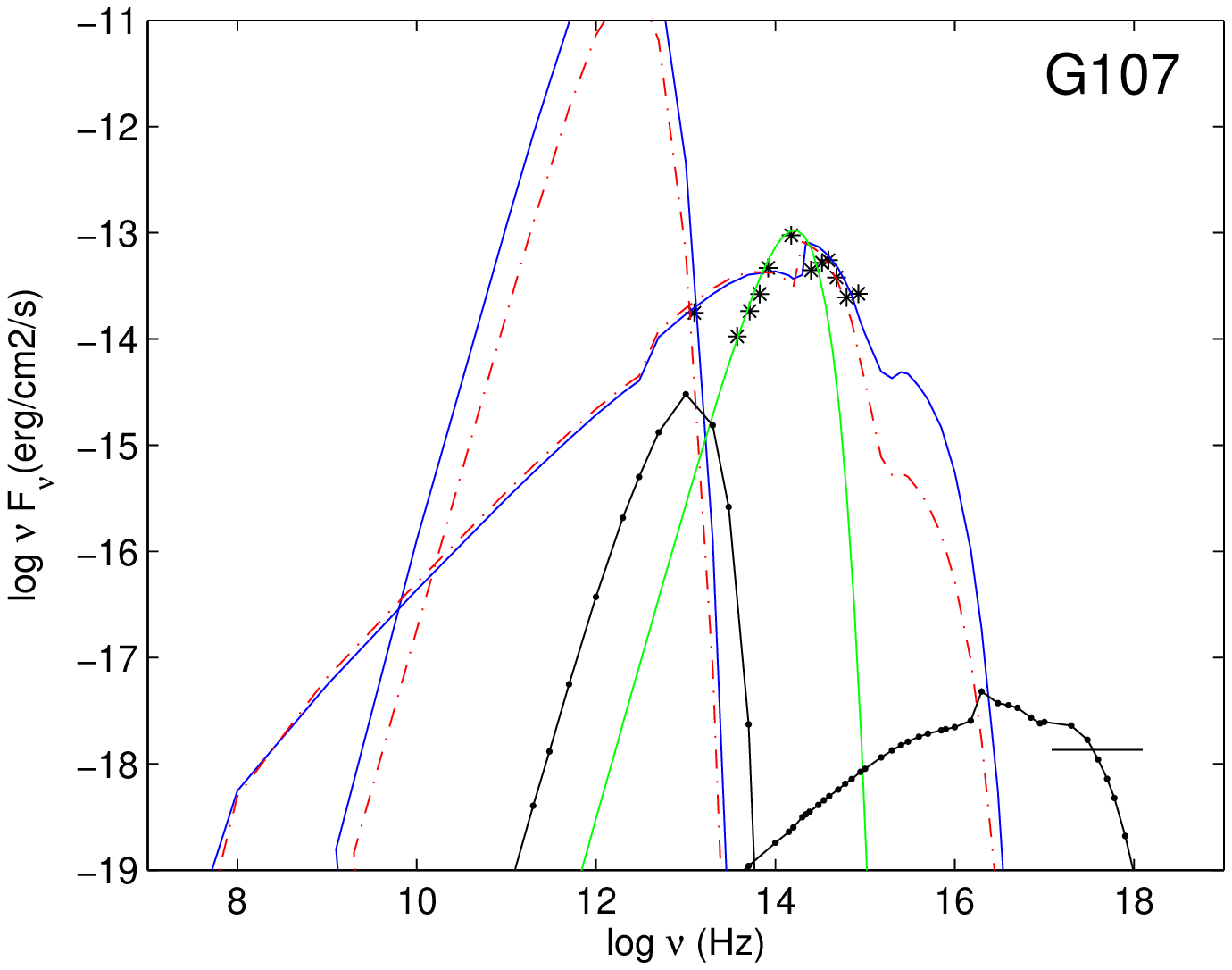}
\caption{The best fit of model calculations to the observed  Ramos Almeida (2013) SB-AGN    
continuum SEDs.
Black asterisks : the data; black horizontal segments : the X-ray flux;  
blue solid lines :  AGN model for G63b;
red dot-dashed lines : SB model for G63b; blue dotted : AGN  
model for G63r; red dotted lines : SB model for G63r
black  lines marked with dots : shock dominated high velocity models ;
green lines : black body radiation from the old star population background ;
magenta  dotted line :  black body flux from the  SB for G90.
}
\end{figure*}

Whether the AGN or the SB dominates in  each galaxy
 was investigated    by  modelling the observed spectra using
   a power-law  and/or a black-body radiation flux,  
respectively. 
The models are  described in Tables 8 and 9.

\begin{table}
\centering
\caption{Physical parameters  in AGN dominated   
models for  Ramos Almeida et al  (2013)}
\begin{tabular}{ccccccccccc} \hline  \hline
\ model &  \Vs   &  \n0 &  $F$       & $D$               & z \\
\       &   \kms & \cm3 & units$^1$  & 10$^{16}$ cm      & - \\  \hline
\ M90   & 700    & 800  &2.        & 0.2                 &1.148\\
\ M26   & 200    & 1300 &2.        & 0.00584             &0.808\\
\ M25   & 250    &1300  &2.        & 0.17                &0.761\\
\ M53   & 350    &1000  &0.8        & 0.37               &0.72\\
\ M107  & 100    &1000  &0.8        & 1.67               &0.67\\
\ M74   & 280    &1600  &2.       & 0.073                &0.551\\
\ M60   & 200    &650   &2.       &0.18                  &0.484\\
\ M63b   & 600    &500   &2.       &0.3         &0.482\\
\ M63r   & 200    &330   &2.       &0.6         &0.482\\
\ M59   & 260    & 600  &0.5       & 0.145     &0.465\\ \hline
\end{tabular}

$^1$ in 10$^{10}$ photons cm$^{-2}$s$^{-1}$eV$^{-1}$ at the Lyman limit

\caption{Physical parameters   in SB  dominated 
models for  Ramos Almeida et al (2013)}
\begin{tabular}{ccccccccccc} \hline  \hline
\ model &  \Vs   &  \n0 &  \Ts  &  $U$   & $D$               & z \\
\       &   \kms & \cm3 & 10$^4$ K    & -  & 10$^{16}$cm    & - \\ \hline
\ Mz90  & 400    & 300  &4.3  & 0.76      & 2.              & 1.148\\
\ Mz26  & 200    & 1300  & 4.  & 0.08     & 0.6            & 0.808\\
\ Mz25  & 200    & 1300  & 4.  & 0.07     & 0.6             & 0.761\\
\ Mz53  & 350    & 1000  & 4.  & 0.11     & 1.              & 0.72\\
\ Mz107 & 100    & 1000   & 4.  & 0.03     &1.              & 0.67 \\
\ Mz74  & 280    & 1000  & 4.  & 3.0      & 100.            & 0.551\\
\ Mz60  & 250    & 750   & 4.  & 0.5      & 0.9             & 0.484\\
\ Mz63b  & 560    & 500   & 4.  &0.5       & 0.6            & 0.482\\
\ Mz63r  & 250    & 400   & 6.  & 60.      & 8              & 0.482\\
\ Mz59  & 280    & 600  & 7.  & 12.       & 9.              & 0.465\\ \hline
\end{tabular}

\caption{Parameters used to model the  Ramos Almeida et al (2013) SEDs}
\begin{tabular}{lccccccl} \hline  \hline
\    & d$^1$   & $\eta_{AGN}$& r$_{AGN}$$^2$ & $\eta_{SB}$ &
r$_{SB}$$^2$ & T$_{bb}$$^3$   \\ \hline
\ G25&3.04  &-12.       & 3.         &-12.5 & 1.7        & 3        \\
\ G26&3.23  & -11.9    & 3.6        &-12.5 & 1.8       & 3      \\
\ G53&2.88  & -12.1    & 2.6        &-12.7 & 1.29       & 2     \\
\ G59&1.86  & -11.6    & 2.95       &-13.5 & 0.33       & 3       \\
\ G60&1.94  & -11.4    & 3.87       & -12.9& 0.69       &  -      \\
\ G63b&1.93  & -11.1    & 5.44       & -12.1& 1.7       &  -       \\
\ G63r&1.93  & -10.9    & 6.8           & -12.5&1.08    &  -       \\
\ G74&2.20  & -11.9    & 2.47       & -12.6& 1.1       & 2       \\
\ G90&4.59  & -13.     & 1.45       & -11.9& 5.1       & 1          \\
\ - &   -     &   -      &   -        &   -  &     -     &0.3     \\
\ G107&2.68 & -11.2    & 6.73       & -11.95& 2.84      & 2        \\ \hline
\end{tabular}

$^1$ in 10$^3$ Mpc ;
$^2$ in kpc ;
$^3$ in 1000 K ;

\end{table}

The  templates  adopted by Ramos Almeida et al in their fig. 2  to explain the  
observed IR data, were
obtained combining the observed  continuum SEDs of
many objects  for  each of the  sample galaxy.
They classified the galaxies as
starburst dominated, starburst contaminated, Seyfert 1, Seyfert 2 and   
normal galaxies.
In Fig. 6 we show  the  SEDs on a scale large  enough ($\nu$
between 10$^7$ and 10$^{19}$ Hz) to contain the X-ray data. 
There is now  common agreement that there is  mutual triggering  
between   SB and   AGN
activity in the galaxies. So in  Fig. 6 diagrams we have included both  
kinds of models.

In Table  10 we summarize the results obtained by modelling the SEDs. In  
column 2 the distances of the galaxies
in Mpc are given.
 The  readjusting factors ($\eta_{AGN}$,  $\eta_{SB}$)  are followed by
the distances of the
emitting nebula from the galaxy centre (r$_{AGN}$), r$_{SB}$), respectively.
The results show that the  distance from the radiation source of the clouds
photoionised by the AGN flux is larger by a factor at least 2 than that  of clouds
photoionised  by the the SB flux. We suggest that the AGN flux reaches the outskirts of the galaxy
while the starburst flux is contained inside smaller regions throughout the ISM.
The radii of the emitting clouds from the active centre
are  similar to those of the   AGN narrow line region (NLR).
Fig. 6 shows that the datum at 24 \mum is always reproduced by dust  
reprocessed radiation.
It is the only datum  at $\lambda$ $>$ 10 \mum
constraining   the dust reradiation peak.
Dust-to-gas ratios by mass $d/g$= 0.0004  were adopted in the  models.
At frequencies referring to wavelengths between $\sim$ 8 and 12 \mum  
the large crater of silicate absorption
at $\lambda$ $\sim$ 10 \mum
and a large absorption by ices and HAC at 6 \mum and 7 \mum
respectively (Spoon et al 2002)  may affect the SED
between  the vertical dashed lines in Fig. 6
(top left panel).

For all the sample galaxies, except for G60, G63 and G90,  bb  
radiation  calculated by a  uniform temperature \Ts ($\sim$ 4$\times$10$^4$K, Table 7)
contributes to the modelling of nearly all the  data.
The old star background temperature ranges between 2000K and 3000 K
for all the galaxies except for G90  where both T = 1000 K and  
T = 300 K  are required to  the best
fit of the NIR and FIR data, respectively (Table 10).
A temperature of 1000 K  can be reached by small dust grains ($<$0.1 \mum, Draine 2003),
while a temperature of 300 K can be achieved  only by 
large grains which could
survive sputtering by a high velocity shock (\Vs$\geq$ 1000-2000 \kms).
We have added in Fig. 6  the X-ray fluxes presented by  Ramos Almeida et al.
Bremsstrahlung from the  high \Vs gas contributes to the
  X-ray emission.
The  high frequency region in the SEDs
is determined by \Vs.
 Therefore we  show in the G60 and G107  diagrams the SED of shock dominated
models calculated  with a high \Vs in order to fit the X-ray data.
In G26 the high shock component is not observed.
Perhaps the broad socket in the profile of  weak  lines is hidden by  
the continuum noise.

For  G53, G59, G63, G74 and G90 the    
SEDs calculated by  \Vs $>$ 200 \kms nicely fit the data,
G25 and G26 are at the limit. These two galaxies show components with  
large FWHM that  translate  to
velocities $>$ 1500 \kms and $>$ 800\kms, respectively  
in the  \Ha and \Hb lines,
indicating  some contribution from the broad line region.
Therefore the spectra  are not well reproduced with the present models.

In particular, if we disentagle the large bump in the near IR range of  
G90, the contribution of
a starburst dominated  bremsstrahlung
with high \Vs and \n0 can be recognized summed up  with the old star  
background black body radiation
  and the reradiation by hot dust. In fact, Ramos Almeida et al classified it as SB  
contaminated, including three different
starburst composite SEDs, one starburst/Seyfert 1 and two starburst/Seyfert 2.
The G63 galaxy shows the contribution of an AGN with
rather high \Vs and
\n0 and the contribution of a SB. 
The other galaxies of the sample show a mixed nature of AGN and SB and  
shocks.

\section{Concluding remarks}

We have modelled the continuum SED   for a sample of GRB hosts presented by  
Kr\"{u}hler et al (2011) at z= 0.543-2.45, of obscured GRB hosts presented by Hunt et al (2014) at 
z= 0.0085-2.086 and of GRB hosts showing high extinction by Sokolov et al (2001)  at z=0.7-1.6.
We  have compared them with  the sample of SLSN hosts observed by Perley et al (2016) at z=0.0395-0.477 
and the SB-AGN by Ramos Almeida et al  (2013) at z=0.465-1.148.
Line and continuum  fluxes are calculated consistently by the  {\sc suma} code which accounts for both
the photoionization from an active source (SB, AGN) and for the shock.
We   confirm that the continuum (bremsstrahlung)  emitted by the same  gaseous clouds which emit 
the line flux contributes to the SED. Therefore, we 
 have selected  from the samples the  objects that
could be  modelled on the basis of the line spectra because modelling the SED
is less constraining.  
As was found in previous works the flux from the background stars and reprocessed radiation 
from dust grains  also shape the continuum SED.

The modelling of the continuum shows that the SED of
 GRB host galaxies in some frequency domains (e.g. the radio) is  similar to that  of
SLSN, SB and AGN even at redshifts higher than local, because it
 follows  the bremsstrahlung trend at relatively low $\nu$. 
Summarizing:

1) The bremsstrahlung  is particularly observable in the radio frenquency range
 if  thermal bremsstrahlung
dominates on the synchrotron radiation created by the Fermi mechanism at the shock front  and in the UV - X-ray
corresponding to gas heated to high ($>10^5$ K) temperatures by the shock.

2) In the IR frequency range the SED is dominated by dust reprocessed radiation even  for very low $d/g$ ($<$0.0004).
Dust reradiation depends on $d/g$ but also on \agr and \Vs. It seldom contributes to the SED at $\nu$$<$ 10$^{10}$ Hz.
The highest frequency limit depends on \Vs. We have found that $d/g$ ranges in obscured GRB hosts between $\leq$0.0001
and 0.032.

3) In the NIR-optical domain the old star population background emerges from the bremsstrahlung. 
Our modelling shows that its contribution is present in  nearly all  the  continua of SLSN hosts with temperatures
  of $\sim$ 6000-8000K,
in most of the SB - AGN galaxies  with temperatures of $\sim$3000K, 
 and in  the GRB hosts
with temperatures of $\sim$4000-8000K. 
It was shown  by previous investigations
(e.g. Contini \& Contini 2003) that the data are nested within bb radiation bumps.
  In  the present analysis  only a few data are available.

4) The direct bb flux corresponding to SB  effective temperatures  are seen in a few objects,
GRB100621A, PTF11hrq and  PTF12dam hosts.

\begin{table}
\caption{The physical parameters  in the different samples}
\begin{tabular}{ccccccccccc} \hline  \hline
\                      &object  &  \Vs     &  \n0    &  \Ts   \\
\                      & -      &   \kms   & \cm3    & 10$^4$ K  \\ \hline
\ Hunt et al           & GRB    & 120-300  & 50-150  & 4-34     \\
\ Kr\"{u}hler et al    & GRB    &130-280   & 80-150  & 5-9      \\
\ Perley et al         &SLSN    & 100-200  & 55-260  & 1-6.5    \\
\ Ramos Almeida et al  &SB      & 100-400  &50-1300  & 4-7      \\
\ Ramos Almeida et al  &AGN     & 100-700  &330-1600 &  -       \\  \hline   
\end{tabular}
\end{table}

The  physical parameters characteristic of the  host galaxies  in the different samples 
are summarized in Table 11.
They indicate that 
the  \Ts which are responsible of gas photoionization  by black body
radiation are    roughly lower for  SLSN hosts  than for the GRB.
The systematic presence of two  radiation sources at $\sim$1-30$\times$10$^4$K and at 3-8$\times$10$^3$K
in most of the objects leads to think 
to some sort of  connection between them, but comparing the SEDs in Figs. 1-6 diagrams  
with those of e.g. symbiotic systems (Angeloni et al 2010),
a different situation is evident.

Even  neglecting the broad line contribution to AGN spectra, the shock velocities   reach
higher values in the SB-AGN galaxies than in the GRB and SLSN hosts.
Preshock densities are definitively higher in the SB-AGN sample   by a factor $\leq$10,
increasing the downstream cooling rate  and reducing the  bremsstrahlung  
for   $\nu$$>$ 10$^{18}$ Hz.

\begin{figure}
\centering
\includegraphics[width=9.6cm]{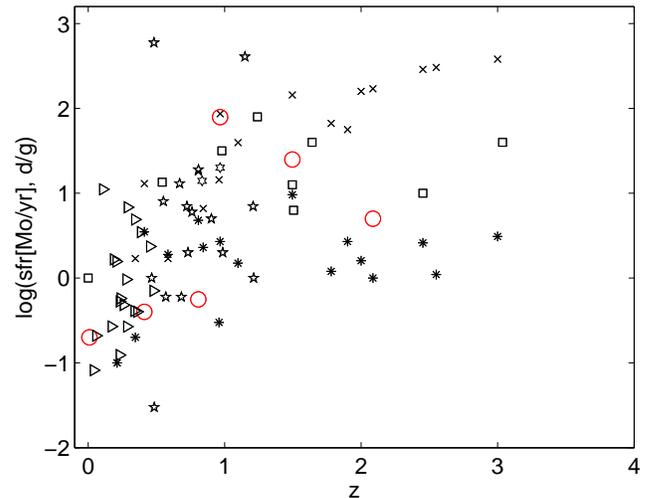}
\caption{Data from Hunt et al (2014)  SFR(UV) (asterisks), 
to the SFR(IR) (cross);  SFR data from Kr\"{u}hler et al (2011) (squares);  
SFR data from Sokolov et al (2001) (hexagons); SFR data from Ramos Almeida et al (2013) (stars)
and from Perley et al (2016) (triangles).
Calculated $d/g$ in units of $(d/g)_0$ (open circles). 
}
\end{figure}

Some objects  deserve a special mention. GRB980703 host extended  clouds are photoionised by  radiation
corresponding to a  relatively high SB effective 
temperature  (\Ts=6.5 10$^4$ K) .  It was found by modelling the continuum SED
that this temperature corresponds to the W-R stars and  leads to the thermal bremsstrahlung which reproduces
the data in the radio range.
Dust reradiation in the GRB980425 host  shows  the lowest  $d/g$ ($<$0.0001).
We report in Fig. 7 the SFR presented by Hunt et al. (2014),  Kr\"{u}hler et al (2011) and Sokolov et al (2011) 
for GRB hosts,
by Perley et al for SLSN hosts and by Ramos Almeida et al for SB-AGN galaxies.
For all of them SFR  roughly increases with z. 
Dust-to-gas ratios   which result  by modelling the observed  SEDs of obscured GRB  hosts,
  follow the SFR  increasing trend.
This suggests that   dust  grains are   destroyed with time towards low z mainly  by sputtering downstream of shock fronts
 throughout the  GRB hosts.

%\section*{Acknowledgements}
%This research  has made use of the NASA/IPAC Extragalactic Database  
%(NED) which is operated by the Jet Propulsion Laboratory, California  
%Institute of Technology, under contract with the National Aeronautics  
%and Space Administration.

% \begin{thebibliography}{99}
\section*{References}

\def\ref{\par\noindent\hangindent 18pt}
%\ref Anders E., Grevesse N. 1989, Geochim. Cosmochim. Acta, 53, 197
\ref Angeloni, R.; Contini, M.; Ciroi, S.; Rafanelli, P.  2010, MNRAS, 402, 2075	
\ref Bell, A.R. 1977, MNRAS, 179, 573
\ref Bruzual, G. \& Charlot, S. 2003, MNRAS, 344, 1000
\ref Contini, M. 2017b, MNRAS, 469,3125
\ref Contini, M. 2017a, MNRAS, 466, 2787
\ref Contini, M. 2016, MNRAS, 460, 3232
\ref Contini, M. 2013, MNRAS, 429, 242
%\ref Contini, M. 2016b, MNRAS, 461, 2374
\ref Contini, M. 2013 ArXiv:1310.3619
\ref Contini, M. \& Viegas, S.M. 2000, ApJ, 535, 721
\ref Contini, M. Viegas, S.M., Prieto, M.A. 2004, MNRAS, 348, 1065
\ref Contini, M, Prieto, M.A., Viegas, S.M. 1998, ApJ,505, 621
\ref Contini, M, Prieto, M.A., Viegas, S.M. 1998, ApJ,492, 511
\ref Contini, M. \& Contini, T. 2003, MNRAS, 342, 299
\ref  Contini, M. \& Contini, T. 2007, AN, 328, 953  
\ref Draine, B.T. 2003 in "The cold Universe", SAAS Fee Advanced Course 32, astro-ph:10304488
\ref Draine, B.T. \& Salpeter, E.E. 1979, ApJ, 231, 438
\ref Draine, B.T. \& Lee, M.M. 1994, ApJ, 285, 89
\ref Dwek, E. \& Cherchneff, I. 2011, ApJ, 727, 63
\ref El\'{i}asd\'{o}ttir, \'{A}. et al 2009, ApJ, 697, 1725 
\ref Galama, T.J. et al  1998, Nature, 395, 670
\ref Graham, J.F., Fruchter, A.S. 2013, ApJ, 774, 119
\ref Grevesse, N., Sauval, A.J. 1998 SSRv, 85, 161
\ref Helou, G. 1986, ApJ, 311, L33
\ref Kr\"{u}hler, J. et al 2011 A\&A, 534, 108
\ref Kr\"{u}hler, J. et al 2015 A\&A, 581, 125
\ref Han, X. H., Hammer, F., Liang, Y. C., Flores, H., Rodrigues, M., Hou, J. L., Wei, J. Y.
 2010, A\&A, 514, 24
\ref Helou, G. 1986, ApJ, 311, L33
\ref Hunt, L.K. et al 2014, A\&A, 565, A112
\ref Iglesias-Paramo et al 2007, ApJ, 670, 279
\ref Levesque, E.M., Berger, E., Kewley, L.J., Bagley, M.M. 2010, ApJ, 139, 694
\ref Michalowski, M.J.  et al 2009  ApJ,693, 347
\ref Michalowski, M.J. , Hjorth, J., Watson, D. 2010  ApJ, 514, A67
\ref Michalowski, M.J. et al 2016 arXiv:1609.01742
\ref Michalowski, M.J. et al 2014, A\&A 562, 70
\ref Nagao, T., Maiolino, R., Marconi, A. 2006, A\&A, 459, 85
\ref Osterbrock, D.E. 1974  in "Astrophysics of Gaseous Nebulae" .  ed. W.H. Freeman and Company , San Francisco
\ref Polletta, M. et al. 2007, ApJ, 663, 81
\ref Perley, D.A. et al 2016, ApJ, 830, 13
\ref Perley, D.A. et al 2013, ApJ, 778, 128
\ref Poonam, C. \& Frail, D.A. 2012, ApJ, 746, 156
\ref Ramos Almeida, C., Rodr\'{i}guez Espinosa, J.M., Acosta-Pulido, J.A. Alonso-Herrero, A.,
P\'{e}rez Garc\'{i}a, A.M, Rodr\'{i}guez-Eugenio, N. 2013, MNRAS, 429, 3449
\ref Sakamoto, T. et al 2011, ApJS, 195, 25
\ref Savaglio, S., Glazerbrook, K., Le Borgne, D. 2009, ApJ, 691, 182
\ref Schlafly, E.F. \& Finkbeiner, D.P. 2011, ApJ, 737, 103 
\ref Silva, L., Granato, G.L., Bressan, A., Danese, L. 1998, ApJ, 509, 103
\ref Sokolov, V.V. et al 2001, A\&A, 372, 438
\ref Sollerman J.; \"{O}stlin, G.; Fynbo, J. P. U.; Hjorth, J.; Fruchter, A.; Pedersen, K.  2005, NewA, 11, 103
\ref Stanek, K.Z. et al 2003, ApJ, 591, L17
\ref Spaans, M. \& Carollo, C.M. 1997 ApJ, 482, L93
\ref Spoon, H. W. W.; Keane, J. V.; Tielens, A. G. G. M.; Lutz, D.; Moorwood, A. F. M.; Laurent, O. 2000, A\&A, 385, 1022
\ref Springel, V., Di Matteo T., Hernsquist, L. 2005, ApJ, 620, L79

\end{document}